\documentclass[review,3p]{elsarticle}

\usepackage{framed,multirow}

\usepackage{amssymb}
\usepackage{latexsym}

\usepackage{url}
\usepackage{xcolor}
\definecolor{newcolor}{rgb}{.8,.349,.1}
\usepackage{bm} % bold
\usepackage{physics}
\usepackage{amsmath}
\usepackage{yhmath}
\newcommand{\volume}{{\ooalign{\hfil$V$\hfil\cr\kern0.08em--\hfil\cr}}}
\usepackage{graphicx}
\usepackage{caption}
\usepackage{subcaption}
\usepackage{float}
\newcommand*{\ldblbrace}{\{\mskip-3mu\{}
\newcommand*{\rdblbrace}{\}\mskip-3mu\}}

\usepackage{cancel}
\usepackage[normalem]{ulem}

\usepackage{tikz}

\usepackage{pifont}% http://ctan.org/pkg/pifont
\usepackage{hyperref}

\journal{Journal of Computational Physics}

\begin{document}

% \verso{J. Brillon and S. Nadarajah}

\begin{frontmatter}

\title{Large Eddy Simulation using Nonlinearly Stable Flux Reconstruction}%
% \tnotetext[tnote1]{This is an example for title footnote coding.}
\author[1]{Julien Brillon}
% \cortext[cor1]{Corresponding author.}
% \author[1]{Given-name2 \snm{Surname2}\fnref{fn1}}
% \fntext[fn1]{This is author footnote for second author.}  
% \author[2]{Given-name3 \snm{Surname3}}
%% Third author's email
\ead{julien.brillon@mail.mcgill.ca}
\author[1]{Siva Nadarajah}
\ead{siva.nadarajah@mcgill.ca}
%\tnoteref{tnote1} % <-- This is an example for title footnote coding.
\address[1]{Department of Mechanical Engineering, McGill University, Montreal, Quebec H3A OC3, Canada}

\begin{abstract}
%%%
The performance of the nonlinearly stable flux reconstruction (NSFR) schemes for resolving subsonic viscous turbulent free-shear flows is investigated. 
% DNS
The schemes are extensively verified for the direct numerical simulation (DNS) of the Taylor-Green Vortex (TGV) problem. 
% ILES
Several under-resolved simulations of the TGV problem are conducted to assess the performance of NSFR for large eddy simulation (LES) that is implicitly filtered (iLES) and fully implicit (ILES). 
%The impact of the choice of flux reconstruction correction parameter $c$,  
Increasing the flux reconstruction correction parameter ensures that NSFR is stable and accurate for ILES while allowing for larger explicit time-steps. 
The entropy-stable schemes implemented with sum-factorization for tensor and Hadamard products are shown to be more cost-effective than classical DG with over-integration. 
The choice of the two-point numerical flux does not impact the solution and the use of standard eddy-viscosity-based sub-grid scale (SGS) models does not yield improvements for the problem considered. 
% pressure dilatation
From the DNS results, the pressure dilatation-based dissipation rate for the nonlinearly stable schemes is consistent with literature when computed from the kinetic energy (KE) budget terms, while spurious oscillations are seen when the term is directly computed. 
The magnitude of these oscillations is significantly lower for a collocated NSFR scheme and are effectively eliminated with the addition of Roe upwind dissipation to the two-point convective numerical flux. 
Therefore, these oscillations are believed to be associated with the treatment of the face terms in nonlinearly stable schemes. 
% oversampling 
It is shown that oversampling the velocity field is necessary for obtaining accurate turbulent kinetic energy (TKE) spectra and eliminates an apparent pile-up of TKE at the smallest resolved scales. 
% DHIT
Lastly, the TKE spectra for a decaying homogeneous isotropic turbulence (DHIT) case are in good agreement with experiment measurements and computational results in the literature.
%%%%
\end{abstract}

\end{frontmatter}

\section{Introduction}
\indent Modern high-performance computing (HPC) technologies have opened the door to high-fidelity turbulent flow simulations of industrial interest in the field of computational fluid dynamics (CFD). For decades, the affordable Reynolds-averaged Navier-Stokes (RANS) approach has been used despite its well-known deficiencies in capturing unsteady flow phenomena such as separated flows present in many engineering applications. 
While the direct numerical simulation (DNS) of the Navier-Stokes equations in which all flow and time scales are resolved remains prohibitively expensive for practical CFD applications, a compromise between RANS and DNS is the large eddy simulation (LES) approach. 
\newline
\indent Higher-order numerical methods are particularly appealing for the large eddy simulation (LES) of turbulent flows, as they show significant potential to improve simulation accuracy at comparable or reduced computational cost \cite{wang2013high}. 
In traditional LES, the unsteady turbulent motion is decomposed into a filtered (resolved) component, representing large-scale energy-containing eddies, and a residual component by use of a low-pass filter. Due to the convenient spectral dissipation properties of high-order methods, this filtering can be performed implicitly by the numerical dissipation of the spatial discretization scheme, effectively reducing the computational costs. This is referred to as implicitly filtered LES (iLES). Furthermore, in traditional LES, the effect of the residual motions on the resolved scales is modelled by a sub-grid scale (SGS) model, representing a residual-stress tensor; which effectively dissipates turbulent kinetic energy (TKE) from the smallest resolved scales to account for the unresolved scales. In general, the truncation error of high-order schemes resembles an additional numerical stress and amounts to an implicit SGS model, having a similar influence as conventional SGS models. This is referred to as implicit LES (ILES) and has the computational advantage of not having to perform traditional filtering operations or evaluations of an SGS model. 
\newline
\indent To handle the complex geometries encountered in industrial applications of CFD, 
numerical methods must be able to handle unstructured grids efficiently. Perhaps the most popular high-order method for unstructured grids is the discontinuous Galerkin (DG) method. The computational domain is discretized into a finite number of elements, and within each element, the solution is locally reconstructed by high-order polynomials of order $p$. The method favourably allows for discontinuities between elements by employing a localized numerical flux at element interfaces, e.g. a Riemann solver. Thus, the required matrix-vector operations can be done efficiently in an element-wise fashion, allowing for the code to be easily parallelized for large simulations. 
First introduced by \citet{reed1973triangular}, classical DG methods operate on the integral form of the conservation equations, requiring numerical quadratures across an element face to compute inter-element fluxes. However, classical DG can be expensive if standard Gaussian quadrature rules are employed. 
The more recent nodal DG (NDG) scheme \cite{hesthaven2007nodal} has gained popularity as it omits quadrature procedures associated with classical DG, representing the solution as nodal values at interpolation points. 
Based on NDG, \citet{kopriva2009implementing} developed the more efficient DG spectral element method (DGSEM) approach. 
Flux reconstruction (FR), is a framework that unifies unstructured high-order methods through the use of correction functions and offers the ability to generate new schemes with favourable accuracy and stability properties \cite{huynh2007flux}. In FR, correction functions of one-degree order higher than the scheme, ``correct'' the flux across the surface, giving the approximate flux both a discontinuous and continuous component \cite{wang2009unifying}. This allows for the largest explicit time-steps amongst high-order methods while maintaining the correct order of accuracy \cite{huynh2007flux,wang2009unifying,jameson2012non,vincent2011new,castonguay2012high}, making FR particularly appealing. A brief history of FR will now be discussed. 
\newline
\indent After being originally proposed by \citet{huynh2007flux}, a generalized approach to FR, known as the lifting collocation penalty (LCP) approach, was developed by \citet{wang2009unifying} in which they considered a ``correction field'' applied to the surface integral, instead of reconstructing the flux across the surface. Due to their similarities, \citet{haga2011high} merged LCP and FR into a common framework: the correction procedure via reconstruction (CPR) approach. \citet{jameson2012non} later proved that the FR surface reconstruction is identical to the CPR correction field, hence `FR' and `CPR' became loosely interchangeable. FR and CPR were further merged into a unified framework of provably linearly stable schemes, in the form of energy stable flux reconstruction (ESFR) schemes, also known as the Vincent-Castonguay-Jameson-Huynh (VCJH) schemes \cite{vincent2011new}. The stability of these ESFR schemes was later extended to linear advection-diffusion problems by \citet{castonguay2013energy}. It was shown that certain ESFR schemes for advection–diffusion problems possess significantly higher explicit time-step limits than DG schemes while maintaining the expected order of accuracy \cite{castonguay2013energy}. In addition, by the use of appropriate correction functions, ESFR schemes can recover other existing high-order schemes, such as the DG method, and the spectral difference (SD) method \cite{kopriva1996conservative,liu2006spectral}. Furthermore, \citet{allaneau2011connections} presented ESFR schemes as a filtered DG scheme in a one-dimensional (1D) setting by relating the ESFR correction functions to a DG lifting operator. This equivalence between ESFR and DG was later extended up to 3D elements and generalized for both weak and strong modal DG formulations by \citet{zwanenburg2016equivalence}. 
\newline
\indent With the ultimate goal of industry-relevant CFD simulations in mind, it is necessary to first assess the performance of these high-order numerical methods for the LES of turbulent flows. Perhaps the most commonly used test case is the Taylor-Green vortex (TGV), which was introduced by \citet{taylor1937mechanism} as a model problem for the analysis of transition and turbulence decay. The viscous TGV problem adopted for compressible flows \cite{drikakis2007simulation} is an ideal flow case as it includes both transitional and decaying turbulent flow regimes, vortex stretching/interaction, and compressibility effects such as pressure dilatation. 
These characteristics make for a challenging problem which has been featured in several international workshops on high-order CFD methods. 
It has been used by many authors for the assessment of various discontinuous high-order methods for the simulation of turbulent flows. 
This includes DGSEM \cite{gassner2013accuracy,diosady2013design,diosady2014dns}, classical DG \cite{chapelier2012final,chapelier2013developpement,chapelier2016development,carton2014assessment,johnsen2013three}, CPR \cite{vermeire2016implicit,navah2020high}, and various ESFR schemes \cite{bull2015simulation}. 
However, using various ESFR schemes, \citet{bull2015simulation} identified that the accuracy of discontinuous high-order methods for the simulation of turbulent flows on coarse meshes required for ILES is limited by the lack of nonlinear stability to handle aliasing errors. 
So far, stabilization in such cases is provided by applying \textit{polynomial de-aliasing}, in which the nonlinearity of the flux is accounted for by increasing the number of numerical quadrature points for the approximation of the integrals \cite{kirby2003aliasing}. However, it has been shown that there are in fact flow cases in which the popular over-integration strategy fails \cite{gassner2016split,moura2017eddy}.\\ 
\indent Currently, the most promising de-aliasing strategy is to use split formulations of the nonlinear terms in the governing equations based on the chain rule. This enables de-aliasing to be built directly into the discretization. 
This was initially introduced by \citet{tadmor1984skew} for finite-volume methods using a weak condition on the numerical flux known as the \textit{Tadmor shuffle condition} to satisfy the entropy condition of \citet{harten1983symmetric}. This was later extended to high-order finite difference stencils by \citet{lefloch2002fully}, and further developed for bounded domains by \citet{fisher2013discretely}, who established the \textit{flux differencing} connection between the summation-by-parts (SBP) framework \cite{fernandez2014review,svard2014review} and the Tadmor shuffle condition to achieve entropy conservation. \citet{gassner2013skew} extended the flux differencing approach to discontinuous high-order methods, which has led to the development of provably stable DG split forms of the compressible Euler equations in three dimensions  \cite{gassner2016split,crean2018entropy,chan2019efficient,fernandez2019entropy}. 
The use of a nonlinearly stable split form DG-scheme for the simulation of viscous turbulent flows was first investigated by \citet{flad2017use}. The authors assessed the performance of a collocated kinetic energy preserving (KEP) split form DG-scheme, identified by \citet{gassner2016split}, for the ILES and iLES of a decaying homogeneous isotropic turbulence (DHIT) case and investigated the role of added Riemann solver and SGS model dissipation. 
The scheme demonstrated superior robustness and the ability to strongly increase fidelity on very coarse grids compared to state-of-the-art ILES DG. 
Adding upwind dissipation proved beneficial for attenuating the pressure dilatation oscillations exhibited by the stable yet dissipation-free KEP DG scheme. 
Demonstrating that the aforementioned accuracy and stability limitations of discontinuous high-order methods on coarse meshes can indeed be overcome with a nonlinearly stable scheme. In addition, the authors reported that the scheme reduced the computational cost by at least a factor of 3 within their code~\cite{flad2017use}. 
Since the work of \citet{flad2017use}, other more complex viscous flow cases have been successfully simulated using nonlinearly stable DG schemes \cite{fernandez2020entropy,parsani2021high,rojas2021robustness,al2021optimized}. 
These promising results encourage further investigations into the role of dissipation mechanisms in the development of LES approaches using provably nonlinearly stable FR schemes given the advantages of FR over DG; such as larger explicit time stepping. 
\newline
\indent In this work, we will investigate the use of the recently introduced framework of generalized provably nonlinearly stable flux reconstruction (NSFR) schemes in split form proposed by \citet{cicchino2022nonlinearly}. 
The NSFR schemes are a high-order finite element method implemented in the Parallel High-Order Library for PDEs (\texttt{PHiLiP}) code \cite{shi2021full}, available at \cite{philip2022code}, which uses \texttt{deal.II} \cite{bangerth2007deal} as its backbone. The NSFR schemes utilize the FR formulation, unifying high-order method framework through the correction parameter $c$, which permits larger allowable explicit time steps than classical DG. The split form scheme utilizes two-point convective fluxes, and is provably nonlinear stable by construction, i.e. entropy stability, yielding a robust scheme that is stable on the coarsest of meshes. 
Contrary to classical DG, the scheme is stable without over-integration for nonlinear problems, and stable without upwind-type dissipation. In the case of turbulence modelling such as iLES, this allows sub-grid scale (SGS) models to function as intended without dissipation interference from upwinding. Furthermore, this allows one to study the effects of different dissipation sources separately. 
The focus of this work is to investigate the performance and accuracy of NSFR schemes for resolved and under-resolved simulations of subsonic shear-free compressible turbulent flows with transition.  
First, an extensive verification of the scheme for the direct numerical simulation (DNS) of the viscous TGV flow case is presented and compared to classical strong DG. Next, the performance of NSFR schemes for iLES and ILES approaches is assessed. 
The accuracy and computational cost of ILES using NSFR schemes vs a strong DG approach are presented for several different degrees of freedom (DOF). 
The impact of different dissipation mechanisms is studied independently for the stable yet dissipation-free baseline NSFR scheme. This includes the choice of flux nodes (i.e. collocated vs uncollocated schemes), the FR correction parameter $c$, different two-point numerical fluxes, and added dissipation from different Riemann solvers and modern SGS models. 
In addition, the effects of oversampling the velocity field for computing the TKE spectra are highlighted.
Lastly, the NSFR scheme is demonstrated for the simulation of a decaying homogeneous isotropic turbulence (DHIT) test case.
To the author's best knowledge, this is the first study on the use of a provably entropy conserving and stable high-order flux reconstruction method for the simulation of viscous turbulent flows. 

\section{Methodology}
\subsection{Compressible Navier-Stokes Equations}\label{sec:navier_stokes_equations}
We consider the compressible Navier-Stokes equations expressed in conservative form for a three-dimensional domain: 
\begin{equation}\label{eq:navier_stokes}
    \pdv{\mathbf{U}}{t} + \pdv{\mathbf{F}^{c}_{i}}{x_{i}} -\pdv{\mathbf{F}^{v}_{i}}{x_{i}} = \bm{0},
\end{equation}
where $\mathbf{U}=[\rho,\rho \text{v}_{1}, \rho \text{v}_{2}, \rho \text{v}_{3}, E]^T$ is the vector of conserved quantities: density, $x$-momentum, $y$-momentum, $z$-momentum, and total energy. The convective $\mathbf{F}^{c}$ and viscous $\mathbf{F}^{v}$ fluxes are given by:
\begin{equation}\label{eq:navier_stokes_convective_and_viscous_fluxes}
\mathbf{F}_{i}^{c}\equiv\mathbf{F}_{i}^{c}\left(\mathbf{U}\right) = 
\begin{bmatrix}
\rho \text{v}_{i}\\ 
\rho \text{v}_{1}\text{v}_{i} + P\delta_{1i}\\ 
\rho \text{v}_{2}\text{v}_{i} + P\delta_{2i}\\ 
\rho \text{v}_{3}\text{v}_{i} + P\delta_{3i}\\ 
\left(E + P\right)\text{v}_{i}
\end{bmatrix},
\quad 
\mathbf{F}_{i}^{v}\equiv\mathbf{F}_{i}^{v}\left(\mathbf{U},\grad\mathbf{U}\right) = 
\begin{bmatrix}
0 \\
\tau_{1i} \\ 
\tau_{2i} \\ 
\tau_{3i} \\ 
\tau_{ij}\text{v}_{j}-q_{i}
\end{bmatrix},
\end{equation}
where $i=1,2,3$ denotes the Cartesian components $(x,y,z)$ respectively, i.e. $\bm{x}=\left[x_{1},x_{2},x_{3}\right]\equiv\left[x,y,z\right]$, while $\text{v}_{i}$ denotes the $i^{\text{th}}$ component of the velocity vector, $\mathbf{v}=\left[\text{v}_{1},\text{v}_{2},\text{v}_{3}\right]$. The viscous stress tensor $\tau_{ij}$ is given by:
\begin{equation}
    \tau_{ij} = \mu\left(\pdv{\text{v}_{i}}{x_{j}}+\pdv{\text{v}_{j}}{x_{i}}\right)+\lambda\pdv{\text{v}_{k}}{x_{k}}\delta_{ij},
\end{equation}
where the second viscosity coefficient $\lambda=-\frac{2}{3}\mu$ by Stokes' hypothesis \cite{stokes2007theories}, and $\mu$ is the dynamic viscosity coefficient. In this work, $\mu=\mu(T)$ is determined by Sutherland's law \cite{sutherland1893lii}:
\begin{equation}\label{eq:sutherlands}
    \mu(T) = \mu_{\text{ref}}\left(\frac{T}{T_{\text{ref}}}\right)^{\frac{3}{2}}\frac{T_{\text{ref}}+T_{s}}{T+T_{s}},
\end{equation}
where $T_{\text{ref}}$ is the reference viscosity, $\mu_{\text{ref}}$ is the corresponding reference viscosity at $T_{\text{ref}}$, and $T_{s}$ is the Sutherland temperature. Since the strain-rate tensor is given by:
\begin{equation}
    S_{ij}=\frac{1}{2}\left(\pdv{\text{v}_{i}}{x_{j}}+\pdv{\text{v}_{j}}{x_{i}}\right),
\end{equation}
we can write the stress-tensor in terms of the strain-rate tensor:
\begin{equation}
    \tau_{ij} = 2\mu\left(S_{ij}-\frac{1}{3}S_{kk}\delta_{ij}\right).
\end{equation}
The heat-flux vector $q_{i}$ is given by:
\begin{equation}
    q_{i} = -\kappa\pdv{T}{x_{i}},
\end{equation}
where $T$ is the temperature, and $\kappa$ is the thermal conductivity:
\begin{equation}
    \kappa = \frac{c_{p}\mu}{\text{Pr}} \equiv \frac{\gamma R\mu}{\left(\gamma-1\right)\text{Pr}},
\end{equation}
where $\text{Pr}$ is the Prandtl number. To close the system of equations, pressure $P$ is determined from the ideal gas law:
\begin{equation}
    P = \rho RT = \left(\gamma-1\right)\left(E-\frac{1}{2}\rho \text{v}_{k}\text{v}_{k}\right),
\end{equation}
where the total energy $E$ is given by:
\begin{equation}
    E = \rho e + \frac{1}{2}\rho \text{v}_{k}\text{v}_{k} \equiv \rho c_{v}T + \frac{1}{2}\rho \text{v}_{k}\text{v}_{k} \equiv \frac{\rho RT}{\gamma-1} + \frac{1}{2}\rho \text{v}_{k}\text{v}_{k} \equiv \frac{P}{\gamma-1} + \frac{1}{2}\rho \text{v}_{k}\text{v}_{k},
\end{equation}
where $e$ is the specific internal energy. 

\subsection{Dimensionless Compressible Navier-Stokes Equations}\label{sec:dimensionless_gov_eqxns}
In the numerical implementation, Eq.(\ref{eq:navier_stokes}) is non-dimensionalized by the density $\rho_{\infty}$, characteristic length $L$, velocity $V_{\infty}$, time $L/V_{\infty}$, temperature $T_{\infty}$, pressure $\rho_{\infty}V_{\infty}^{2}$, dynamic viscosity $\mu_{\infty}$, and speed of sound $c_{\infty}=\sqrt{\gamma RT_{\infty}}$. Thus, we introduce the non-dimensional quantities denoted by superscript $(\cdot)^{*}$ as: 
\begin{equation}
    \rho^{*} = \frac{\rho}{\rho_{\infty}},\quad x_{i}^{*}=\frac{x_{i}}{L},\quad 
    \text{v}_{i}^{*}=\frac{\text{v}_{i}}{V_{\infty}},\quad t^{*}=\frac{t}{L/V_{\infty}},\quad T^{*}=\frac{T}{T_{\infty}}, \quad 
    P^{*}=\frac{P}{\rho_{\infty}V_{\infty}^{2}}, \quad  \mu^{*}=\frac{\mu}{\mu_{\infty}}, \quad c^{*}=\frac{c}{c_{\infty}}.
\end{equation}
Substituting the above into Eq.(\ref{eq:navier_stokes}) and omitting all the asterisks for clarity yields the non-dimensionalized Navier-Stokes equations: 
\begin{equation}\label{eq:navier_stokes_nondim}
    \pdv{\mathbf{U}}{t} + \pdv{\mathbf{F}^{c}_{i}}{x_{i}} -\pdv{\mathbf{F}^{v}_{i}}{x_{i}} = \bm{0},
\end{equation}
where $\mathbf{U}=[\rho,\rho \text{v}_{1}, \rho \text{v}_{2}, \rho \text{v}_{3}, E]^T$ is the vector of non-dimensional conserved quantities: density, $x$-momentum, $y$-momentum, $z$-momentum, and total energy. The non-dimensional convective $\mathbf{F}^{c}$ and viscous $\mathbf{F}^{v}$ fluxes take the same form as Eq.(\ref{eq:navier_stokes_convective_and_viscous_fluxes}), where the dimensionless viscous stress-tensor $\tau_{ij}$ is given by:
\begin{equation}\label{eq:dimensionless_viscous_stress_tensor}
    \tau_{ij} = \frac{2\mu}{\text{Re}_{\infty}}\left(S_{ij}-\frac{1}{3}S_{kk}\delta_{ij}\right)
\end{equation}
and the dimensionless heat-flux vector $q_{i}$ is given by:
\begin{equation}
    q_{i} = -\left(\left(\gamma-1\right)\text{M}_{\infty}^{2}\text{Re}_{\infty}\text{Pr}\right)^{-1}\mu\pdv{T}{x_{i}}.
\end{equation}
The above dimensionless Navier-Stokes equations are described by the Prandtl number $\text{Pr}$, Reynolds number $\text{Re}_{\infty}$, and Mach number $\text{M}_{\infty}$:
\begin{equation}\label{eq:nondim_parameters}
    \text{Pr} = \frac{\mu c_{p}}{\kappa} \equiv \frac{\mu \gamma R}{\kappa \left(\gamma-1\right)},
    \quad 
    \text{Re}_{\infty} = \frac{\rho_{\infty}V_{\infty}L}{\mu_{\infty}},
    \quad 
    \text{M}_{\infty} = \frac{V_{\infty}}{c_{\infty}}=\frac{V_{\infty}}{\sqrt{\gamma RT_{\infty}}}.
\end{equation}
Furthermore, the dimensionless equation of state becomes:
\begin{equation}
    P = \frac{\rho T}{\gamma \text{M}_{\infty}^{2}}.
\end{equation}
The dimensionless Sutherland's law becomes:
\begin{equation}
    \mu = \left(\frac{1 + T_{s}^{*}}{T+T_{s}^{*}}\right)T^{\frac{3}{2}}, \quad T_{s}^{*} = \frac{T_{s}}{T_{\infty}},
\end{equation}
where $T_{s}=110.4\text{K}$. In this work, we chose $T_{\infty}=T_{\text{ref}}=273.15\text{K}$ and $\mu_{\infty}=\mu_{\text{ref}}=1.716\times10^{-5}~\text{kg}\cdot\text{m}^{-1}\text{s}^{-1}$ for consistency with Sutherland's law, and chose the standard values for ideal air: $\text{Pr}=0.71$, $\gamma=1.4$.

\subsection{Implicitly Filtered LES Equations}
For iLES, the dissipation by the sub-grid (i.e. unresolved) scales are accounted for by using a sub-grid scale (SGS) model. This effectively adds an additional dissipative flux (or ``SGS flux'') $\mathbf{F}^{\text{sgs}}_{i}$ to the Navier-Stokes equations:
\begin{equation}\label{eq:iLES_equation}
    \frac{\partial \mathbf{U}}{\partial t} + \pdv{\mathbf{F}^{c}_{i}}{x_{i}} - \pdv{\mathbf{F}^{v}_{i}}{x_{i}} -\pdv{\mathbf{F}^{\text{sgs}}_{i}}{x_{i}} = \bm{0},
\end{equation}
where the SGS flux is given by:
\begin{equation}\label{eq:iLES_SGS_flux}
 \mathbf{F}^{\text{sgs}}_{i}\left(\mathbf{U},\grad\mathbf{U}\right) = \begin{bmatrix}0 & \tau^{\text{sgs}}_{1i} & \tau^{\text{sgs}}_{2i} & \tau^{\text{sgs}}_{3i} & \tau^{\text{sgs}}_{ij}\text{v}_{j}-q^{\text{sgs}}_{i}\end{bmatrix}^{T},
\end{equation}
where $i=1,2,3$ denotes the Cartesian components $(x,y,z)$ respectively. Under the eddy-viscosity assumption \cite{pope2001turbulent}, the SGS stress tensor $\tau^{\text{sgs}}_{ij}$ is written as:
\begin{equation}\label{eq:SGS_stress_tensor}
    \tau_{ij}^{\text{sgs}} = 2\rho\nu_{t}\left(S_{ij} - \frac{1}{3}S_{kk}\delta_{ij}\right),
\end{equation}
where $\nu_{t}$ is the turbulent kinematic eddy viscosity. 
Thus, the resultant SGS heat-flux, $q^{\text{sgs}}_{i}$, is expressed as:
\begin{equation}
    q^{\text{sgs}}_{i} = -\kappa_{t}\pdv{T}{x_{i}},
\end{equation}
where the turbulent heat conductivity $\kappa_{t}$ is obtained from $\nu_{t}$:
\begin{equation}
    \kappa_{t} = \frac{c_{p}\rho\nu_{t}}{\text{Pr}_{t}},
\end{equation}
where $\text{Pr}_{t}$ is the turbulent Prandtl number. To close the above set of equations for iLES, $\nu_{t}$ is modelled by an SGS model. 

\subsection{Dimensionless Implicitly Filtered LES Equations}
Applying the methodology of section \ref{sec:dimensionless_gov_eqxns} to Eq.(\ref{eq:iLES_equation}) and omitting all the asterisks for clarity yields the non-dimensionalized iLES equations:
\begin{equation}
    \frac{\partial \mathbf{U}}{\partial t} + \pdv{\mathbf{F}^{c}_{i}}{x_{i}} - \pdv{\mathbf{F}^{v}_{i}}{x_{i}} -\pdv{\mathbf{F}^{\text{sgs}}_{i}}{x_{i}} = \bm{0},
\end{equation}
where $\mathbf{F}^{c}_{i}$ and $\mathbf{F}^{v}_{i}$ take the same form as Eq.(\ref{eq:navier_stokes_convective_and_viscous_fluxes}), while the non-dimensional SGS flux takes the same form as Eq.(\ref{eq:iLES_SGS_flux}). 
The non-dimensional SGS stress tensor $\tau^{\text{sgs}}_{ij}$ has the same form as Eq.(\ref{eq:SGS_stress_tensor}), and 
the non-dimensional SGS heat-flux $q^{\text{sgs}}_{i}$ is expressed as:
\begin{equation}
    q^{\text{sgs}}_{i} = -\left(\left(\gamma-1\right)\text{M}_{\infty}^{2}\text{Pr}_{t}\right)^{-1}\rho\nu_{t}\pdv{T}{x_{i}}.
\end{equation}
The non-dimensional turbulent kinematic eddy viscosity $\nu_{t}$ takes the same form as the dimensional form, unless otherwise specified in this work. In this work, $\text{Pr}_{t}$ is assumed to be constant and equal to 0.6 as done in \cite{plata2019performance}.

\subsection{Dissipation Rate Components}\label{subsec:dissipation_rate_components}
\subsubsection{Compressible Flow}
Following \citet{schranner2015assessing}, we can split the transport of $E$ given by Eq.(\ref{eq:navier_stokes}) into the transport of internal energy $E_{\text{int.}}=\rho e$ and kinetic energy $E_{\text{kin.}}=\rho \text{v}_{k}\text{v}_{k}/2$: 
\begin{align}
    \pdv{E_{\text{kin.}}}{t} + \pdv{E_{\text{kin.}}\text{v}_{i}}{x_{i}} + \text{v}_{i}\pdv{P}{x_{i}}-\text{v}_{j}\pdv{\tau_{ij}}{x_{i}} & = 0 \label{eq:kinetic_energy_transport} \\
    \pdv{E_{\text{int.}}}{t} + \pdv{E_{\text{int.}}\text{v}_{i}}{x_{i}} + P\pdv{\text{v}_{i}}{x_{i}}-\tau_{ij}\pdv{\text{v}_{j}}{x_{i}} + \pdv{q_{i}}{x_{i}} & = 0 \label{eq:internal_energy_transport}
\end{align}
e.g. $E=E_{\text{int.}}+E_{\text{kin.}}$, rewriting Eq.(\ref{eq:kinetic_energy_transport}) as:
\begin{equation}
    \pdv{E_{\text{kin.}}}{t} + \pdv{E_{\text{kin.}}\text{v}_{i}}{x_{i}} + \pdv{P\text{v}_{i}}{x_{i}}-P\pdv{\text{v}_{i}}{x_{i}}-\pdv{\tau_{ij}\text{v}_{j}}{x_{i}}+\tau_{ij}\pdv{\text{v}_{j}}{x_{i}} = 0,
\end{equation}
integrating over entire domain and applying the divergence theorem, we have:
\begin{equation}
    \pdv{}{t}\iiint_{\volume} E_{\text{kin.}}d\volume + \iint_{A}E_{\text{kin.}}\text{v}_{i}\hat{n}_{i}dA + \iint_{A}P\text{v}_{i}\hat{n}_{i}dA-\iiint_{\volume}P\pdv{\text{v}_{i}}{x_{i}}d\volume-\iint_{A}\tau_{ij}\text{v}_{j}\hat{n}_{i}dA+\iiint_{\volume}\tau_{ij}\pdv{\text{v}_{j}}{x_{i}}d\volume = 0.
\end{equation}
For periodic boundary conditions, the flux terms (i.e. area integrals) vanish \cite{castiglioni2015characterizaion},
\begin{equation}
    \pdv{}{t}\iiint_{\volume} E_{\text{kin.}}d\volume -\iiint_{\volume}P\pdv{\text{v}_{i}}{x_{i}}d\volume +\iiint_{\volume}\tau_{ij}\pdv{\text{v}_{j}}{x_{i}}d\volume = 0.
\end{equation}
Dividing by the volume of the domain $\volume$, we can write: 
\begin{equation}
    \pdv{E_{K}}{t} = 
    \left\langle P\pdv{\text{v}_{i}}{x_{i}}\right\rangle - \left\langle \tau_{ij}\pdv{\text{v}_{j}}{x_{i}}\right\rangle,
\end{equation}
where 
\begin{equation}\label{eq:volumetric_averaged_integrated_kinetic_energy}
    E_{K}=\left\langle E_{\text{kin.}}\right\rangle
\end{equation}
and $\left\langle \cdot\right\rangle$ denotes volumetric averaging over the entire domain, i.e.:
\begin{equation}
    \left\langle \cdot\right\rangle = \frac{1}{\volume}\iiint_{\volume} \left(\cdot\right)d\volume.
\end{equation}
Similarly, from Eq.(\ref{eq:internal_energy_transport}) we can write:
\begin{equation}
    \pdv{E_{I}}{t} = 
    -\left\langle P\pdv{\text{v}_{i}}{x_{i}}\right\rangle + \left\langle \tau_{ij}\pdv{\text{v}_{j}}{x_{i}}\right\rangle,
\end{equation}
where $E_{I}=\left\langle E_{\text{int.}}\right\rangle$. We can write the dimensionless viscous dissipation rate term as:
\begin{equation}\label{eq:viscous_dissipation_strain_rate_terms}
    \left\langle \tau_{ij}\pdv{\text{v}_{j}}{x_{i}}\right\rangle = \frac{2}{\text{Re}_{\infty}}\left\langle \mu S_{ij}S_{ij}\right\rangle -\frac{2}{3\text{Re}_{\infty}}\left\langle \mu\left(\pdv{\text{v}_{k}}{x_{k}}\right)^{2} \right\rangle.
\end{equation}
Alternatively, we can express the viscous dissipation term as\footnote{See \ref{appendix:derivation_viscous_dissipation_rate_terms_compressible} for proof of Eq.(\ref{eq:viscous_dissipation_strain_rate_terms}-\ref{eq:viscous_dissipation_deviatoric_strain_rate_terms}).}:
\begin{equation}\label{eq:viscous_dissipation_deviatoric_strain_rate_terms}
    \left\langle \tau_{ij}\pdv{\text{v}_{j}}{x_{i}}\right\rangle = \frac{2}{\text{Re}_{\infty}}\left\langle \mu S_{ij}^{d}S_{ij}^{d}\right\rangle, 
\end{equation}
where the deviatoric strain-rate tensor $S_{ij}^{d}$ is given by:
\begin{equation}\label{eq:deviatoric_strain_rate_tensor}
    S_{ij}^{d}=S_{ij}-\frac{1}{3}S_{kk}\delta_{ij}.
\end{equation}
Thus, we have: 
\begin{equation}\label{eq:dissipation_rate_components_compressible_flow}
    \pdv{E_{K}}{t} = 
    \underbrace{\left\langle P\pdv{\text{v}_{i}}{x_{i}}\right\rangle}_{\displaystyle\varepsilon_{P}} - \underbrace{\frac{2}{\text{Re}_{\infty}}\left\langle \mu S_{ij}^{d}S_{ij}^{d}\right\rangle}_{\displaystyle\varepsilon_{v}},
\end{equation}
where $\varepsilon_{P}$ and $\varepsilon_{v}$ represent the pressure and viscous dissipation rate components, respectively.

\subsubsection{Incompressible Flow}
For an incompressible flow, density is constant therefore the divergence of the velocity field is zero, i.e. $S_{kk}=0$, thus $S_{ij}^{d}=S_{ij}$ and $\varepsilon_{P}=0$. Thus, for an incompressible flow with \textit{constant viscosity} ($\mu^{*}=1$), and periodic boundary conditions (vanishing surface integrals), Eq.(\ref{eq:dissipation_rate_components_compressible_flow}) becomes\footnote{See \ref{appendix:derivation_viscous_dissipation_rate_terms_incompressible} for proof of Eq.(\ref{eq:incompressible_vorticity_identity}).}:
\begin{align}
    \pdv{E_{K}}{t} & = - \frac{2}{\text{Re}_{\infty}}\left\langle \mu S_{ij}S_{ij}\right\rangle \\
    & = - \frac{1}{\text{Re}_{\infty}}\left\langle \mu\omega_{k}\omega_{k} + 2\mu\pdv{}{x_{j}}\left(\text{v}_{i}\pdv{\text{v}_{j}}{x_{i}}\right)\right\rangle \label{eq:incompressible_vorticity_identity} \\
    & = - \frac{\mu}{\text{Re}_{\infty}}\left\langle \omega_{k}\omega_{k}\right\rangle - \frac{2\mu}{\text{Re}_{\infty}}\cancelto{0}{\iint_{A}\left(\text{v}_{i}\pdv{\text{v}_{j}}{x_{i}}\right)\hat{n}_{k}dA}\\
    & = - \frac{\mu}{\text{Re}_{\infty}}\left\langle \omega_{k}\omega_{k}\right\rangle\label{eq:numerical_viscosity_origin}\\
    & = - \frac{1}{\text{Re}_{\infty}}\left\langle \omega_{k}\omega_{k}\right\rangle,
\end{align}
where the vorticity $\omega_{k}$ is given by:
\begin{equation}
\omega_{k}=\epsilon_{ijk}\pdv{\text{v}_{j}}{x_{i}}.
\end{equation}
Omitting all asterisks for clarity, the non-dimensional volume-averaged enstrophy, $\tilde{\zeta}$ is:
\begin{equation}\label{eq:integrated_enstrophy_proper}
    \tilde{\zeta} = \frac{1}{2}\left\langle\omega_{k}\omega_{k}\right\rangle.
\end{equation}
Substituting $\tilde{\zeta}$, we have:
\begin{align}\label{eq:KE_budget_incompressible_flow}
    \pdv{E_{K}}{t} & = - \underbrace{\frac{2\tilde{\zeta}}{\text{Re}_{\infty}}}_{\displaystyle\varepsilon_{\zeta}},
\end{align}
where $\varepsilon_{\zeta}$ is the viscous dissipation rate based on the enstrophy.

\subsection{Sub-Grid Scale Modelling}
Throughout the various SGS models considered in this work, the filter width for local element $e$, denoted as $\Delta_{e}$, is computed as the cube root of the local element volume $\volume_{e}^{1/3}$ divided by local number of solution points $(\text{p}_{e}+1)$ \cite{duan2024calibrating}, where $\text{p}_{e}$ is the local polynomial order of the solution in element $e$, i.e.:
\begin{equation}\label{eq:filter_width}
    \Delta_{e} = \Delta_{e}\left(\text{p}_{e}\right) = \frac{\volume_{e}^{1/3}}{\text{p}_{e}+1}.
\end{equation}
In this work, all elements have the same volume and local polynomial order. Hence, the subscript $e$ will be dropped from the filter width $\Delta$ in the following sections for simplicity.

\subsubsection{Smagorinsky Model}
The standard Smagorinsky eddy-viscosity model is constructed by assuming that the small scales are in equilibrium, so that energy production and dissipation are in balance \cite{smagorinsky1963general}. It is expressed as:
\begin{equation}\label{eq:standard_smagorinsky_model}
\nu_{t} = \left(C_{S}\Delta\right)^{2}\sqrt{2S_{ij}S_{ij}},
\end{equation}
where $C_{S}$ is the Smagorinsky coefficient; set to $0.10$ in this work based on \citet{deardorff1970numerical}. 

\subsubsection{Shear-Improved Smagorinsky Model}
The shear-improved Smagorinsky model \cite{leveque2007shear} modifies the well-established Smagorinsky model by subtracting the magnitude of the mean strain-rate tensor $\overline{S}_{ij}$ from the magnitude of the instantaneous strain-rate tensor. It is expressed as:
\begin{equation}
\nu_{t} = \left(C_{S}\Delta\right)^{2}\left(\sqrt{2S_{ij}S_{ij}}-\sqrt{2\overline{S}_{ij}\overline{S}_{ij}}\right).
\end{equation}
In this work, the mean strain-rate tensor is computed as:
\begin{equation}
    \overline{S}_{ij} = \left\langle S_{ij}\right\rangle_{e},
\end{equation}
where $\left\langle \cdot\right\rangle_{e}$ denotes volumetric averaging over the domain of local element $e$, i.e.:
\begin{equation}\label{eq:element_volume_average}
    \left\langle \cdot\right\rangle_{e} = \frac{1}{\volume_{e}}\iiint_{\volume_{e}} \left(\cdot\right)d\volume_{e}.
\end{equation}

\subsubsection{High-Pass Filtered SGS}\label{subsubsec:hpf}
So far, all the above SGS models can be viewed as functions of the conservative solution vector $\bm{U}$, gradient of the conservative solution vector $\grad\bm{U}$, and $\text{p}_{e}$ from Eq.(\ref{eq:filter_width}), i.e.:
\begin{equation}
    \nu_{t} \equiv \nu_{t}\left(\text{p}_{e},\mathbf{U},\grad\mathbf{U}\right).
\end{equation}
In the work of \citet{chapelier2016development} on the variational multi-scale (VMS) approach in the context of DG discretizations, a separation of resolved scales is accomplished by splitting the polynomial bases into two parts: (1) polynomial degrees less than $\text{p}^{L}_{e}+1$ representing large scales, and (2) the remaining representing the smallest resolved scales. Based on this separation of scales, we obtain the high-pass filtered (HPF) velocity field in this work by the following sequence of steps:
\begin{enumerate}
    \item Convert $\bm{U}$ and $\grad\bm{U}$ to their primitive variable forms $\bm{W}$ and $\grad\bm{W}$ where $\mathbf{W}=[W_{0},W_{1},W_{2},W_{3},W_{4}]^T=[\rho,\text{v}_{1},\text{v}_{2},\text{v}_{3},P]^T$ is the vector of non-dimensional primitive quantities: density, $x$-velocity, $y$-velocity, $z$-velocity, and pressure.
    \item Project the $\bm{W}$ and $\grad\bm{W}$ from Lagrange solution nodes to Legendre solution nodes, obtaining $\bm{W}^{\text{Leg.}}$ and $\grad\bm{W}^{\text{Leg.}}$.
    \item For the velocities only (i.e. $W_{i}^{\text{Leg.}}$ and $\pdv{W_{i}^{\text{Leg.}}}{x_{j}}$ where $i,j=1,2,3$): Truncate (i.e. filter out) modes in the large scale space, i.e. those with an index less than $p^{{L}}_{e}+1$ for element $e$ (indexing starts from zero), to obtain only the modes in the small-scale space.
    \item Reconstruct the solution and corresponding gradient on Legendre nodes, obtaining the high-pass filtered primitive solution and gradient on Legendre nodes, i.e. $\widehat{\bm{W}}^{\text{Leg.}}$ and $\widehat{\grad\bm{W}}^{\text{Leg.}}$
    \item Convert the modally-filtered primitive solution $\widehat{\bm{W}}^{\text{Leg.}}$ and corresponding gradient $\widehat{\grad\bm{W}}^{\text{Leg.}}$ to conservative form $\widehat{\bm{U}}^{\text{Leg.}}$ and $\widehat{\grad\bm{U}}^{\text{Leg.}}$
    \item Project both the modally-filtered conservative solution $\widehat{\bm{U}}^{\text{Leg.}}$ and gradient $\widehat{\grad\bm{U}}^{\text{Leg.}}$ from Legendre nodes to Lagrange nodes, obtaining the high-pass filtered solution $\widehat{\bm{U}}$ and gradient $\widehat{\grad\bm{U}}$ on the Lagrange solution nodes
\end{enumerate}
The above operation on the solution will be denoted by $\widehat{\left(\cdot\right)}$. Thus, the high-pass filtered solution and gradient is $\widehat{\bm{U}}$ and $\widehat{\grad\bm{U}}$, respectively. Finally, an HPF SGS model will be denoted as:
\begin{equation}
    \nu^{\text{HPF}}_{t} = \nu_{t}\left(\text{p}_{e},\widehat{\bm{U}},\widehat{\grad\bm{U}}\right).
\end{equation}
In this work we consider both an HPF Smagorinsky model and HPF shear-improved Smagorinsky model. 

\subsubsection{Dynamic Smagorinsky Model}
The dynamic Smagorinsky model is expressed as:
\begin{equation}
\nu_{t} = C_{D}\Delta^{2}\sqrt{2S_{ij}S_{ij}},
\end{equation}
where the dynamic Smagorinsky coefficient $C_{D}$ for element $e$ is given by \cite{lilly1992proposed,blazek2001cfd}:
\begin{equation}
    C_{D} = -\frac{1}{2}\frac{\left\langle L_{ij}M_{ij}\right\rangle_{e}}{\left\langle M_{ij}M_{ij}\right\rangle_{e}},
\end{equation}
where the operation $\left\langle \cdot\right\rangle_{e}$ is given by Eq.(\ref{eq:element_volume_average}). The $L_{ij}$ and $M_{ij}$ tensors are given by \cite{lilly1992proposed,blazek2001cfd,flad2017use}:% similar to \cite{spyropoulos1996evaluation} without density
\begin{equation}\label{eq:L_matrix}
    L_{ij} = \widehat{\text{v}_{i}\text{v}_{j}} - \widehat{\text{v}}_{i}\widehat{\text{v}}_{j},
\end{equation}
\begin{equation}\label{eq:M_matrix}
    M_{ij} = \Delta^{2}\widehat{m_{ij}\left(\mathbf{U},\grad\mathbf{U}\right)} - \widehat{\Delta}^{2}m_{ij}\left(\widehat{\mathbf{U}},\widehat{\grad\mathbf{U}}\right),
\end{equation}
where $\widehat{\left(\cdot\right)}$ denotes the application of the filter operation similar to that detailed in Section \ref{subsubsec:hpf}, and
\begin{equation}
    m_{ij}\left(\mathbf{U},\grad\mathbf{U}\right) = \sqrt{2S_{ij}\left(\mathbf{U},\grad\mathbf{U}\right)S_{ij}\left(\mathbf{U},\grad\mathbf{U}\right)}\left(S_{ij}\left(\mathbf{U},\grad\mathbf{U}\right)-\frac{1}{3}S_{kk}\left(\mathbf{U},\grad\mathbf{U}\right)\delta_{ij}\right),
\end{equation}
where the test-filter width $\widehat{\Delta}$ for element $e$ is given by:
\begin{equation}
    \widehat{\Delta} = \widehat{\Delta}_{e} = \Delta_{e}\left(\text{p}_{e}^{\text{test}}\right) = \frac{\volume_{e}^{1/3}}{\text{p}_{e}^{\text{test}}+1}.
\end{equation}
Clipping of the model constant was used to limit its range to $C_{D}\in\left[0,C_{S}^{2}\right]$. The lower limit value of zero prevents anti-dissipation, while the upper limit $C_{S}^{2}$ was chosen so that it is not more dissipative than the standard Smagorinsky given by Eq.(\ref{eq:standard_smagorinsky_model}) with $C_{S}=0.10$ used in this work. In addition, $\text{p}_{e}^{\text{test}}$ is set to  $\text{p}_{e}^{L}$ for the filter operation $\widehat{\left(\cdot\right)}$ in Eq.(\ref{eq:L_matrix},\ref{eq:M_matrix}). 

\subsubsection{Low Reynolds Number Correction}
To account for low Reynolds number effects that can occur for transitional flows, \citet{meyers2006model} proposed the following correction to the eddy-viscosity:
\begin{equation}\label{eq:lrnc_dimensional}
    \nu^{\star}_{t} = \sqrt{\nu_{t}^{2}+\nu^{2}}-\nu,
\end{equation}
where $\nu^{\star}_{t}$ is the corrected eddy-viscosity and $\nu$ is the kinematic viscosity $\nu=\mu/\rho$. Substituting Eq.(\ref{eq:lrnc_dimensional}) into Eq.(\ref{eq:iLES_equation}), applying the methodology of section \ref{sec:dimensionless_gov_eqxns}, and omitting all the asterisks for clarity yields the dimensionless low Reynolds number correction (LRNC):
\begin{equation}\label{eq:lrnc_nondimensional}
    \nu^{\star}_{t} = \sqrt{\nu_{t}^{2}+\left(\frac{\nu}{\text{Re}_{\infty}}\right)^{2}}-\frac{\nu}{\text{Re}_{\infty}},
\end{equation}
where the kinematic viscosity $\nu=\mu/\rho$. 

\subsection{Nonlinearly Stable Flux Reconstruction Schemes}\label{subsec:NSFR-discretization}
\indent In this section, we will present the NSFR discretization \cite{cicchino2022nonlinearly,cicchino2022provably,cicchino2023discretely,cicchino2024scalable} of the Navier-Stokes equations from Section~\ref{sec:navier_stokes_equations} following the notation and work of \citet{cicchino2024thesis}. To do so, we re-write Eq.(\ref{eq:navier_stokes}) as~\cite{cicchino2024thesis}:
\begin{equation}\label{eq:alex_navier_stokes}
    \pdv{\bm{W}^{T}}{t} + \nabla\cdot\bm{f}_{c}\left(\bm{W}\right)^{T} - \nabla\cdot\bm{f}_{v}\left(\bm{W},\nabla\bm{W}\right)^{T} = \bm{0}^{T},
\end{equation}
where $\bm{W},~\bm{f}_{c},\bm{f}_{v}$ respectively correspond to $\bm{U},~\bm{F}^{c},~\bm{F}^{v}$ defined in Section \ref{sec:navier_stokes_equations}. To obtain the discretization, we first represent the physical domain $\bm{\Omega}$, where $\bm{x}\in\bm{\Omega}$, as $M$ non-overlapping elements $\bm{\Omega}_{m}$, where the domain is represented by the union of the elements, i.e. 
\begin{equation}
    \bm{\Omega} \simeq \bm{\Omega}^h:= \bigcup_{m=1}^M \bm{\Omega}_m,
\end{equation}
where $m$ represents the $m$-th element and $\bm{\Omega}^h$ is the discrete domain. The physical coordinates within each element are denoted as $\bm{x}_{m}\in\Omega_{m}$. Next, the physical solution $\bm{W}$ is approximated by the global discrete solution $\bm{u}^{h}=\bm{u}^{h}(\bm{x},t)$ which is constructed from the direct sum of each local (element-wise) approximation $\bm{u}^{h}_{m}(\bm{x},t)$:
\begin{equation}\label{eq:discrete_solution_nsfr}
    \bm{W}=\bm{W}\left(\bm{x},t\right) \simeq \bm{u}^{h}(\bm{x},t):=\bigcup_{m=1}^M\bm{u}^{h}_{m}(\bm{x},t).
\end{equation}
Next we write Eq.(\ref{eq:alex_navier_stokes}) as a system of first-order equations in the discretized domain:
\begin{align}
    &\pdv{\bm{u}^{h}}{t} + \nabla\cdot\bm{f}_{c}\left(\bm{u}^{h}\right)^{T} - \nabla\cdot\bm{f}_{v}\left(\bm{u}^{h},\bm{q}^{h}\right)^{T} = \bm{0}^{T}\label{eq:primary_simple}\\
    &\bm{q}^{h}=\nabla\bm{u}^{h}\label{eq:aux_simple}
\end{align}
where the above two equations are referred to as the primary and auxiliary equations, respectively. The auxiliary solution $\bm{q}^{h}$ is defined similar to Eq.(\ref{eq:discrete_solution_nsfr}). On each element $m$, both $\bm{u}^{h}$ and $\bm{q}^{h}$ are represented as a linear combination of $N_{p}$ linearly independent modal polynomial basis functions $\bm{\chi}_{m}(\bm{x})$ of polynomial order $p$, where $N_{p}:=(p+1)^{3}$ for 3D problems. Hence, the local solutions (i.e. on an element $m$) for a \textit{single state} are expressed as:
\begin{equation}
    u^{h}_{m}(\bm{x},t):=\sum_{i=1}^{N_{p}}\hat{u}_{m,i}(t)\chi_{m,i}(\bm{x}), \quad q^{h}_{m}(\bm{x},t):=\sum_{i=1}^{N_{p}}\hat{q}_{m,i}(t)\chi_{m,i}(\bm{x})
\end{equation}
where $\hat{\bm{u}}_{m}$ and $\hat{\bm{q}}_{m}$ are the modal coefficients of each solution on element $m$. The polynomial basis functions for the solutions are defined as
\begin{equation}\label{eq:solution_bases}
     \bm{\chi}_m(\bm{x}) := [\chi_{m,1}(\bm{x}), \ \chi_{m,2}(\bm{x}), \  \dots, \  \chi_{m,N_p}(\bm{x})]
     =\bm{\chi}(x)\otimes \bm{\chi}(y)\otimes \bm{\chi}(z).
\end{equation}
where $\otimes$ denotes a tensor product. To obtain a system of equations in terms of modal coefficients, fluxes in physical space are transformed using a flux basis $\bm{\phi}_{m}$ expressed similarly to Eq.(\ref{eq:solution_bases}). Furthermore, for each element, the physical coordinates $\bm{x}_{m}$ are mapped to the standard reference space,
\begin{equation}\label{eq:reference_space}
    \bm{\xi}^r := \left\{ [\xi \text{, } \eta \text{, }\zeta]:-1\leq \xi\text{, }\eta\text{, }\zeta\leq1 \right\}, 
\end{equation}
by mapping shape functions $\bm{\Theta}_{m}$:
\begin{equation}\label{eq:mapping_function}
\bm{x}_{m}:=\bm{\Theta}_{m}\left(\bm{\xi}^{r}\right)
\end{equation}
where superscript $r$ denotes reference space. Each element has volume nodes $\bm{\xi}_{v}^{r}$ and facet nodes $\bm{\xi}_{f}^{r}$. 
The mapping to reference space introduces metric terms such as the metric Jacobian and metric co-factor matrix to handle transformation between physical and reference space efficiently; refer to~\cite{cicchino2022provably} for more details. 
In this work, the reference coordinates $\bm{\xi}^r$ are chosen as Gauss-Lobatto-Legendre (GLL) nodes for the solution bases, whereas either GLL or Gauss-Legendre (GL) nodes are used for the flux bases. 
The impact of the choice of flux nodes is investigated in Section~\ref{sec:collocation_effects}.
\newline \indent Substituting Eq.(\ref{eq:solution_bases}) and Eq.(\ref{eq:mapping_function}) in Eq.(\ref{eq:aux_simple}), performing integration-by-parts twice, evaluating bilinear forms with quadrature rules as detailed in~\cite{cicchino2022provably}, we arrive at the strong formulation of the auxiliary equation~\cite{cicchino2022provably}:
\begin{equation}\label{eq:NSFR_viscous_aux}
\left(\bm{M}_{m} +\Tilde{\bm{K}}_{m}\right)\bm{q}_{m}^T = \bm{\chi}(\bm{\xi}_{v}^{r})^{T} \bm{\mathcal{W}}\nabla^{r} \bm{\phi}\left(\bm{\xi}_{v}^{r}\right)\hat{\bm{u}}_{m}^{r}(t)^{T} + \sum_{f=1}^{N_{f}}\sum_{k=1}^{N_{fp}}\bm{\chi}(\bm{\xi}_{fk}^{r})^{T} \mathcal{W}_{fk} \left[\hat{\bm{n}}^{r} \left(\bm{u}_{m}^{*,r} - \bm{\phi}\left(\bm{\xi}_{fk}^{r}\right)\hat{\bm{u}}_{m}^{r}(t)^{T}\right)\right],
\end{equation}
where $\bm{M}_{m}$ is the mass matrix for element $m$, with volume integral and surface integral terms on the right hand side. 
Omitting the full derivation provided by \citet{cicchino2024thesis}, from Eq.(\ref{eq:primary_simple}) we arrive at the strong formulation of the primary equation in split form that is provably entropy stable:
\begin{multline}\label{eq:NSFR_viscous_primary}
  \left(\bm{M}_{m} +\bm{K}_{m}\right)\frac{d}{dt} \hat{\bm{u}}_{m}(t)^{T} = - \underbrace{\left[\bm{\chi}(\bm{\xi}_{v}^{r})^{T}\bm{\chi}(\bm{\xi}_{f}^{r})^{T}\right] \left[\left(\Tilde{\bm{Q}}-\Tilde{\bm{Q}}^{T}\right) \odot \bm{\mathcal{F}}_{m,c}^{r} \right]\bm{1}^T}_{\text{(a)}} - \underbrace{\sum_{f=1}^{N_{f}}\sum_{k=1}^{N_{fp}}\bm{\chi}(\bm{\xi}_{fk}^{r})^{T} \mathcal{W}_{fk} \left[\hat{\bm{n}}^{r} \cdot \bm{f}_{m,c}^{*,r}\right]}_{\text{(b)}}\\
  +\underbrace{\bm{\chi}(\bm{\xi}_{v}^{r})^{T} \bm{\mathcal{W}}\nabla^{r} \bm{\phi}\left(\bm{\xi}_{v}^{r}\right) \cdot \hat{\bm{f}}^{r}_{m,v}(t)^{T}}_{\text{(c)}}+\underbrace{\sum_{f=1}^{N_{f}}\sum_{k=1}^{N_{fp}}\bm{\chi}(\bm{\xi}_{fk}^{r})^{T} \mathcal{W}_{fk} \left[\hat{\bm{n}}^{r} \cdot \left( \bm{f}_{m,v}^{*,r} - \bm{\phi}\left(\bm{\xi}_{f,k}\right) \hat{\bm{f}}^{r}_{m,v}(t)^T\right)\right]}_{\text{(d)}}.
\end{multline}
Thus, the NSFR discretization for convective-viscous problems is two-part: (1) the auxiliary equation given by Eq.(\ref{eq:NSFR_viscous_aux}) and (2) the primary equation given by Eq.(\ref{eq:NSFR_viscous_primary}). 
The energy stable flux reconstruction (ESFR) correction operator matrix for the primary equation $\bm{K}_{m}$ and auxiliary equation $\Tilde{\bm{K}}_{m}$ contain the correction parameter $c$ which enables larger explicit time-step and the recovery of other existing high-order methods. In this work, $\Tilde{\bm{K}}_{m}$ always uses $c=c_{DG}$ \cite{quaegebeur2019cnf,quaegebeur2019ipbr2f}, whereas $\bm{K}_{m}$ uses $c=c_{DG}$ unless otherwise specified. The performance of the different $c$ choices for NSFR are investigated in Section \ref{sec:correction_parameter_tgv}; \citet{bull2015simulation} investigated this using ESFR schemes.
Term (a) of Eq.(\ref{eq:NSFR_viscous_primary}) represents the volume term of the convective flux in split form, where $\Tilde{\mathcal{Q}}-\Tilde{\mathcal{Q}}^T$ is the general skew-symmetric stiffness operator of Chan~\cite{chan2018discretely}, $\odot$ denotes a Hadamard product, the two-point convective reference fluxes are stored in matrix $\bm{\mathcal{F}}^r_{c,m}$, defined as:
\begin{equation}
     {  \left( \mathcal{F}^r_{c,m}\right)_{ij}} = 
     {\mathbf{f}_{c}\left(\Tilde{\mathbf{u}}_m(\bm{\xi}_{i}^r),\Tilde{\mathbf{u}}_m(\bm{\xi}_{j}^r)\right)}
     {\left(
       \frac{1}{2}\left(\mathbf{C}_m(\bm{\xi}_{i}^r)+ \mathbf{C}_m(\bm{\xi}_{j}^r) \right)\right)},\:\forall\: 1\leq i,j\leq N_v + N_{fp}
       \label{eq:two-pt-flux-matrix}
\end{equation}
where $\mathbf{C}_m$ is the metric Jacobian cofactor matrix formulated as described in~\cite[Section 3.2]{cicchino2022provably}. Note that all tensor products $\otimes$ and Hadamar products $\odot$ are evaluated using sum factorization~\cite{cicchino2024scalable}. In this work, the two-point flux $\bm{f_{c}}$ in Eq.(\ref{eq:two-pt-flux-matrix}) is chosen as the Ismail and Roe two-point flux~\cite{ismail2009affordable} which is entropy conservative and described in the next section. The performance using other existing two-point fluxes is explored in Section~\ref{sec:tpflux}.
Crucially, the two-point convective flux in Eq.(\ref{eq:two-pt-flux-matrix}) is evaluated using entropy-projected variables $\Tilde{\mathbf{u}}_m$, per~\cite[Eq.(42)]{cicchino2023discretely}. This procedure, first proposed by Chan~\cite{chan2018discretely}, is key in enabling entropy stability on uncollocated solution and flux or integration nodes. 
Term (b) of Eq.(\ref{eq:NSFR_viscous_primary}) represents the convective flux at the element faces, computed using a numerical flux $\bm{f}_{m,c}^{*,r}$, where quadrature weights for integration are stored in $\bm{\mathcal{W}}$ and $\hat{\bm{n}}^{r}$ is the normal vector. Term (c) of Eq.(\ref{eq:NSFR_viscous_primary}) represents the volume term of the viscous flux, while term (d) represents the viscous flux at element faces in which the viscous numerical flux for element $m$ in reference space $\bm{f}_{m,v}^{*,r}$ is evaluated using the symmetric interior penalty (SIPG) method of \citet{hartmann2008numerical}. 
For complete details and proofs related to the NSFR discretization Eq.(\ref{eq:NSFR_viscous_aux}-\ref{eq:NSFR_viscous_primary}), we direct the reader \cite{cicchino2024thesis} and the references therein, namely: \cite{cicchino2022nonlinearly,cicchino2022provably,cicchino2023discretely}. 

\subsubsection{Two-Point Flux}
As previously mentioned, in this work we use the NSFR schemes of  \citet{cicchino2022nonlinearly,cicchino2022provably} that are provably nonlinearly stable by the use of split-formulations with the entropy conservative, Ismail and Roe (IR) split form convective two-point flux \cite{ismail2009affordable}. The IR numerical two-point flux 
in the $x$, $y$, and $z$ directions is given by:
\begin{equation}\label{eq:two-point-generic}
    \bm{f}_{c} = \left[F_{IR}^{\#},~G_{IR}^{\#},~H_{IR}^{\#}\right],
\end{equation}
where each component is computed using the left state $\bm{U}_{L}$ and right state $\bm{U}_{R}$ as:
\begin{equation}\label{eq:two-point-ir}
    F_{IR}^{\#}(\bm{U}_{L},\bm{U}_{R}) = 
    \begin{bmatrix}
    \hat{\rho}\hat{u}\\
    \hat{\rho}\hat{u}^{2}+\hat{p}_{1}\\
    \hat{\rho}\hat{u}\hat{v}\\
    \hat{\rho}\hat{u}\hat{w}\\
    \hat{\rho}\hat{u}\hat{h}
    \end{bmatrix}, \quad G_{IR}^{\#}(\bm{U}_{L},\bm{U}_{R}) = 
    \begin{bmatrix}
    \hat{\rho}\hat{v}\\
    \hat{\rho}\hat{v}\hat{u}\\
    \hat{\rho}\hat{v}^{2}+\hat{p}_{1}\\
    \hat{\rho}\hat{v}\hat{w}\\
    \hat{\rho}\hat{v}\hat{h}
    \end{bmatrix}, \quad H_{IR}^{\#}(\bm{U}_{L},\bm{U}_{R}) = 
    \begin{bmatrix}
    \hat{\rho}\hat{w}\\
    \hat{\rho}\hat{w}\hat{u}\\
    \hat{\rho}\hat{w}\hat{v}\\
    \hat{\rho}\hat{w}^{2}+\hat{p}_{1}\\
    \hat{\rho}\hat{w}\hat{h}
    \end{bmatrix}
\end{equation}
\begin{equation}
    \hat{\rho}=\ldblbrace z_{1}\rdblbrace z_{5}^{\ln},\quad \hat{u}=\frac{\ldblbrace z_{2}\rdblbrace}{\ldblbrace z_{1}\rdblbrace},\quad \hat{v}=\frac{\ldblbrace z_{3}\rdblbrace}{\ldblbrace z_{1}\rdblbrace},\quad \hat{w}=\frac{\ldblbrace z_{4}\rdblbrace}{\ldblbrace z_{1}\rdblbrace},\quad \hat{p}_{1}=\frac{\ldblbrace z_{5}\rdblbrace}{\ldblbrace z_{1}\rdblbrace},\quad 
\end{equation}
\begin{equation}
    \hat{h}=\frac{\gamma\hat{p}_{2}}{\hat{\rho}\left(\gamma-1\right)}+\frac{1}{2}\left(\hat{u}^{2}+\hat{v}^{2}+\hat{w}^{2}\right),\quad \hat{p}_{2}=\frac{\gamma+1}{2\gamma}\frac{z_{5}^{\ln}}{z_{1}^{\ln}}+\frac{\gamma-1}{2\gamma}\hat{p}_{1}
\end{equation}
\begin{equation}
    \mathbf{z} = \begin{bmatrix} z_{1} & z_{2} & z_{3} & z_{4} & z_{5} \end{bmatrix}^{T} = \begin{bmatrix}
    \sqrt{\displaystyle\frac{\rho}{P}} 
    & \text{v}_{1}\sqrt{\displaystyle\frac{\rho}{P}} 
    & \text{v}_{2}\sqrt{\displaystyle\frac{\rho}{P}} 
    & \text{v}_{3}\sqrt{\displaystyle\frac{\rho}{P}} 
    & \sqrt{\rho P}
    \end{bmatrix}^{T}
\end{equation}
where $\left\ldblbrace\cdot\right\rdblbrace$ denotes the mean:
\begin{equation}\label{eq:mean_operator}
    \left\ldblbrace\cdot\right\rdblbrace = \frac{\left(\cdot\right)_{L}+\left(\cdot\right)_{R}}{2},
\end{equation}
and $\left(\cdot\right)^{\ln}$ denotes the logarithmic mean:
\begin{equation}\label{eq:log_mean}
    \left(\cdot\right)^{\ln} = \frac{\left(\cdot\right)_{L}-\left(\cdot\right)_{R}}{\ln\left(\left(\cdot\right)_{L}\right)-\ln\left(\left(\cdot\right)_{R}\right)}.
\end{equation}
To compute Eq.(\ref{eq:log_mean}), the numerically stable procedure described by \citet{ismail2009affordable} is used. 

\subsection{Temporal Discretization}
The solution is advanced in time using the strong stability preserving third-order accurate Runge Kutta (SSP-RK3) explicit time advancement method \cite{shu1988efficient}, and a typical advection-based CFL condition is used for computing the time step on the fly. For convective-based CFL condition in a cube domain of side length $2\pi$:
\begin{equation}\label{eq:approximate_grid_spacing}
    \Delta t = \text{CFL}\frac{\widetilde{\Delta x}}{\lambda_{\text{max}}},\quad \widetilde{\Delta x}=\frac{2\pi}{\left(DOF\right)^{1/3}},\quad \lambda_{\text{max}}=\text{max}\left(\left|\bm{\textbf{v}}\right|+c\right),%\quad L=2\pi
\end{equation}
where $\lambda_{\text{max}}$ is the maximum wavespeed throughout the entire domain with speed of sound $c=\sqrt{\gamma RT}$. In this work, we use a $\text{CFL}=0.10$ unless otherwise mentioned.

\subsection{Addition of Riemann Solver Dissipation}
To add the Riemann solver dissipation to our NSFR schemes, we simply add the \textit{upwind term} of Eq.(\ref{eq:lxf-flux}, \ref{eq:roe-flux}) to the convective split form two-point flux, i.e. Eq.(\ref{eq:two-point-generic}), when evaluating the flux at element faces.
\subsubsection{Lax-Friedrichs}
The Lax-Friedrichs (LxF) numerical flux, $\bm{f}^{*}_{c}$, approximating the solution of the Riemann problem between constant states $L$ and $R$, is given as: 
\begin{equation}\label{eq:lxf-flux}
    \bm{f}^{*}_{c} = \frac{1}{2}\left(\bm{F}^{c}_{L}+\bm{F}^{c}_{R}\right) \underbrace{- \frac{1}{2}\text{max}\left(\lambda_{\text{max}}^{L},\lambda_{\text{max}}^{R}\right)\left(\bm{U}_{R}-\bm{U}_{L}\right)}_{\text{Upwind Term}},
\end{equation}
where $\bm{F}^{c}_{L}=\bm{F}^{c}(\bm{U}_{L})$, $\bm{F}^{c}_{R}=\bm{F}^{c}(\bm{U}_{R})$, and the maximum convective eigenvalues at states $L$ and $R$, $\lambda^{L/R}$ are given by:
\begin{equation}
    \lambda_{\text{max}}^{L/R}\equiv\lambda_{\text{max}}\left(\bm{U}_{L/R}\right) = \left|\text{v}_{i}\hat{n}_{i}\right| + c.
\end{equation}
The above is referred to as the \textit{global Lax-Friedrichs recipe} \cite{cockburn1998runge}. \textbf{Note:} Thus, LxF can be anti-dissipative if $\left(\bm{U}_{R}-\bm{U}_{L}\right)_{i}<0$, causing the upwind term to be positive. 
\subsubsection{Roe}\label{subsubsec:roe_flux}
\indent The classic Roe (i.e. Roe-Pike \cite{roe1985efficient}) numerical flux, $\bm{f}^{*}_{c}$, approximating the solution of the Riemann problem between constant states $L$ and $R$, is given as (e.g. \cite{toro2013riemann,osswald2016l2roe}):
\begin{equation}\label{eq:roe-flux}
    \bm{f}^{*}_{c} = \frac{1}{2}\left(\bm{F}^{c}_{L}+\bm{F}^{c}_{R}\right) \underbrace{- \frac{1}{2}\sum_{i=1}^{m}\tilde{\alpha}_{i}|{\tilde{\lambda}_{i}}|\bm{K}^{(i)}}_{\text{Upwind Term}},
\end{equation}
where $\bm{F}^{c}_{L}=\bm{F}^{c}(\bm{U}_{L})$, $\bm{F}^{c}_{R}=\bm{F}^{c}(\bm{U}_{R})$. The the eigenvalues or wave speeds $\tilde{\lambda}_{i}$, corresponding eigenvectors $\tilde{\bm{K}}^{(i)}$, and averaged wave strengths $\tilde{\alpha}_{i}$ are given by:
\begin{equation}
    \tilde{\lambda}_{1} = \tilde{\text{v}}_{n}-\tilde{a},\quad
    \tilde{\lambda}_{2}=\tilde{\lambda}_{3}=\tilde{\lambda}_{4} = \tilde{\text{v}}_{n},\quad 
    \tilde{\lambda}_{5} = \tilde{\text{v}}_{n}+\tilde{a},
\end{equation}
\begin{equation}
    \bm{K^{(1)}} = \bm{K^{(5)}} = \begin{bmatrix}1\\\tilde{\textbf{v}}\mp\tilde{a}\bm{n}\\\tilde{H}\mp\tilde{\text{v}}_{n}\tilde{a}\end{bmatrix},
    \hspace{0.3cm}
    \bm{K^{(2)}} = \begin{bmatrix}1\\\tilde{\textbf{v}}\\\frac{1}{2}\tilde{\textbf{v}}\cdot\tilde{\textbf{v}}\end{bmatrix},
    \hspace{0.3cm}
    \bm{K^{(3)}} = \begin{bmatrix}0\\\bm{t}_{1}\\\tilde{\text{v}}_{t_1}\end{bmatrix},
    \hspace{0.3cm}
    \bm{K^{(4)}} = \begin{bmatrix}0\\\bm{t}_{2}\\\tilde{\text{v}}_{t_2}\end{bmatrix},
\end{equation}
\begin{equation}\label{eq:roe-pike}
    \tilde{\alpha}_{1} = \frac{\Delta P -\tilde{\rho}\tilde{a}\Delta \text{v}_{n}}{2\tilde{a}^{2}},
    \hspace{0.3cm}
    \tilde{\alpha}_{2} = \Delta\rho - \frac{\Delta P}{\tilde{a}^{2}},
    \hspace{0.3cm}
    \tilde{\alpha}_{3} = \tilde{\rho}\Delta \text{v}_{t_{1}},
    \hspace{0.3cm}
    \tilde{\alpha}_{4} = \tilde{\rho}\Delta \text{v}_{t_{2}},
    \hspace{0.3cm}
    \tilde{\alpha}_{5} = \frac{\Delta P +\tilde{\rho}\tilde{a}\Delta \text{v}_{n}}{2\tilde{a}^{2}}.
\end{equation}
where $\tilde{\left(\cdot\right)}$ denotes the Roe averages given by:
\begin{equation}
    \tilde{\left(\cdot\right)} = \frac{\sqrt{\rho_{L}}\left(\cdot\right)_{L}+\sqrt{\rho_{R}}\left(\cdot\right)_{R}}{\sqrt{\rho_{L}}+\sqrt{\rho_{R}}}
\end{equation}
except for $\tilde{\rho}=\sqrt{\rho_{L}\rho_{R}}$. In Eq.(\ref{eq:roe-pike}), $\Delta \left(\cdot\right)=\left(\cdot\right)_{R}-\left(\cdot\right)_{L}$, $a$ is the local speed of sound, $\text{v}_{n}$ is the velocity in the normal direction $\bm{n}$, while $\text{v}_{t_{1}}$ and $\text{v}_{t_{2}}$ are the velocities in the direction of arbitrary orthogonal unit tangential vectors $\bm{t}_{1}$ and $\bm{t}_{2}$ spanning the plane with normal $\bm{n}$ \cite{osswald2016l2roe}. 

\subsubsection{Low-Dissipation Roe}
\indent \citet{flad2017use} have demonstrated that adding tailored Riemann solver dissipation, using the low-dissipation Roe (L$^{2}$Roe) flux \cite{osswald2016l2roe}, to an energy stable split form DG scheme for ILES provides a mechanism to damp pressure-dilatation fluctuations driven by the discontinuous nature of the DG approach. Therefore, we wish to investigate the use of the same Riemann solver with our NSFR schemes for ILES. The L$^{2}$Roe flux will now be briefly introduced. \\
\indent The L$^{2}$Roe flux modifies the classical Roe flux given by Eq.(\ref{eq:roe-flux})-Eq.(\ref{eq:roe-pike}) to address the accuracy issues arising in low Mach number flow simulations. This is done by scaling the jumps in all discrete velocity components (normal and tangential) within the numerical flux function; where, it only affects the numerical dissipation, and not the actual evaluation of physical convective fluxes $\bm{F}^{c}_{L/R}$. The jumps in normal and tangential velocities, $\Delta \text{v}_{n}$ and $\Delta \text{v}_{t_i}$, are scaled as follows:
\begin{equation}
    \left(\Delta \text{v}_{n,t_{i}}\right)^{*} = \begin{cases}z\Delta \text{v}_{n,t_{i}} & ssw_{L}=ssw_{R}=0\\\Delta \text{v}_{n,t_{i}}&\text{else}\end{cases},\hspace{0.6cm}z=\min\left[1,\max\left(M_{L},M_{R}\right)\right],
\end{equation}
where $z$ is the blending factor, $M_{L/R}$ is the Mach number at the left and right states, and $ssw_{L/R}$ represents a shock-indicator; see \citet{osswald2016l2roe} for more details.

\subsection{Discontinuous Galerkin Method}
Following the notation presented in Section~\ref{subsec:NSFR-discretization}, the classical DG strong formulation of the auxiliary equation can be written as:
\begin{equation}\label{eq:strong_DG_viscous_aux}
\bm{M}_{m}\bm{q}_{m}^T = \bm{\chi}(\bm{\xi}_{v}^{r})^{T} \bm{\mathcal{W}}\nabla^{r} \bm{\phi}\left(\bm{\xi}_{v}^{r}\right)\hat{\bm{u}}_{m}^{r}(t)^{T} + \sum_{f=1}^{N_{f}}\sum_{k=1}^{N_{fp}}\bm{\chi}(\bm{\xi}_{fk}^{r})^{T} \mathcal{W}_{fk} \left[\hat{\bm{n}}^{r} \left(\bm{u}_{m}^{*,r} - \bm{\phi}\left(\bm{\xi}_{fk}^{r}\right)\hat{\bm{u}}_{m}^{r}(t)^{T}\right)\right],
\end{equation}
and for the primary equation, we have: 
\begin{multline}\label{eq:strong_DG_viscous_primary}
  \bm{M}_{m}\frac{d}{dt} \hat{\bm{u}}_{m}(t)^{T} = -\underbrace{\bm{\chi}(\bm{\xi}_{v}^{r})^{T} \bm{\mathcal{W}}\nabla^{r} \bm{\phi}\left(\bm{\xi}_{v}^{r}\right) \cdot \hat{\bm{f}}^{r}_{m,c}(t)^{T}}_{\text{(a)}} -\underbrace{\sum_{f=1}^{N_{f}}\sum_{k=1}^{N_{fp}}\bm{\chi}(\bm{\xi}_{fk}^{r})^{T} \mathcal{W}_{fk} \left[\hat{\bm{n}}^{r} \cdot \left( \bm{f}_{m,c}^{*,r} - \bm{\phi}\left(\bm{\xi}_{f,k}\right) \hat{\bm{f}}^{r}_{m,c}(t)^T\right)\right]}_{\text{(b)}}\\
  +\underbrace{\bm{\chi}(\bm{\xi}_{v}^{r})^{T} \bm{\mathcal{W}}\nabla^{r} \bm{\phi}\left(\bm{\xi}_{v}^{r}\right) \cdot \hat{\bm{f}}^{r}_{m,v}(t)^{T}}_{\text{(c)}}+\underbrace{\sum_{f=1}^{N_{f}}\sum_{k=1}^{N_{fp}}\bm{\chi}(\bm{\xi}_{fk}^{r})^{T} \mathcal{W}_{fk} \left[\hat{\bm{n}}^{r} \cdot \left( \bm{f}_{m,v}^{*,r} - \bm{\phi}\left(\bm{\xi}_{f,k}\right) \hat{\bm{f}}^{r}_{m,v}(t)^T\right)\right]}_{\text{(d)}}.
\end{multline}
Term (a) of Eq.\textbf{}(\ref{eq:strong_DG_viscous_primary}) represents the volume term of the convective flux, while term (b) represents the convective flux at element faces in which the convective numerical flux for element $m$ in reference space $\bm{f}^{*,r}_{m,c}$ is evaluated using the classic Roe numerical flux presented in Section~\ref{subsubsec:roe_flux}. To suppress the aliasing errors associated with integrating the nonlinear reference convective flux $\hat{\bm{f}}^{r}_{m,c}$, additional quadrature nodes are used; this is referred to as \textit{over-integration} \cite{kirby2003aliasing}. In this work, we use an additional $\text{p}+1$ quadrature nodes yielding a total of $2(\text{p}+1)$ quadrature nodes for the compressible Navier-Stokes equations \cite{beck2014high}, where the flux bases $\bm{\phi}$ are collocated over quadrature nodes. The performance of this de-aliasing strategy in comparison to that of the NSFR schemes is investigated in Section~\ref{subsubsec:dealiasing_comparison}. All other remaining terms have the same definition as those in Eq.(\ref{eq:NSFR_viscous_primary}). 
\section{Numerical Results}
\subsection{Taylor-Green Vortex}
In this section, we will evaluate the performance of the NSFR schemes for the ILES and iLES of shock-free turbulent flows using the viscous Taylor-Green vortex (TGV) problem at $\text{Re}_{\infty}=1600$. 
This case has been widely used as a benchmark case for turbulent flow simulations in the absence of shocks and walls. 
The flow is initially laminar, characterized by large-scale vortices, which interact with each other to produce small-scale vortices generating a 3D turbulent energy cascade, until reaching a peak dissipation, at which point the turbulent flow freely decays. 
Therefore, the flow exhibits two key phases: (1) the transition from laminar to a fully turbulent flow, and (2) the decay of homogeneous isotropic turbulence. To the author's best knowledge, this is the first investigation of provably entropy stable flux reconstruction schemes for both flow regimes at the time of this article. \newline 
\indent The initialization for the problem will now be presented. Applying the non-dimensionalization presented in Section \ref{sec:dimensionless_gov_eqxns}, and omitting all asterisks for clarity, the flow has the following isothermal initialization in a non-dimensional triply-periodic domain $x_{i}\in[-\pi,\pi]^{3}$:
\begin{align}
    P\left(x_{i},0\right) &= \frac{1}{\gamma M_{\infty}^{2}}+\frac{1}{16}\left(\cos\left(2x_{1}\right)+\cos\left(2x_{2}\right)\right)\left(\cos\left(2x_{3}\right)+2\right),\\[1mm]
    \text{v}_{1}\left(x_{i},0\right) &= \sin\left(x_{1}\right)\cos\left(x_{2}\right)\cos\left(x_{3}\right),\\[1mm]
    \text{v}_{2}\left(x_{i},0\right) &= -\cos\left(x_{1}\right)\sin\left(x_{2}\right)\cos\left(x_{3}\right),\\[1mm]
    \text{v}_{3}\left(x_{i},0\right) &= 0,\\[1mm]
    \rho\left(x_{i},0\right)&= \gamma M_{\infty}^{2}P\left(x_{i},0\right),
\end{align}
where the gas constant $\gamma=1.4$, the free-stream Mach number $M_{\infty}=0.1$, and the characteristic length scale $L=1$.  \newline 
\indent For evaluating the performance of the scheme, unsteady quantities of interest will now be presented. Omitting all asterisks for clarity, the non-dimensional dissipation rate $\varepsilon$, i.e. the rate at which kinetic energy decays, is taken as the negative of the production rate:
\begin{equation}\label{eq:dissipation_rate}
    \varepsilon = -\frac{\text{d}E_{K}}{\text{d}t}, 
\end{equation}
where $\varepsilon$ is computed from $E_{K}$ using second-order accurate finite-difference; specifically we are using \texttt{numpy.gradient} which accounts for a non-constant time-step that comes with using an advection-based CFL condition to determine the time-step. For the results presented in this work, we define the non-dimensional volume-averaged enstrophy (omitting all asterisks for clarity) as:
\begin{equation}\label{eq:integrated_enstrophy}
    \zeta = \frac{1}{2\left\langle\rho\right\rangle}\left\langle\rho\omega_{k}\omega_{k}\right\rangle \approx \frac{1}{2}\left\langle\rho\omega_{k}\omega_{k}\right\rangle,
\end{equation}
where we have made the approximation that $\left\langle\rho\right\rangle\approx1$ since the flow is nearly incompressible. From Eq.(\ref{eq:numerical_viscosity_origin}) and Eq.(\ref{eq:dissipation_rate}), we can write:
\begin{equation}
    \varepsilon = \frac{\nu\rho}{\text{Re}_{\infty}}\left\langle\omega_{k}\omega_{k}\right\rangle,
\end{equation}
using Eq.(\ref{eq:integrated_enstrophy}), we have
\begin{equation}
    \varepsilon \approx \frac{2\nu\zeta}{\text{Re}_{\infty}},
\end{equation}
from which we can define the non-dimensional (omitting all asterisks) effective kinematic viscosity, $\nu_{\text{effective}}$, as:
\begin{equation}
    \nu_{\text{effective}} \approx \frac{\text{Re}_{\infty}\varepsilon}{2\zeta},
\end{equation}
which is non-dimensionalized by the free-stream kinematic viscosity $\nu_{\infty}=\mu_{\infty}/\rho_{\infty}$. Thus, the effective viscosity provides a relative ratio of the measure of the dissipation observed by the simulation (i.e.``effective'' or ``numerical'') to the actual viscosity in the flow. Thus providing a measure of simulation accuracy, as done in \cite{van2011comparison,manzanero2018role}; e.g. for $\nu_{\text{effective}}<1$, this indicates that the scheme is less dissipative than it should be according to the physical viscosity of the flow. 

For all results presented in this section, a polynomial degree of $\text{p}=5$, and a $\text{CFL}=0.1$ were used, unless explicitly stated otherwise. Furthermore, the turbulent kinetic energy (TKE) spectra, $E(\kappa,t)$, where $\kappa$ is the wavenumber, are computed using the open-source \texttt{TurboGenPY} code \cite{saad2017scalable}. Although not shown in this work, as a sanity check, the authors found that the \texttt{TurboGenPY} code yielded the same spectra as that generated by the open-source code of \citet{navah2020high}. The authors also found that including overlapping nodes from the high-order grid element interfaces did not yield a different spectrum when compared to averaging values at overlapping nodes including unique points in space. As noted by \citet{carton2014assessment}, the Fourier transform of the high-order method (HOM) results needs a finer sampling resolution than that provided by the high-order solution nodes. As done in \cite{carton2014assessment}, a twice denser sampling has been used for all results in this work. This means that the HOM flow field passed to the Fourier transform (i.e. spectra) code is written out to (i.e. ``sampled from'') a grid where in each element, the solution is interpolated to a set of equidistant nodes containing twice the number of solution nodes in each direction, i.e. written to $2(\text{p}+1)$ equidistant nodes in each direction $n_{\text{equid.}}^{1\text{D}}=2(\text{p}+1)$. From \citet[Eq.(13.46)]{pope2001turbulent}, the maximum resolved wavenumber is:
\begin{equation}
    \kappa_{\text{max}} = \frac{\pi}{h_{\text{max}}},
\end{equation}
where for high-order discontinuous element based methods, the approximate uniform grid spacing is:
\begin{equation}
    h_{\text{max}} = \frac{L}{N_{el.}\text{p}}.
\end{equation}
Taking $\kappa_{\text{max}}$ as the theoretical cut-off wavenumber $\kappa_{c}$, and substituting $L=2\pi$ for this problem, we can write:
\begin{equation}\label{eq:cutoff_wavenumber}
    \kappa_{c} = \frac{N_{el.}\text{p}}{2}\equiv\frac{\left(\text{Unique DOF}\right)^{1/3}}{2},
\end{equation}
as in \cite{carton2014assessment}. For all results presented in this section, the TKE spectra are plotted up to $\kappa_{c}$ (unless specified otherwise), i.e. values past this wavenumber are truncated. Furthermore, the grid cut-off wavenumber $\kappa_{\text{grid}}$ is:
\begin{equation}
    \kappa_{\text{grid}} = \frac{N_{el.}}{2}.
\end{equation}

\subsubsection{Verification by Direct Numerical Simulation}\label{subsec:DNS}
\begin{figure}[H]
\begin{subfigure}{.332\textwidth}
    \centering
    \includegraphics[width=0.99\linewidth]{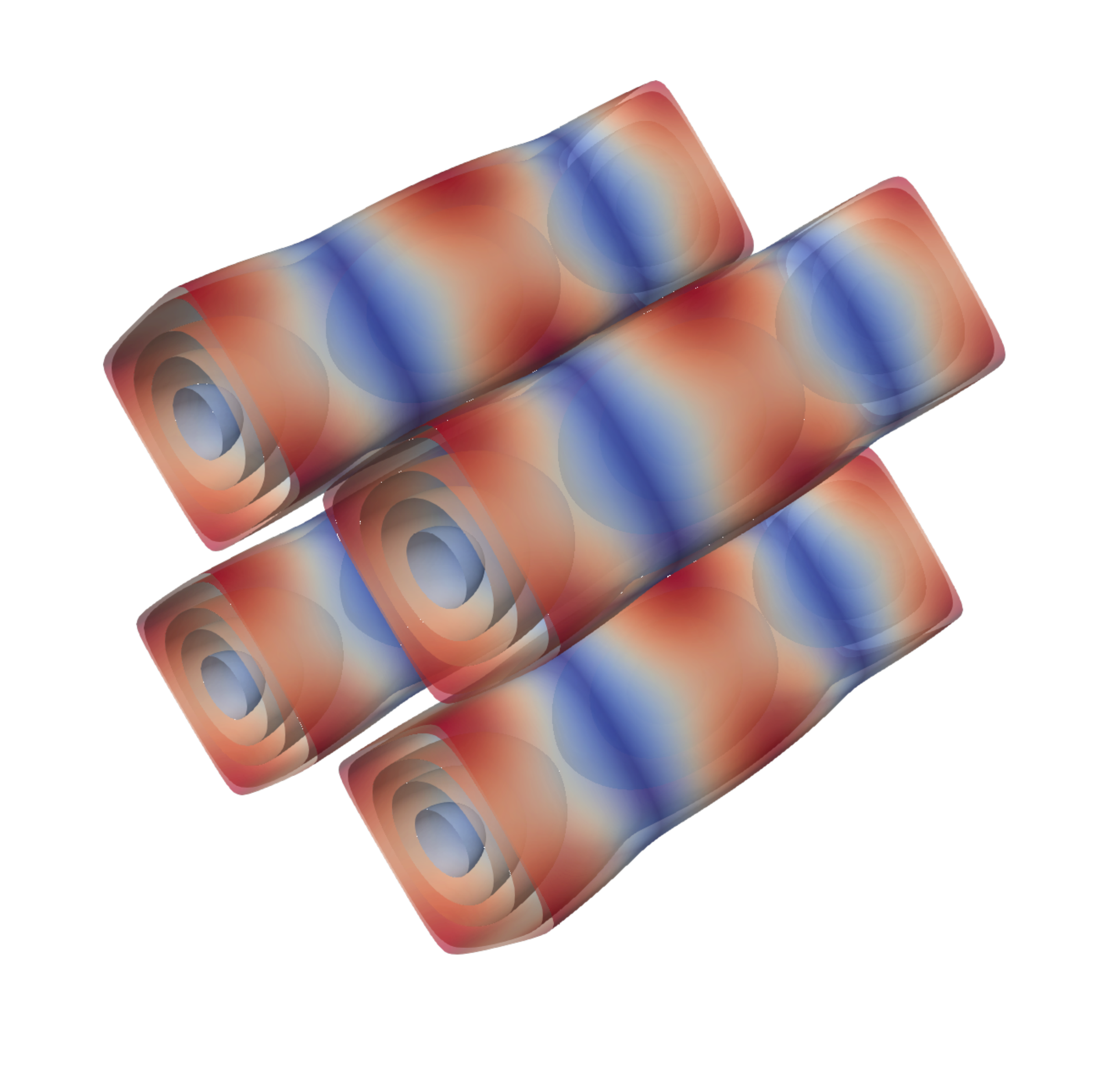}
    \caption{$t^{*}=0$}
\end{subfigure}%
\begin{subfigure}{.332\textwidth}
    \centering
    \includegraphics[width=0.99\linewidth]{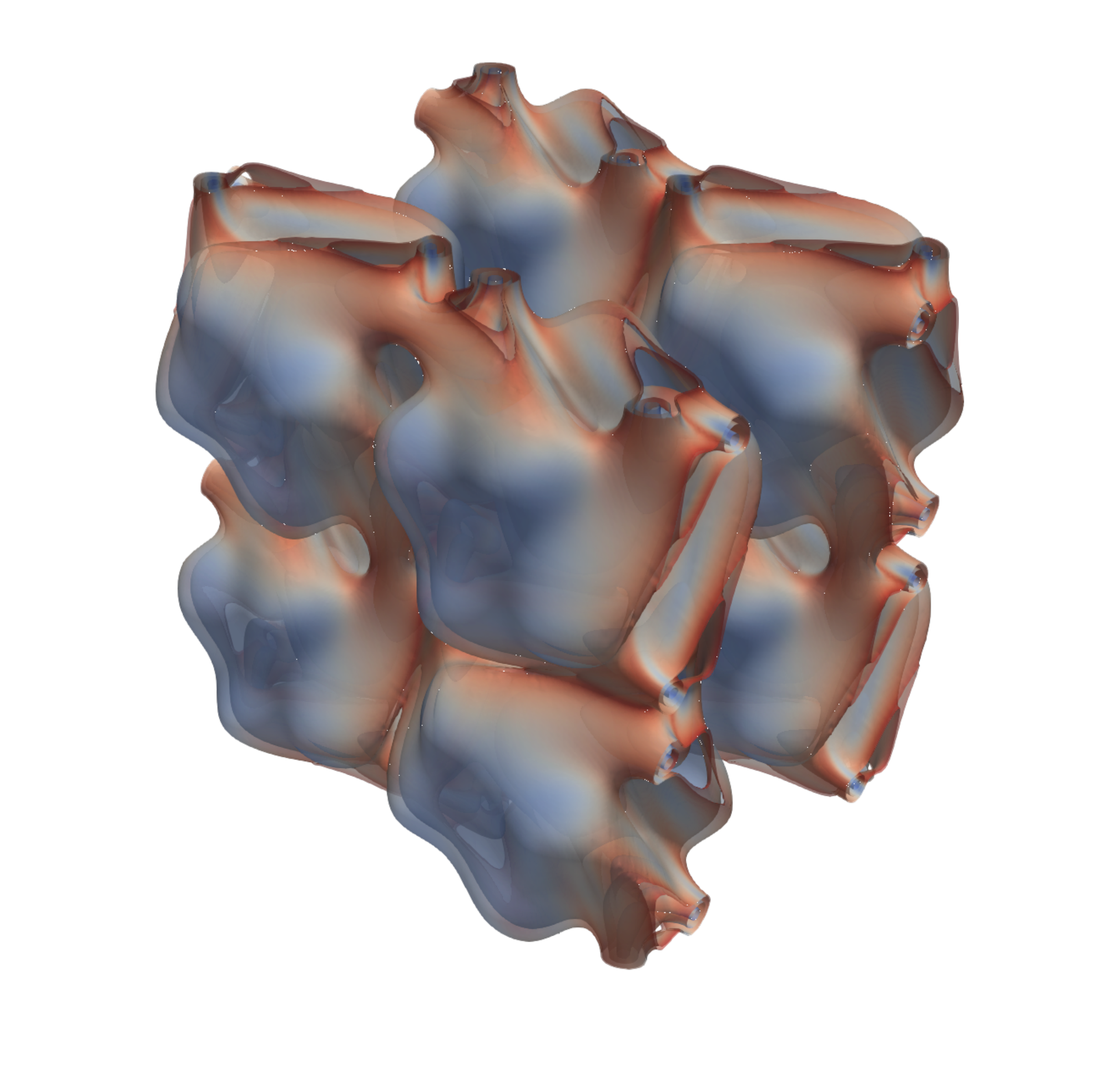}
    \caption{$t^{*}=4$}
\end{subfigure}%
\begin{subfigure}{.332\textwidth}
    \centering
    \includegraphics[width=0.99\linewidth]{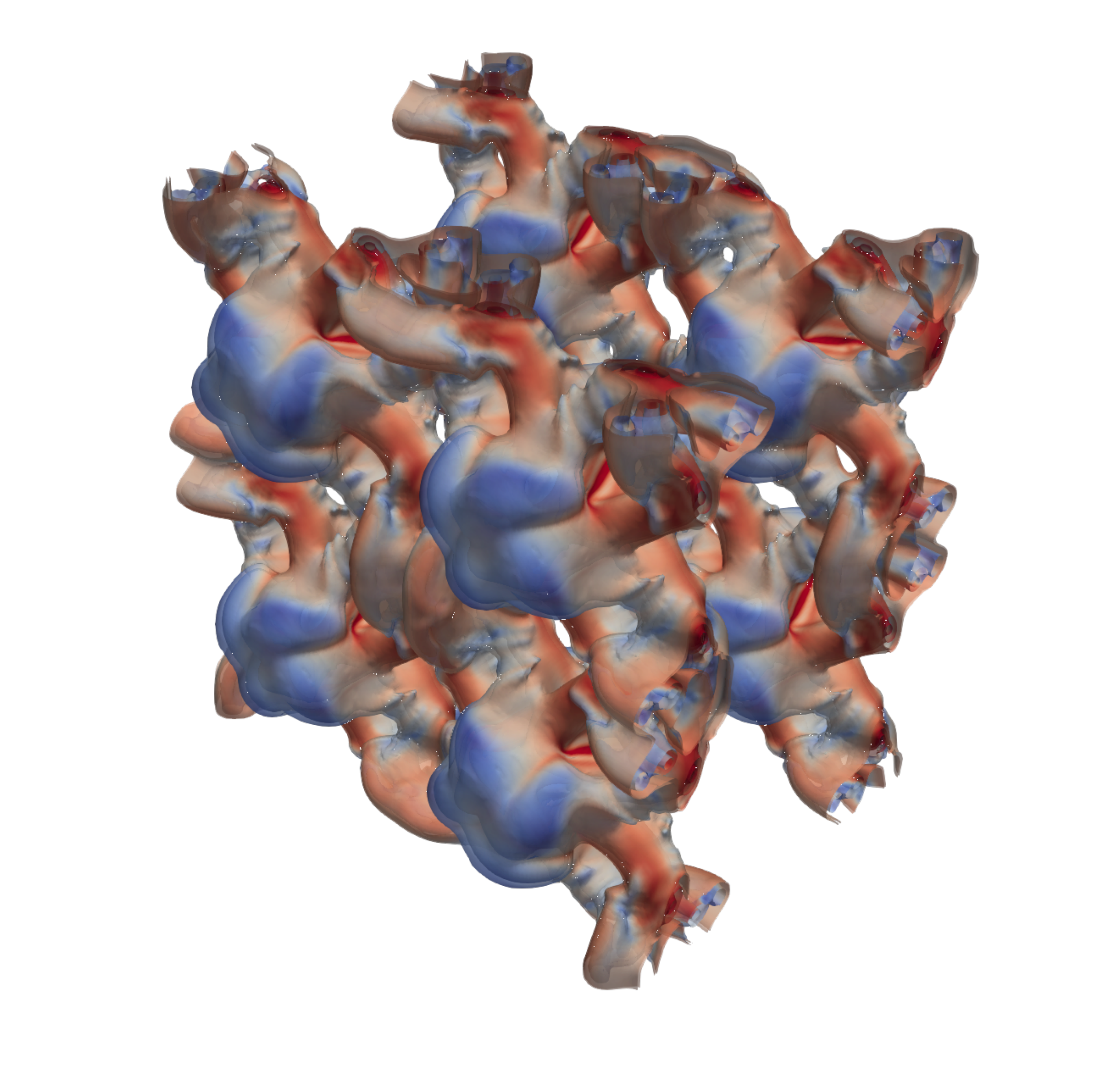}
    \caption{$t^{*}=8$}
\end{subfigure}
\begin{subfigure}{.332\textwidth}
    \centering
    \includegraphics[width=0.99\linewidth]{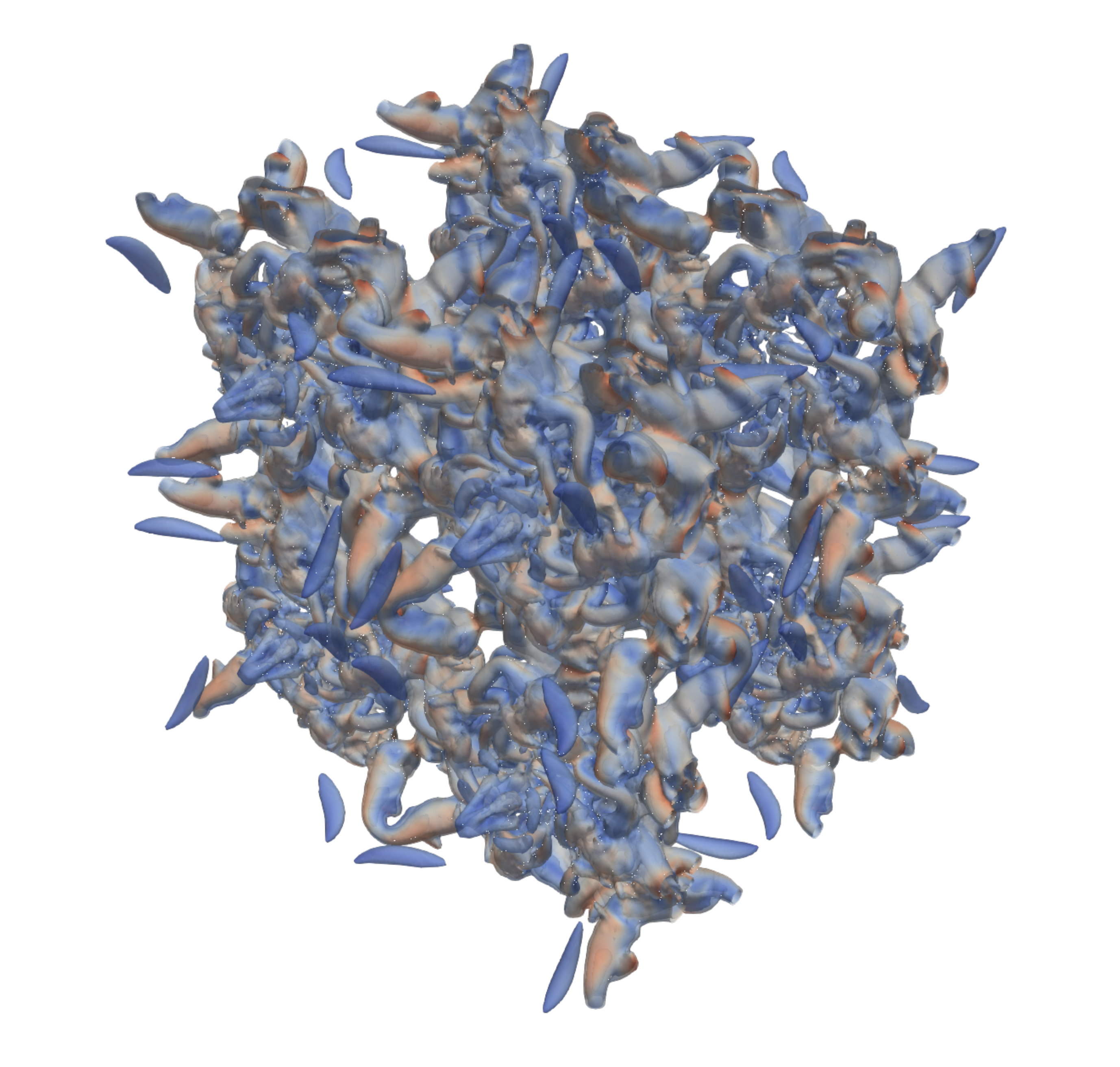}
    \caption{$t^{*}=12$}
\end{subfigure}%
\begin{subfigure}{.332\textwidth}
    \centering
    \includegraphics[width=0.99\linewidth]{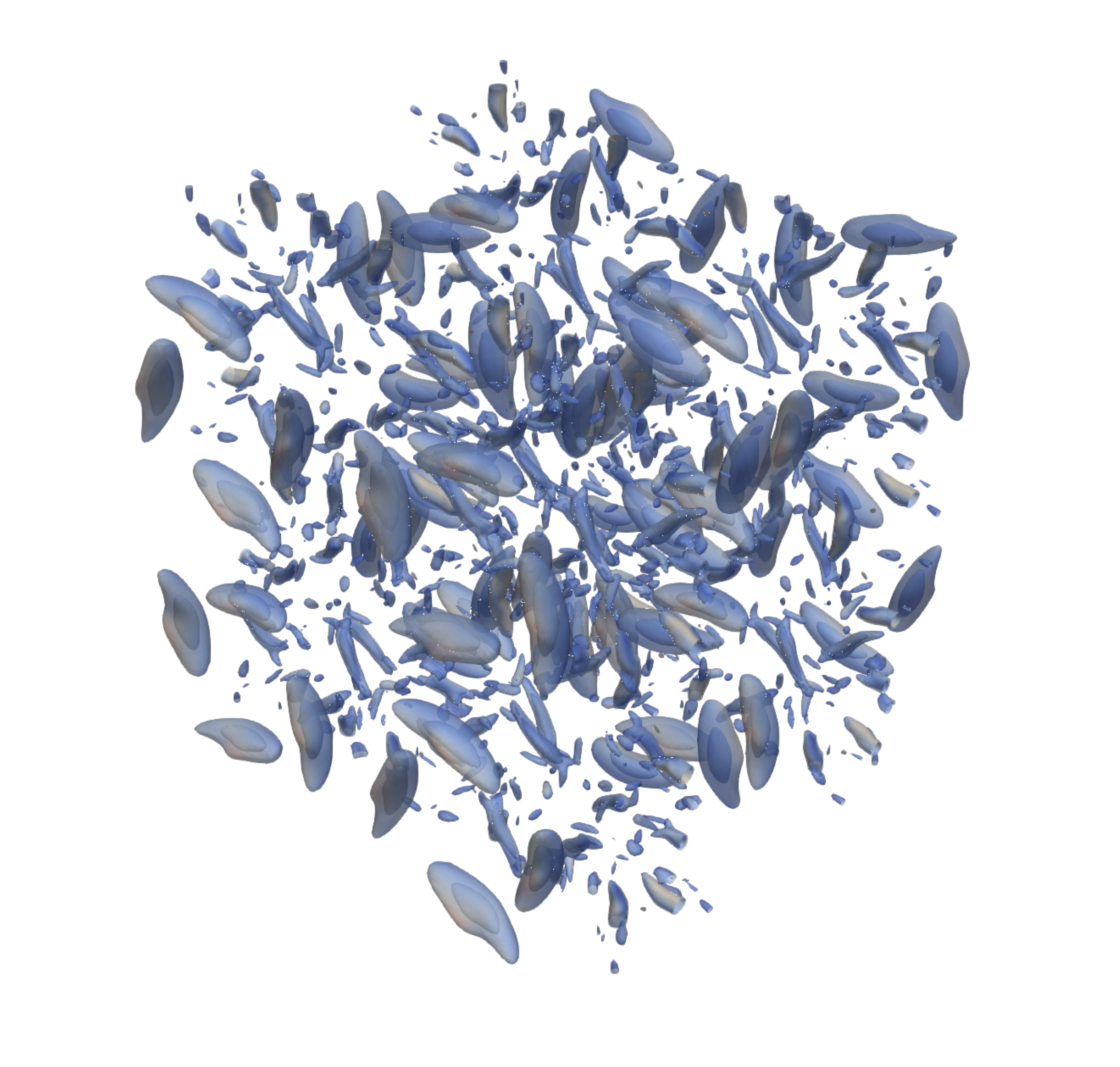}
    \caption{$t^{*}=16$}
\end{subfigure}%
\begin{subfigure}{.332\textwidth}
    \centering
    \includegraphics[width=0.99\linewidth]{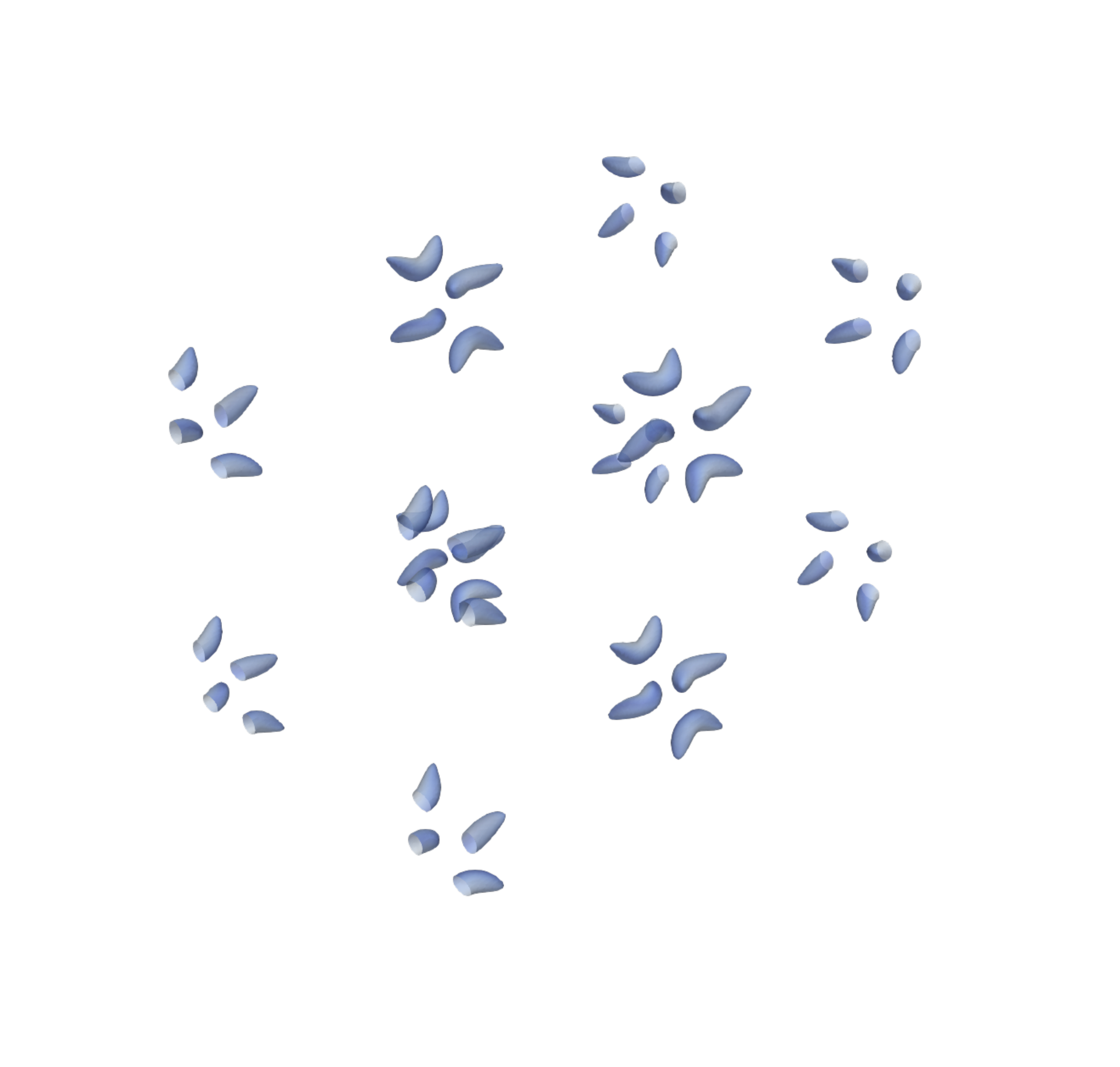}
    \caption{$t^{*}=20$}
\end{subfigure}
\caption{Isocontours of pressure coloured by Mach number for the viscous TGV at $\text{Re}_{\infty}=1600$ with $256^{3}$ DOF using p$7$ $c_{DG}$ NSFR.IR-GL}
\label{fig:tgv-visualization-paraview}
\end{figure}

Isosurfaces of pressure coloured by Mach number are shown in Fig.~\ref{fig:tgv-visualization-paraview} at $t^{*}=0,~4,~8,~12,~16~\text{and}~20$ for a direct numerical simulation (DNS) using a polynomial degree p$7$ $c_{DG}$ NSFR scheme with the Ismail-Roe (IR) two-point flux, and GL flux nodes and GLL solution nodes (i.e. uncollocated), denoted as $c_{DG}$ NSFR.IR-GL for short, with $32^{3}$ elements for a total of $256^{3}$ DOF. These illustrate the evolution of the flow previously mentioned: starting with a smooth initialization of large vortical structures at $t^{*}=0$, which begin to interact through viscous forces at $t^{*}=4$, producing smaller scale structures by $t^{*}=8$ right before reaching a peak dissipation rate at which point the flow is fully turbulent and has started to decay by $t^{*}=12$ until $t^{*}=20$. 

\begin{figure}[H]
\begin{subfigure}{.495\textwidth}
    \centering
    \includegraphics[width=0.99\linewidth]{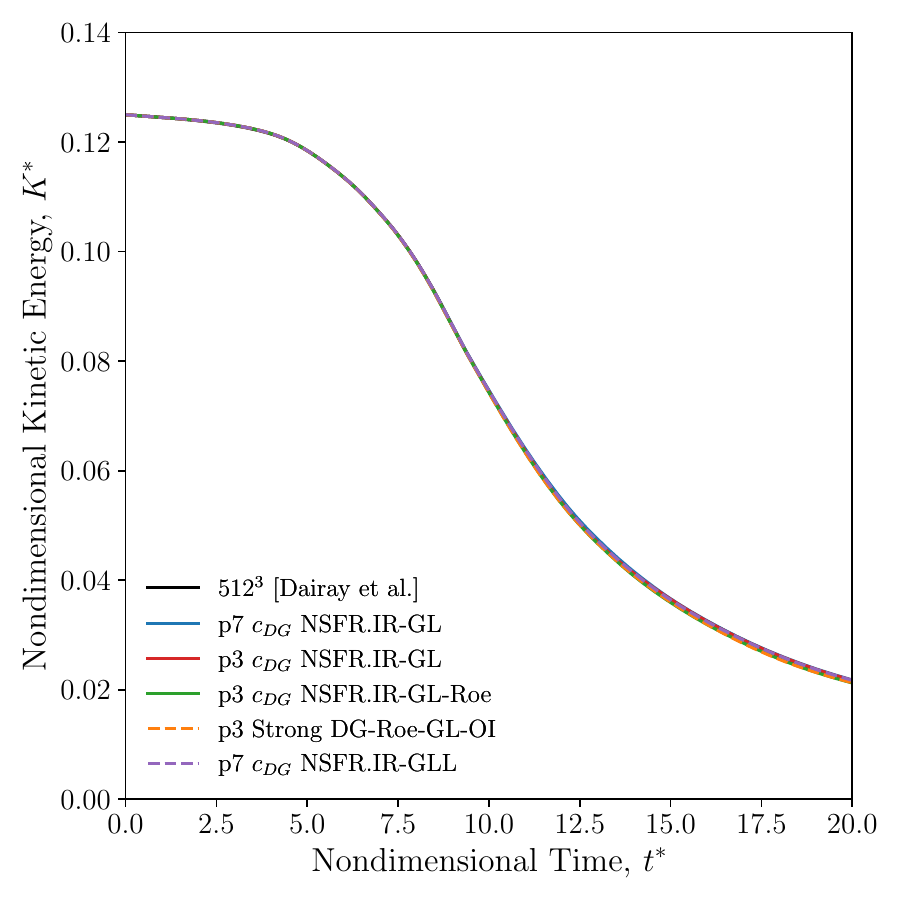}
    \caption{Kinetic energy vs time}
\end{subfigure}%
\begin{subfigure}{.495\textwidth}
    \centering
    \includegraphics[width=0.99\linewidth]{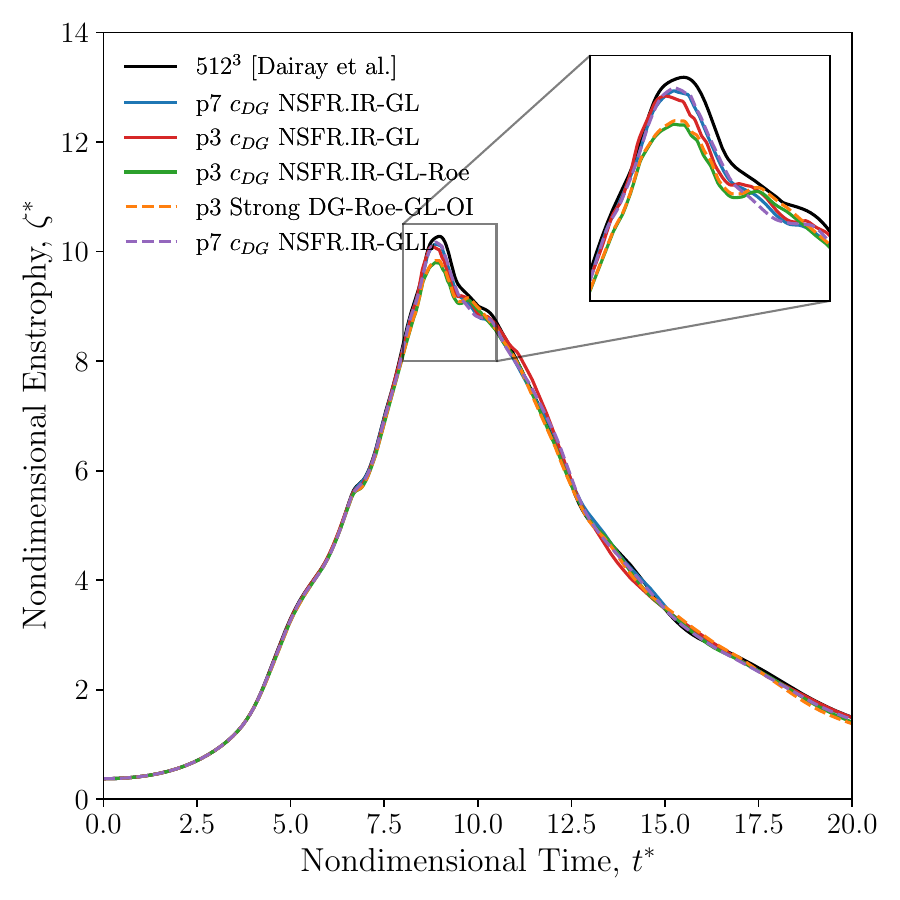}
    \caption{Enstrophy vs time}
\end{subfigure}
\begin{subfigure}{.495\textwidth}
    \centering
    \includegraphics[width=0.99\linewidth]{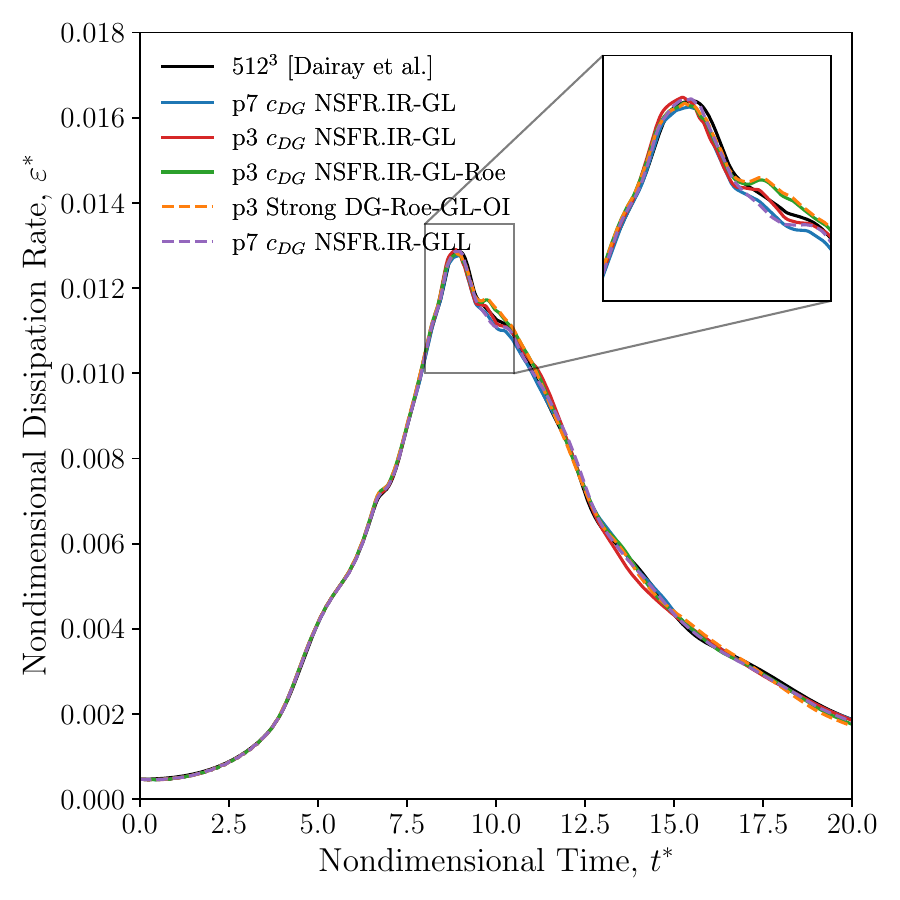}
    \caption{Dissipation rate vs time}
\end{subfigure}%
\begin{subfigure}{.495\textwidth}
    \centering
    \includegraphics[width=0.99\linewidth]{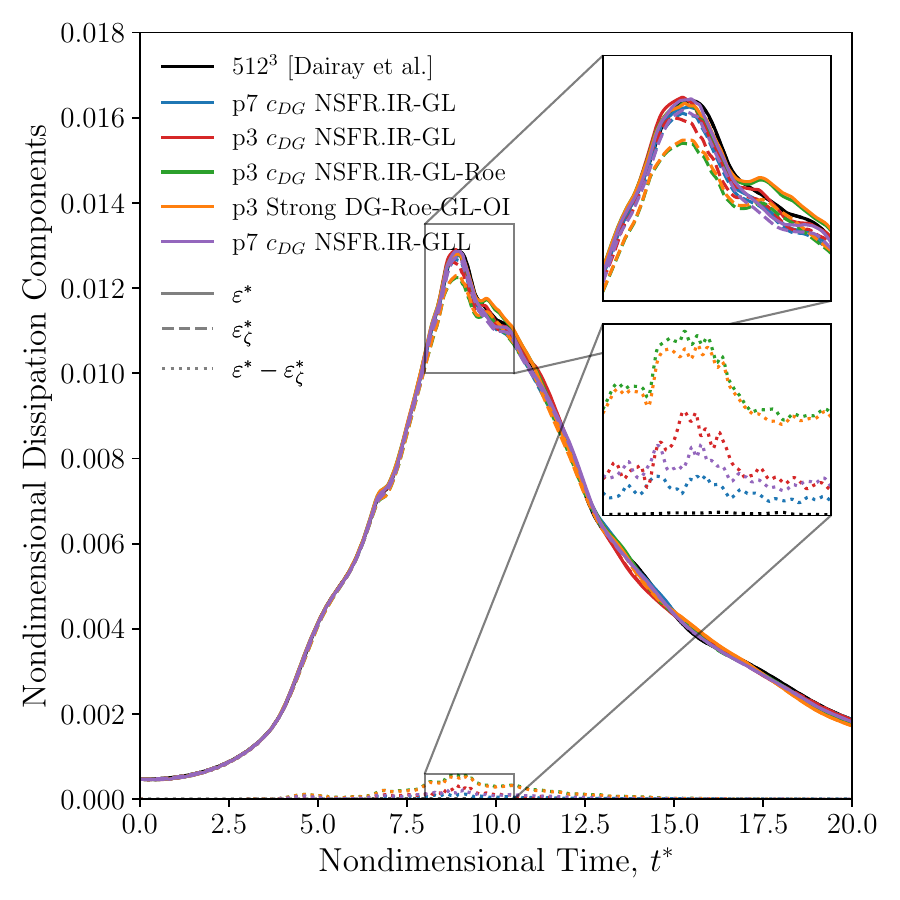}
    \caption{Dissipation components}
\end{subfigure}%
\caption{Temporal evolution of kinetic energy, enstrophy, dissipation rate, and dissipation components for the viscous TGV at $\text{Re}_{\infty}=1600$ with $256^{3}$ DOF using various NSFR schemes and strong~DG-Roe-GL}
\label{fig:verification-quantities}
\end{figure}

Plots for the time evolution of nondimensional kinetic energy $E_{K}$ Eq.(\ref{eq:volumetric_averaged_integrated_kinetic_energy}), dissipation rate $\varepsilon$ Eq.(\ref{eq:dissipation_rate}), enstrophy $\zeta$ Eq.(\ref{eq:integrated_enstrophy}), and dissipation components $\left(\varepsilon,~\varepsilon_{\zeta},~\varepsilon-\varepsilon_{\zeta}\right)$ are shown in Fig.~\ref{fig:verification-quantities} for several different schemes with a reference result for comparison. The different schemes considered are tabulated below:
\begin{table}[H]
\centering
\begin{tabular}{c|c|c|c|c|c|c}
    \hline
    Notation in legends & Scheme & Flux & Sol. & Conv. & Visc. & Over-int. \\
    \hfill & \hfill & Nodes & Nodes & \# Flux & \# Flux & \hfill \\
    \hline
    \hline
    $c_{DG}$ NSFR.IR-GL & $c_{DG}$ NSFR & GL & GLL & Ismail-Roe & SIPG & No\\
    \hfill & \hfill & \hfill & \hfill & two-point flux & \hfill & \hfill\\
    \hline
    $c_{DG}$ NSFR.IR-GLL & $c_{DG}$ NSFR & GLL & GLL & Ismail-Roe & SIPG & No\\
    \hfill & \hfill & \hfill & \hfill & two-point flux & \hfill & \hfill\\
    \hline
    $c_{DG}$ NSFR.IR-GL-Roe & $c_{DG}$ NSFR & GL & GLL & Ismail-Roe & SIPG & No\\
    \hfill & \hfill & \hfill & \hfill & two-point flux & \hfill & \hfill\\
    \hfill & \hfill & \hfill & \hfill & + Roe upwind & \hfill & \hfill\\
    \hline
    Strong DG Roe-GL-OI & Strong DG & GL & GLL & Roe & SIPG & Yes, $2(\text{p}+1)$\\
    \hline
\end{tabular}
\caption{Summary of numerical schemes employed for the DNS of viscous TGV}
\label{tab:dns_schemes_table}
\end{table}
The reference result included in the plots for comparison is that of \citet{dairay2017numerical} with $512^{3}$ DOF using a compact 6th-order finite-difference scheme to solve the incompressible Navier-Stokes equations. In terms of general performance, from Fig.~\ref{fig:verification-quantities}, the kinetic energy and dissipation rate are in good agreement with the reference DNS result. Furthermore, the $c_{DG}$ NSFR.IR-GL and $c_{DG}$ NSFR.IR-GLL schemes have demonstrated stability without any added dissipation (e.g. from upwinding, over-integration, or a turbulence model) for the successful DNS of a viscous free-shear flow. The peak enstrophy is slightly lower than expected, although higher DOF may be required to resolve it for this given scheme. For a short period in the decaying phase of the flow, the schemes with Roe dissipation (p$3$ $c_{DG}$ NSFR.IR-GL-Roe, and p$3$ Strong DG-Roe-GL-OI) exhibit higher dissipation than the reference result. Adding Roe dissipation reduces the peak attainable enstrophy value as seen in Fig.~\ref{fig:verification-quantities}(b). Overall, $c_{DG}$ NSFR.IR-GL p$3$ and p$7$ are similar however p$7$ has smaller numerical error $\varepsilon-\varepsilon_{\zeta}$ as seen in Fig.~\ref{fig:verification-quantities}(d). 

\begin{figure}[H]
\begin{subfigure}{.495\textwidth}
    \centering
    \includegraphics[width=0.99\linewidth]{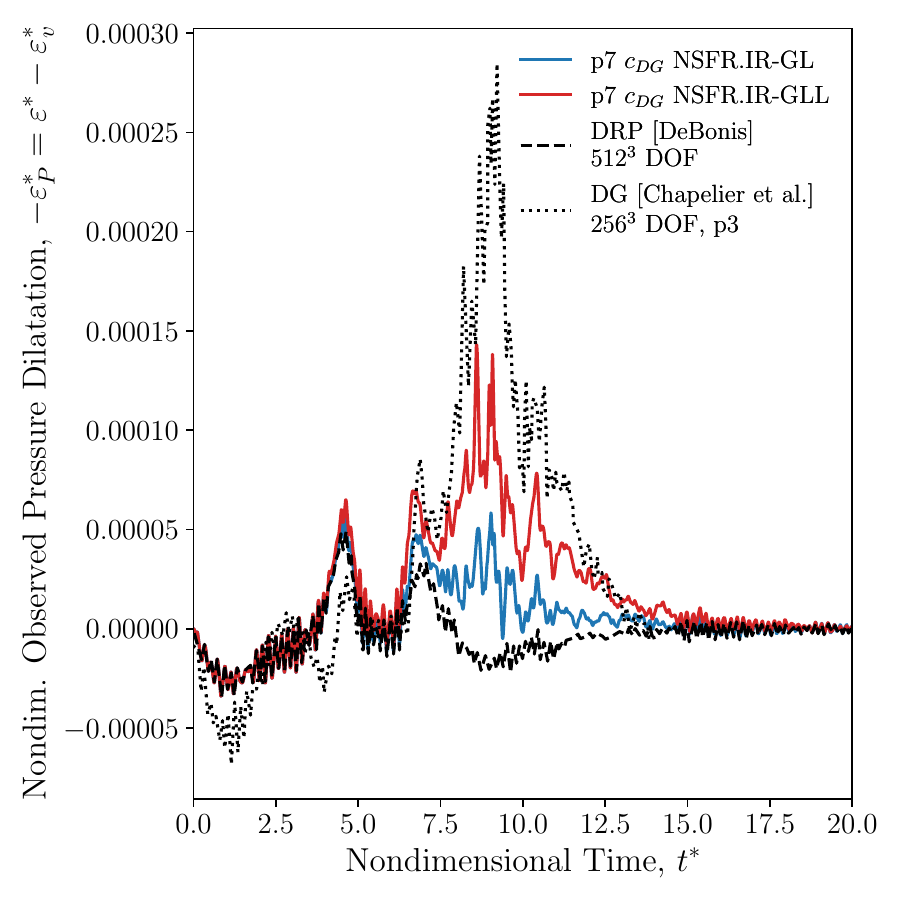}
    \caption{Observed pressure dilatation}
\end{subfigure}%
\begin{subfigure}{.495\textwidth}
    \centering
    \includegraphics[width=0.99\linewidth]{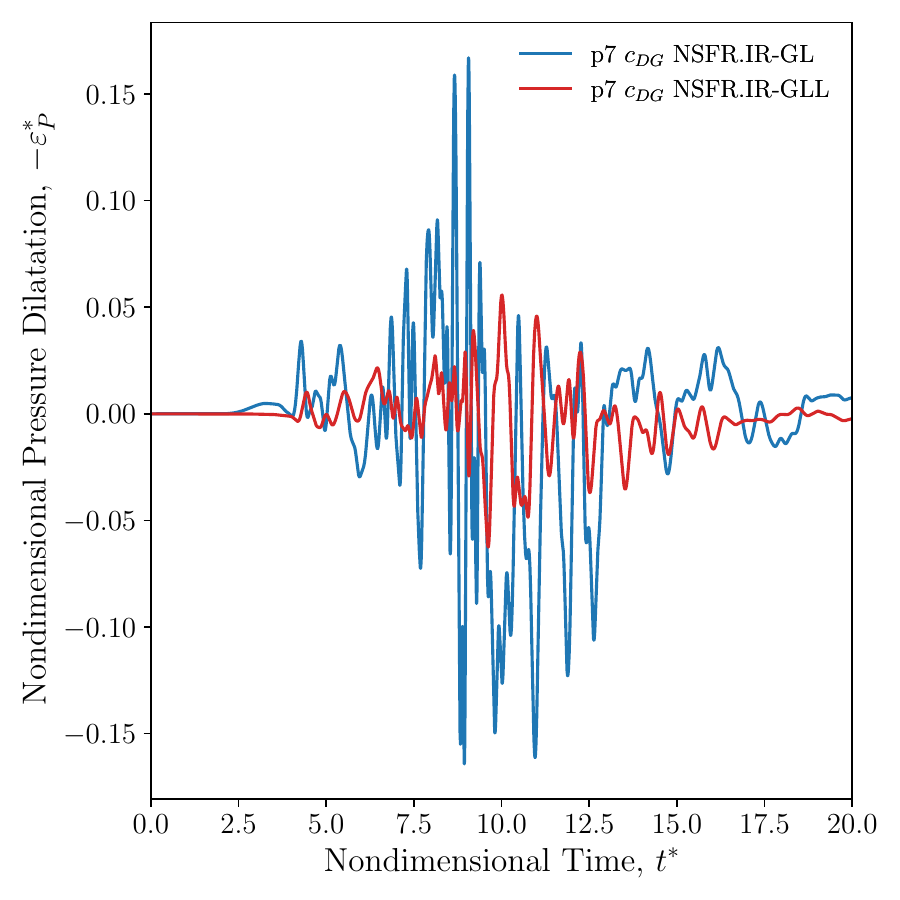}
    \caption{Pressure dilatation}
\end{subfigure}
\caption{Temporal evolution of observed and computed pressure dilation for the viscous TGV at $\text{Re}_{\infty}=1600$ with $256^{3}$ DOF using various p$7$ NSFR schemes}
\label{fig:verification-quantities_pressure_dilatation}
\end{figure}

\begin{figure}[H]
\begin{subfigure}{.495\textwidth}
    \centering
    \includegraphics[width=0.99\linewidth]{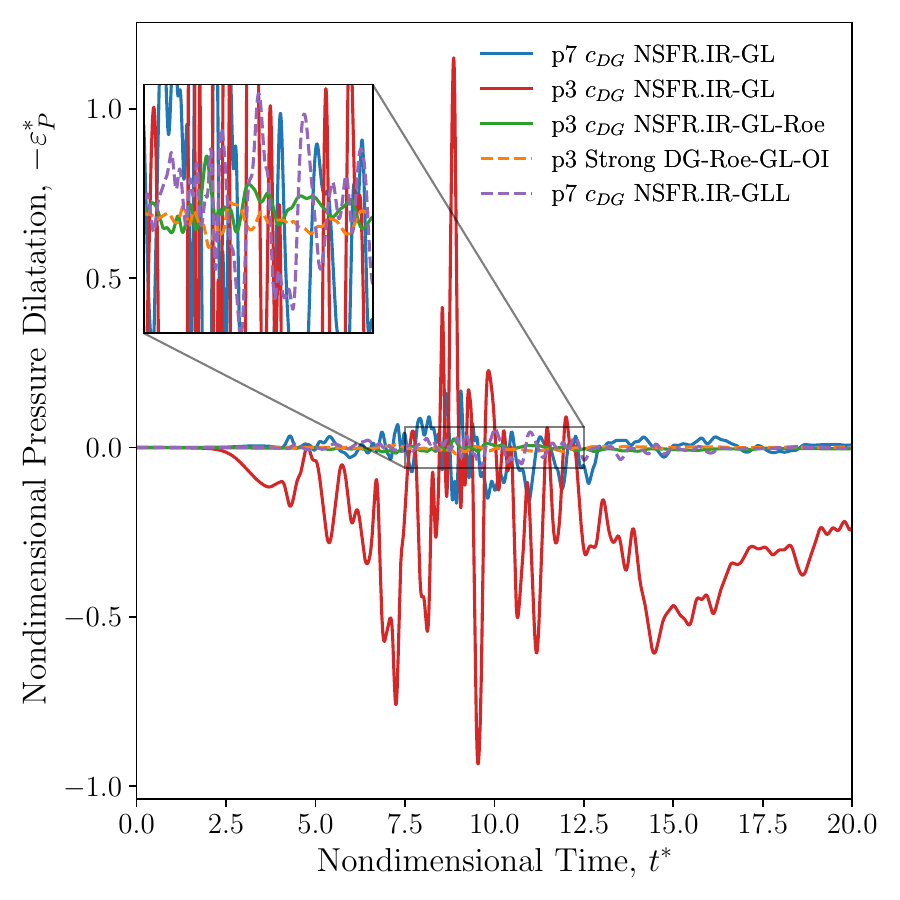}
    \caption{Pressure dilatation}
\end{subfigure}%
\begin{subfigure}{.495\textwidth}
    \centering
    \includegraphics[width=0.99\linewidth]{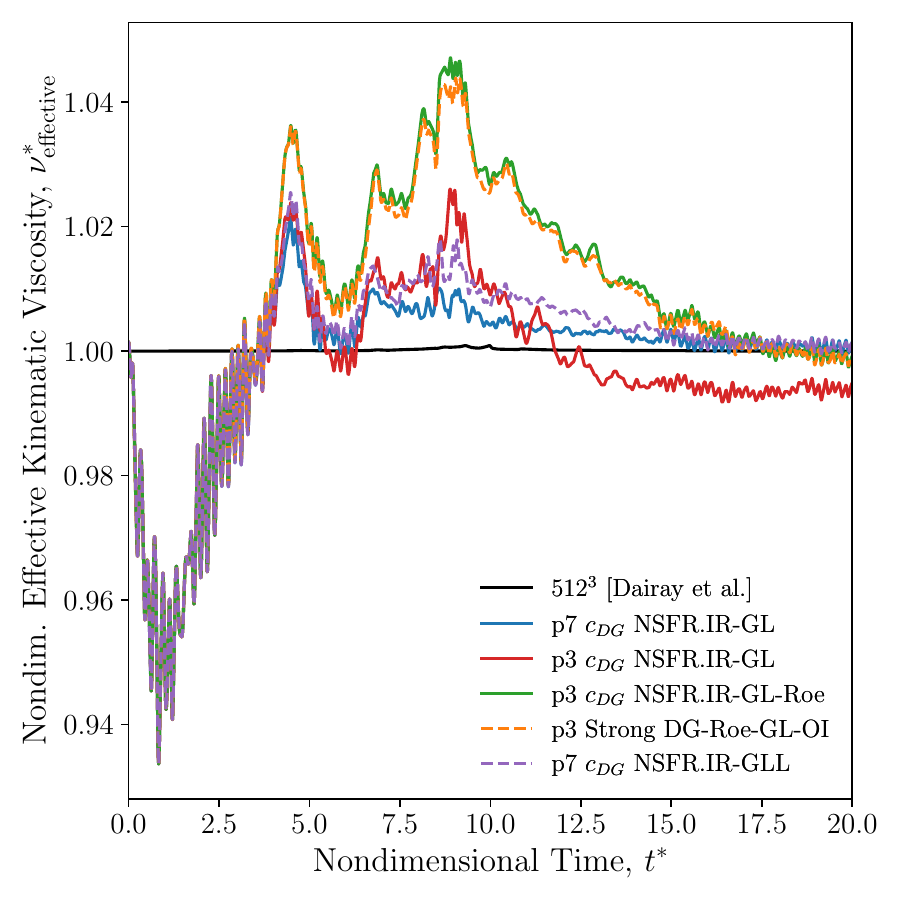}
    \caption{Effective viscosity}
\end{subfigure}
\caption{Temporal evolution of pressure dilatation and effective viscosity for the viscous TGV at $\text{Re}_{\infty}=1600$ with $256^{3}$ DOF using various NSFR schemes and strong~DG-Roe-GL-OI}
\label{fig:verification-quantities_numerical_viscosity}
\end{figure}

In terms of compressibility effects, we now turn to the pressure dilatation dissipation rate $\varepsilon_{P}$. It was found that, the pressure dilatation when computed from Eq.(\ref{eq:dissipation_rate_components_compressible_flow}) with Eq.(\ref{eq:dissipation_rate}), i.e. in the following manner from the right-hand-side terms:
\begin{equation}\label{eq:observed_pressure_dissipation}
    -\varepsilon_{P} \approx \varepsilon - \varepsilon_{v}
\end{equation}
yields quite different results as to directly computing $\varepsilon_{P}$ from the node-wise pressure and velocity divergence; this can be considered the ``observed'' pressure dilatation. For a DNS simulation, the expression of Eq.(\ref{eq:observed_pressure_dissipation}) has little numerical error due to the high DOF and is therefore applicable. Note: We have chosen to present the negative of $\varepsilon_{P}$ in Eq.(\ref{eq:observed_pressure_dissipation}) to be consistent with the literature (e.g. \cite{debonis2013solutions,chapelier2012final}). 
The temporal evolution of the observed pressure dilatation dissipation rate, $-\varepsilon_{P}=\varepsilon-\varepsilon_{v}$, is presented in Fig.~\ref{fig:verification-quantities_pressure_dilatation}(a). The results are compared to two reference results: (1) \citet{chapelier2012final} at $256^3$ DOF using standard DG (OI) with LxF BR2 RK3 with p$3$ and $64^3$ elements, and (2) \citet{debonis2013solutions} at $512^{3}$ DOF using a dispersion relation preserving (DRP) scheme which is a 4th-order accurate finite-difference method. We see that for the $256^3$ DOF p$7$ results for both collocated (NSFR.IR-GLL) and uncollocated (NSFR.IR-GL) schemes lie in-between the two reference results for approximately $t^{*}\in(7,15)$, while matching the DRP $512^{3}$ DOF at all other times. In the intermediate stage of the flow $t^{*}\in(7,15)$, all schemes are significantly less dissipative than the reference DG $256^{3}$ DOF p$3$ result, with the uncollocated p$7$ scheme being the least dissipative of those considered in this study. The less dissipative nature of the NSFR schemes compared to the reference standard DG result is attributed to the absence of upwind dissipation and over-integration. 
Figure~\ref{fig:verification-quantities_pressure_dilatation}(b) shows the result when $\varepsilon_{P}$ is computed directly from the node-wise pressure and velocity divergence. For all schemes, the magnitude was over-predicted and the root issue is suspected to be related to the representation of the velocity divergence term as hypothesized by \citet{bull2015simulation}. Nevertheless, the fact that the observed $\varepsilon_{p}$ is in agreement with reference results verifies that this issue is independent from the ability of the scheme to produce physically correct results as it is isolated to the on-the-fly post-processing of $\varepsilon_{p}$. 
However, there are still important observations to be made from the directly computed $\varepsilon_{p}$ results. From comparing the two p$7$ results in Fig.~\ref{fig:verification-quantities_pressure_dilatation}(b), we see that a collocated scheme (NSFR.IR-GLL) exhibits a significantly lower fluctuation magnitude. From the p$3$ results shown in Fig.~\ref{fig:verification-quantities_numerical_viscosity}(a), it is evident that the lower polynomial order leads to larger fluctuations compared to the higher-order, while the dissipation from the Riemann solver effectively dampens these fluctuations. The latter remark was also found by \citet{flad2017use} who ran a decaying homogeneous isotropic turbulence case using a kinetic energy preserving split form DG-scheme with and without added upwind-dissipation from the low-dissipation Roe Riemann solver. 

The temporal evolution of effective viscosity is presented in Fig.~\ref{fig:verification-quantities_numerical_viscosity}(b). We see that all schemes are within 4.5\% of the physical viscosity of an incompressible flow. Since the physical viscosity we are comparing to would be that of an incompressible flow from which this expression was derived, we can expect some error given the results are for the compressible Navier-Stokes equations. All schemes are similar for $t^{*}<5$, after this point, the collocated p$7$ scheme yields the smallest error (closest to 1), closely followed by the uncollocated p$7$ and p$3$ (NSFR.IR-GL) schemes, while the schemes with upwind-dissipation exhibit the largest errors and are nearly identical. In addition, for $t^{*}>5$  the non-dimensional effective viscosity is greater than 1 at all times except for the p$3$ uncollocated dissipation free scheme (NSFR.IR-GL). Meaning that for $t^{*}>5$, only this scheme was ever more dissipative than it should be according to the physical viscosity of the flow; this occurred briefly around $t^{*}\approx6$ and for almost the entire decay flow phase $t^{*}>11$. 

\begin{figure}[H]
    \centering
    \includegraphics[width=0.495\linewidth]{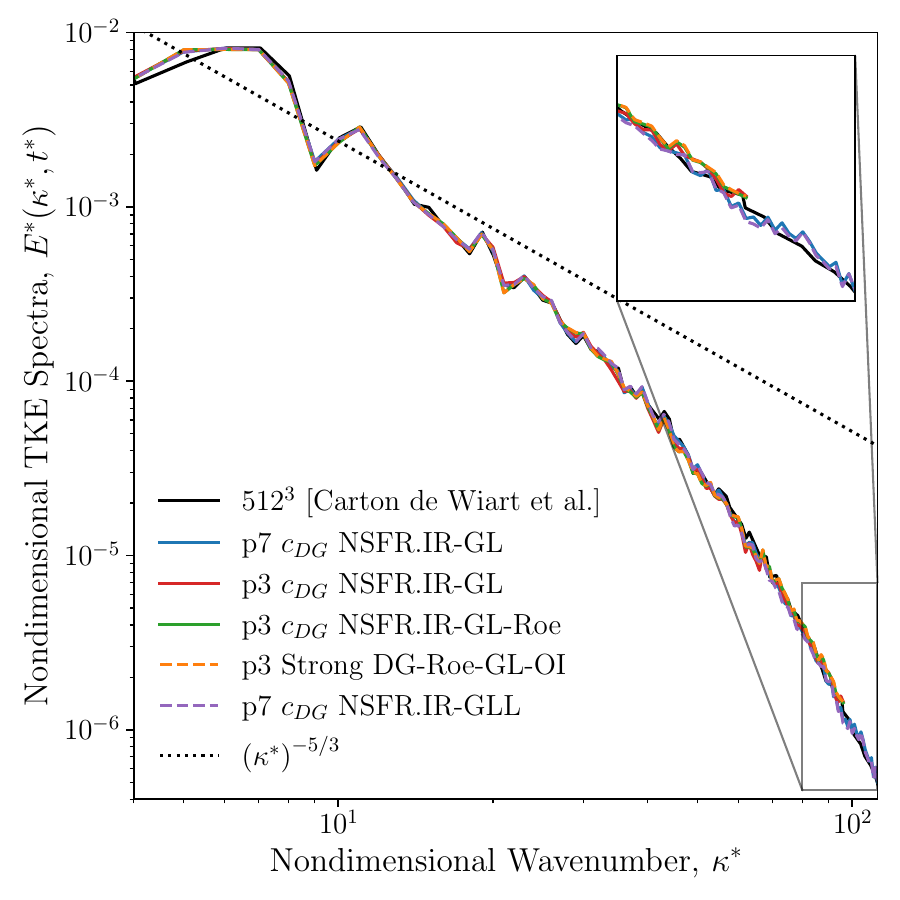}
\caption{Nondimensional TKE spectra of viscous TGV at $\text{Re}_{\infty}=1600$ using various NSFR schemes and strong~DG-Roe-GL-OI with $256^{3}$ DOF plotted until the cut-off wavenumber $\kappa_{c}$. For this DOF, $\kappa_{c}=96$ for the p$3$ schemes, and $\kappa_{c}=122$ for the p$7$ schemes.}
\label{fig:verification-spectra}
\end{figure}

\begin{figure}[H]
\begin{subfigure}{.495\textwidth}
    \centering
    \includegraphics[width=0.99\linewidth]{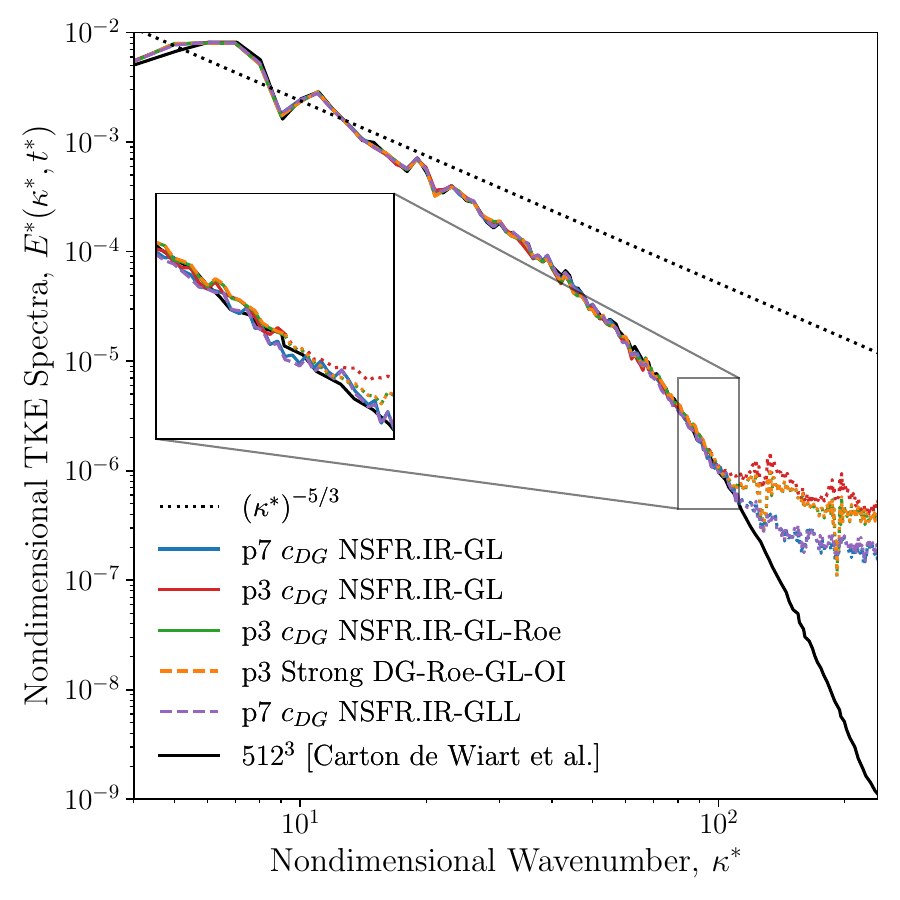}
    \caption{Based on the oversampled flow field, $n_{\text{equid.}}^{1\text{D}}=2(\text{p}+1)$}
\end{subfigure}%
\begin{subfigure}{.495\textwidth}
    \centering
    \includegraphics[width=0.99\linewidth]{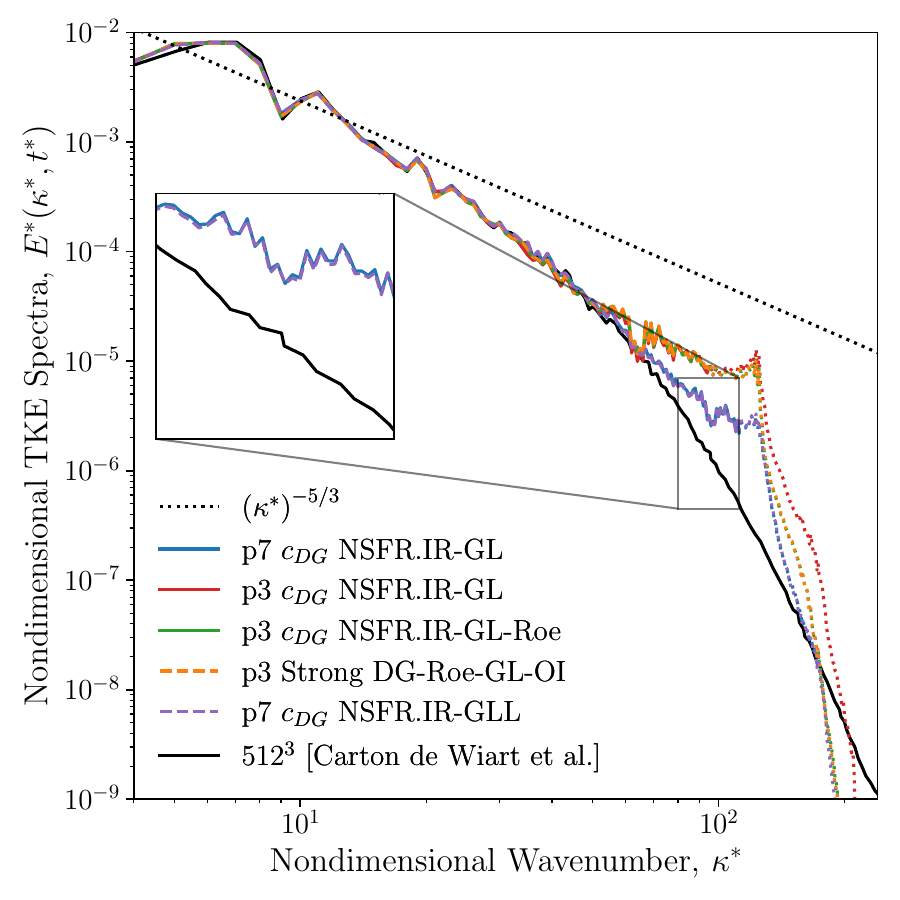}
    \caption{Without oversampling, $n_{\text{equid.}}^{1\text{D}}=\text{p}+1$}
\end{subfigure}
\caption{Nondimensional TKE spectra of viscous TGV at $\text{Re}_{\infty}=1600$ using various NSFR schemes and strong~DG-Roe-GL-OI with $256^{3}$ DOF plotted until the cut-off wavenumber $\kappa_{c}$ (solid lines) and unresolved region $\kappa>\kappa_{c}$ (dotted lines). For this DOF, $\kappa_{c}=96$ for the p$3$ schemes, and $\kappa_{c}=122$ for the p$7$ schemes.}
\label{fig:verification-spectra-oversampling-effects}
\end{figure}

\begin{figure}[H]
\begin{subfigure}{.495\textwidth}
    \centering
    \includegraphics[width=0.99\linewidth]{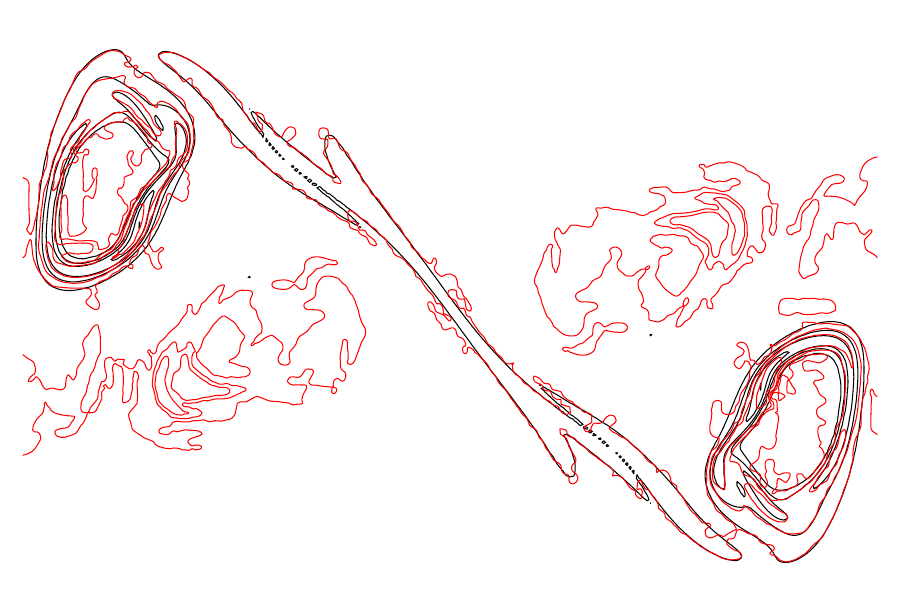}
    \caption{p$7$ $c_{DG}$ NSFR.IR-GL}
\end{subfigure}%
\begin{subfigure}{.495\textwidth}
    \centering
    \includegraphics[width=0.99\linewidth]{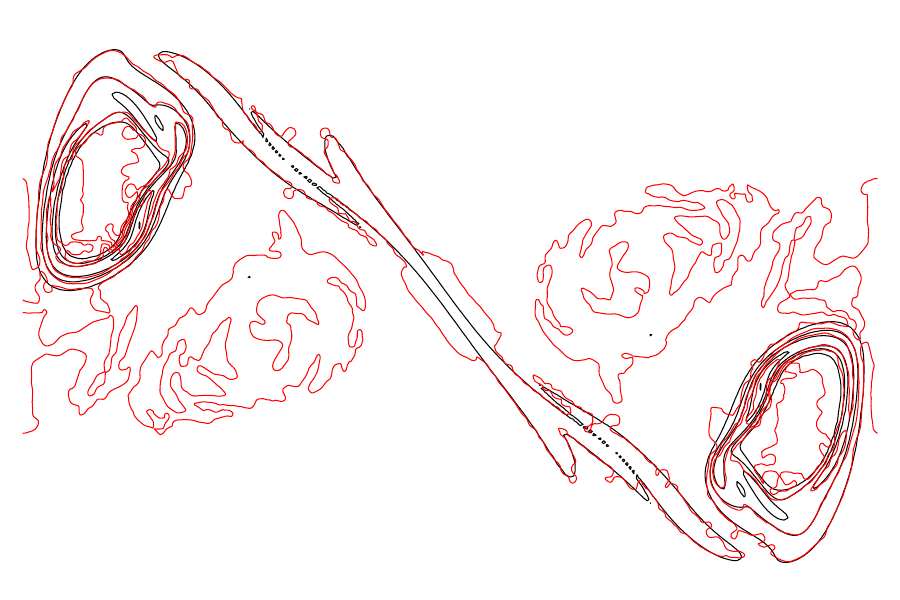}
    \caption{p$7$ $c_{DG}$ NSFR.IR-GLL}
\end{subfigure}
\begin{subfigure}{.495\textwidth}
    \centering
    \includegraphics[width=0.99\linewidth]{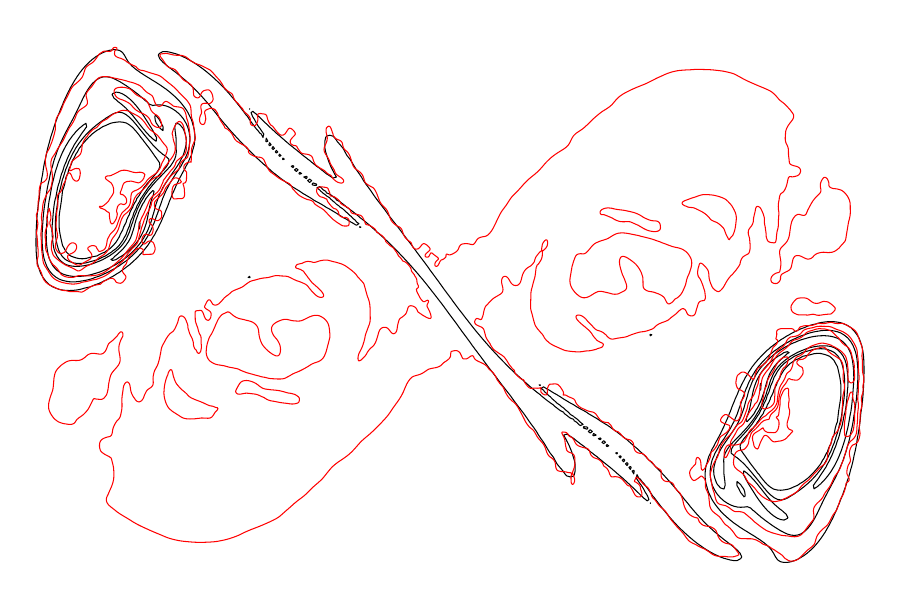}
    \caption{p$3$ $c_{DG}$ NSFR.IR-GL}
\end{subfigure}%
\begin{subfigure}{.495\textwidth}
    \centering
    \includegraphics[width=0.99\linewidth]{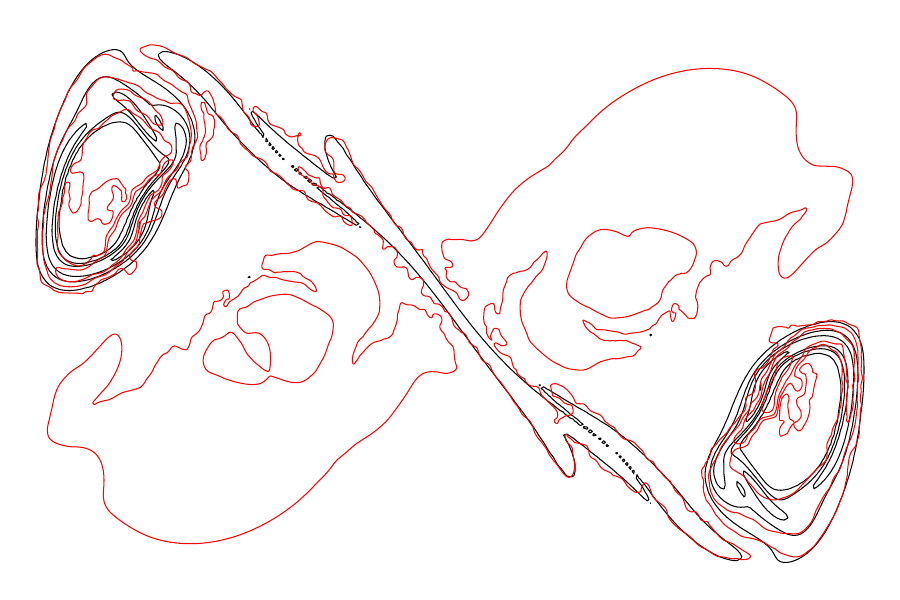}
    \caption{p$3$ $c_{DG}$ NSFR.IR-GL-Roe}
\end{subfigure}
\begin{subfigure}{.99\textwidth}
    \centering
    \includegraphics[width=0.495\linewidth]{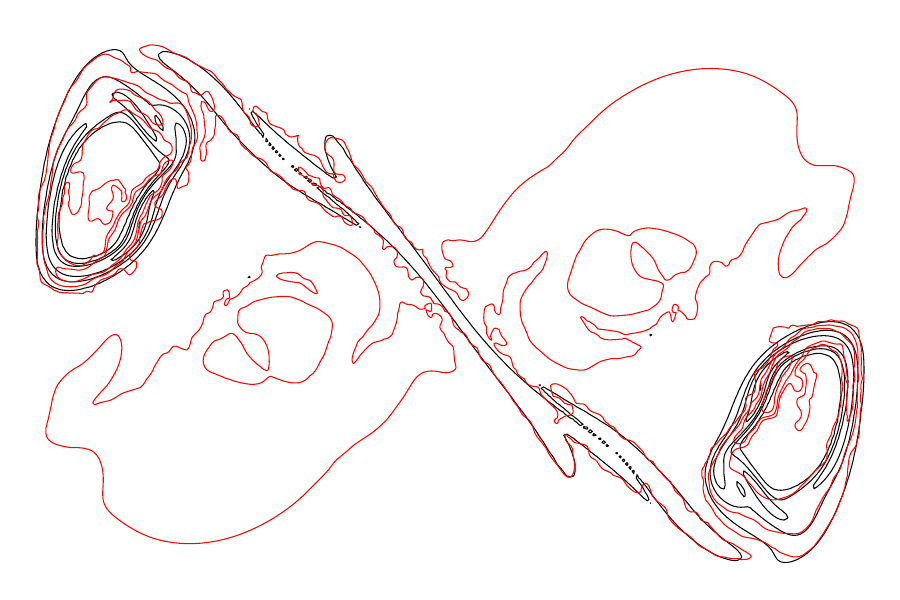}
    \caption{p$3$ Strong DG-Roe-GL-OI}
\end{subfigure}%
\caption{Contours of nondimensional vorticity magnitude $\left|\bm{\omega}\right|=\left[1,~5,~10,~20,~30\right]$ at time $t^{*}=8$ in the plane $x^{*}=0$ (one quadrant shown due to symmetry) for the viscous TGV at $\text{Re}_{\infty}=1600$ with $256^{3}$ DOF (red lines) using various NSFR schemes and strong DG-Roe-GL-OI with reference $512^{3}$ DOF PS results (black lines) \cite{van2011comparison}}
\label{fig:verification-vorticity-magnitude}
\end{figure}

\indent Figure~\ref{fig:verification-spectra} shows the turbulent kinetic energy (TKE) spectra at $t^{*}=9.0$ compared to a pseudo-spectral computation on a $512^3$ grid generated by \citet{carton2014assessment} as a reference.
The TKE spectra for each result are truncated at the respective cut-off wavenumber $\kappa_{c}$ tabulated below:
\begin{table}[H]
\centering
\begin{tabular}{c|c|c|c|c|c}
    \hline
    p & \# Elements & \# Numerical DOF, $N_{\text{el.}}(\text{p}+1)^{3}$ & \# Unique DOF, $N_{\text{el.}}\text{p}^{3}$ & $\kappa_{c}$ & $\kappa_{\text{grid}}$ \\
    \hline
    $3$ & $64^{3}$ & $256^{3}$ & $192^{3}$ & $96$ & $32$\\
    $7$ & $32^{3}$ & $256^{3}$ & $224^{3}$ & $122$ & $16$\\
    \hline
\end{tabular}
\caption{Resolution details of high-order DOF configurations considered for the DNS of viscous TGV}
\label{tab:dns_dofs_table}
\end{table}
\noindent where $\kappa_{\text{grid}}$ is the grid cut-off wavenumber. Comparing the p$7$ to the p$3$, the p$7$ schemes marginally outperform the p$3$ schemes as shown in the zoomed section of the highest wavenumbers. For all results, the spectra are nearly identical to the reference $512^3$ result, capturing all length scales in the flow up to $\kappa_{c}$. Without oversampling the velocity field, this would not have been immediately evident -- the effects of oversampling for the DNS results are illustrated in Fig.~\ref{fig:verification-spectra-oversampling-effects}. We see that the apparent pile-up of TKE approaching $\kappa_{c}$ in Fig.~\ref{fig:verification-spectra-oversampling-effects}(b) is completely reduced, as also observed by \citet{carton2014assessment}, and shifted beyond $\kappa_{c}$ to the unresolved region $\kappa>\kappa_{c}$ of the spectra when oversampling is used.
\indent Lastly, Fig.~\ref{fig:verification-vorticity-magnitude} shows the contours of the instantaneous non-dimensional vorticity magnitude $\left|\bm{\omega}\right|$ (omitting asterisks for clarity) in the $x^{*}=0$ plane at time $t^{*}=8$. In each plot, the $512^{3}$ pseudo-spectral (PS) results of \citet{van2011comparison} are included (solid black contours) for reference; these results were made available by the DLR at \url{http://www.as.dlr.de/hiocfd/case_c3.5.pdf}. For both the $256^{3}$ results and PS reference, the contour lines are defined for values $\left|\bm{\omega}\right|=\left[1,~5,~10,~20,~30\right]$ as done in \cite{van2011comparison}. We see that for all results, the vorticity magnitude contours are in good agreement with the reference DNS result of \citet{van2011comparison}, capturing the location of large vortical structures with the presence of one large additional structure that is absent from the PS results. The discrepancy was also seen in the DNS results of \citet{bull2015simulation} using p$3$ FR-NDG with $256^{3}$ DOF, and those of \citet{chapelier2012inviscid} using p$3$ DG with $256^{3}$ DOF. This is believed to be due to the lower resolution ($256^{3}$ DOF) nature of the results in comparison to the $512^{3}$ DOF PS reference. \citet{carton2014assessment} performed a convergence study on this quantity and showed that for p$3$ DG, $128^{3}$ elements yielding $512^{3}$ DOF were required to fully capture the vorticity contours of the PS method. Furthermore, by comparing the p$7$ and p$3$ results, we can see that the size of this feature is greatly diminished with a higher order of accuracy, suggesting this is simply a low-resolution artifact. Therefore, suggesting that even higher degrees of freedom are required to fully resolve this spatial quantity. In terms of the performance of the different schemes presented, the higher polynomial degree schemes outperform their low-order counterparts by capturing more small-scale flow features, while the uncollocated scheme (NSFR.IR-GL) captures even more vortical structures due to the higher integration accuracy compared to the collocated scheme. In addition, we see that the presence of upwind dissipation dampens the flow field and that the contours of these upwind-including schemes are nearly identical.

\subsubsection{Performance for under-resolved simulations}
\begin{figure}[H]
\begin{subfigure}{.495\textwidth}
    \centering
    \includegraphics[width=0.99\linewidth]{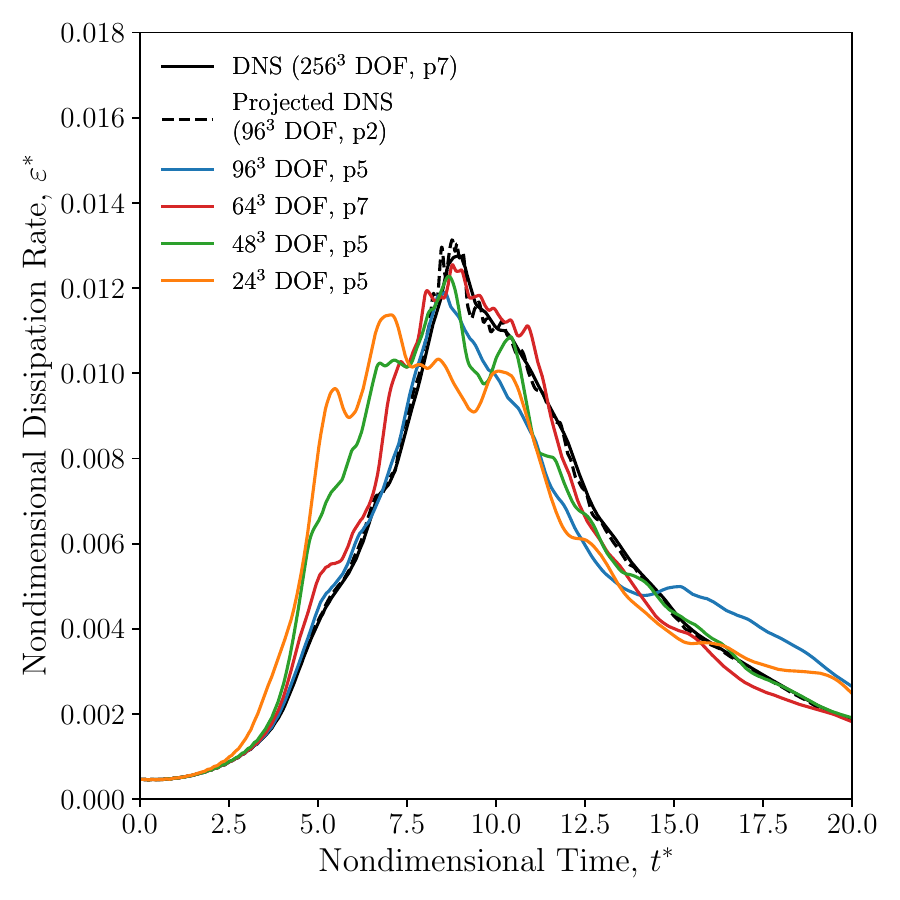}
    \caption{Dissipation rate vs time}
\end{subfigure}%
\begin{subfigure}{.495\textwidth}
    \centering
    \includegraphics[width=0.99\linewidth]{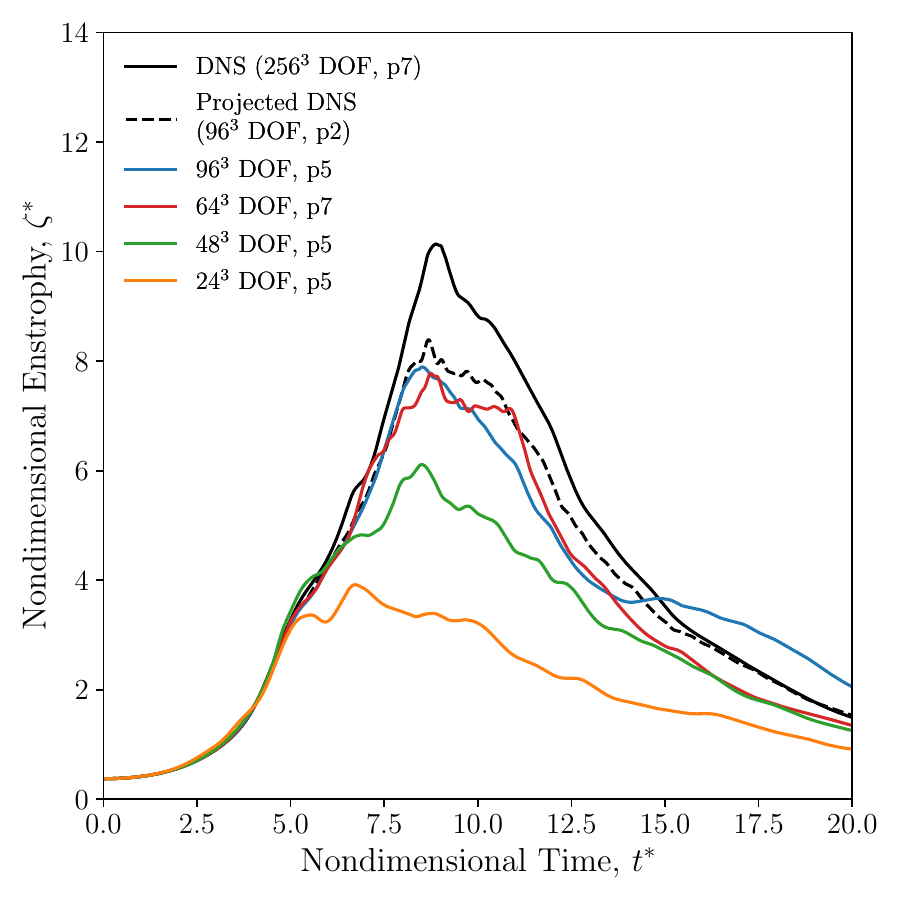}
    \caption{Enstrophy vs time}
\end{subfigure}%
\caption{Temporal evolution of dissipation rate and enstrophy for the viscous TGV at $\text{Re}_{\infty}=1600$ using $c_{DG}$ NSFR.IR-GL with various DOF}
\label{fig:cDG-nsfr-convergence-quantities}
\end{figure}

\begin{figure}[H]
\begin{subfigure}{.495\textwidth}
    \centering
    \includegraphics[width=0.99\linewidth]{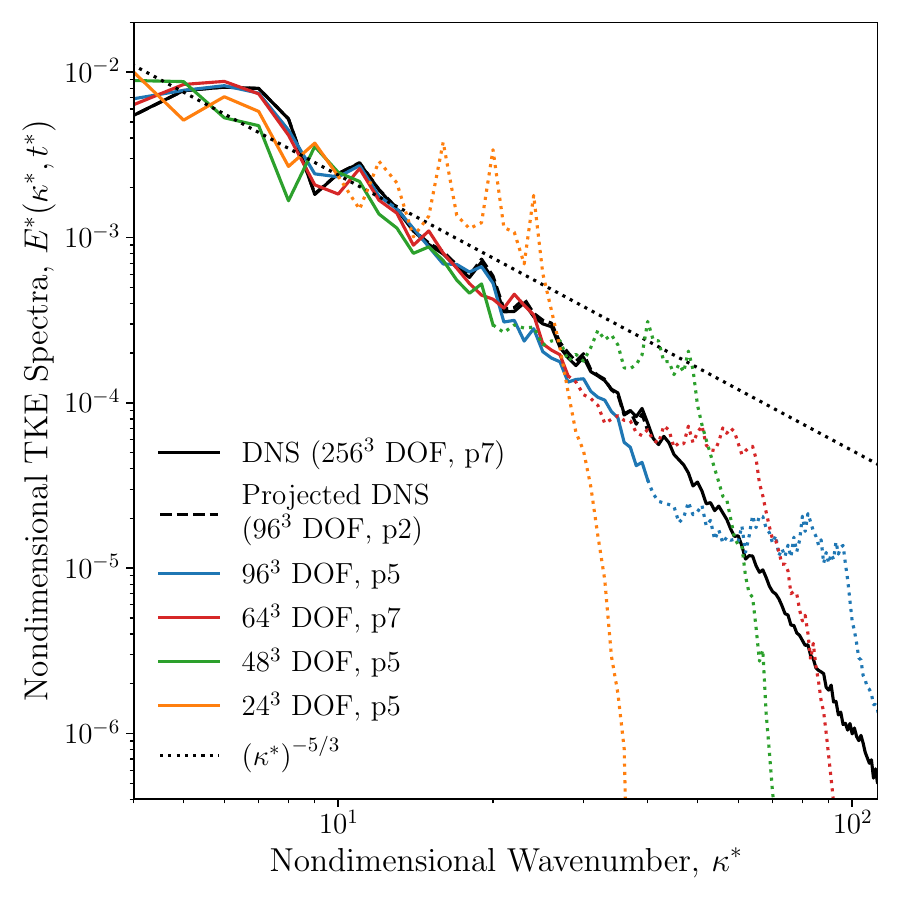}
    \caption{Based on the oversampled flow field}
\end{subfigure}%
\begin{subfigure}{.495\textwidth}
    \centering
    \includegraphics[width=0.99\linewidth]{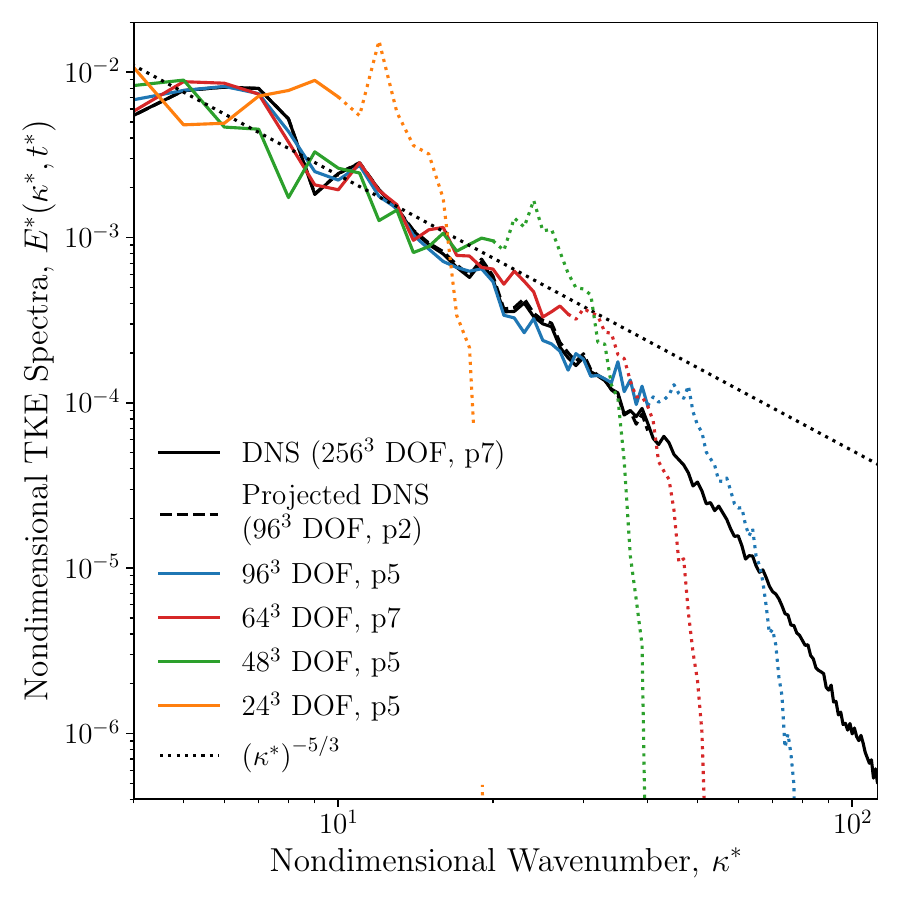}
    \caption{Without oversampling the flow field}
\end{subfigure}%
\caption{Nondimensional turbulent kinetic energy spectra at $t^{*}=9$ of viscous TGV at $\text{Re}_{\infty}=1600$ using $c_{DG}$ NSFR.IR-GL with $96^3$, $64^3$, $48^3$, and $24^3$ DOF plotted until their respective cut-off wavenumbers: $\kappa_{c}=40,~28,~20,~10$ (solid lines) with the unresolved region $\kappa>\kappa_{c}$ (dotted lines)}
\label{fig:cDG-nsfr-convergence-tke_spectra}
\end{figure}

\begin{figure}[H]
\begin{subfigure}{.495\textwidth}
    \centering
    \includegraphics[width=0.99\linewidth]{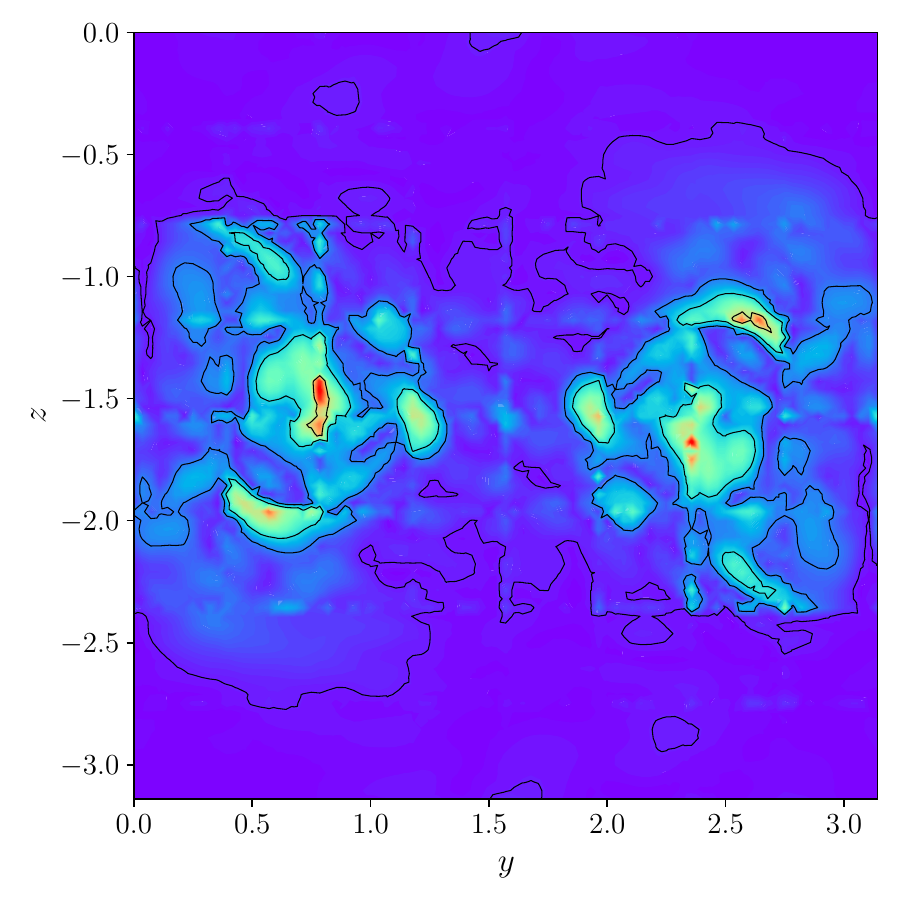}
    \caption{Oversampled flow field, $n_{\text{equid.}}^{1\text{D}}=2(\text{p}+1)$}
\end{subfigure}%
\begin{subfigure}{.495\textwidth}
    \centering
    \includegraphics[width=0.99\linewidth]{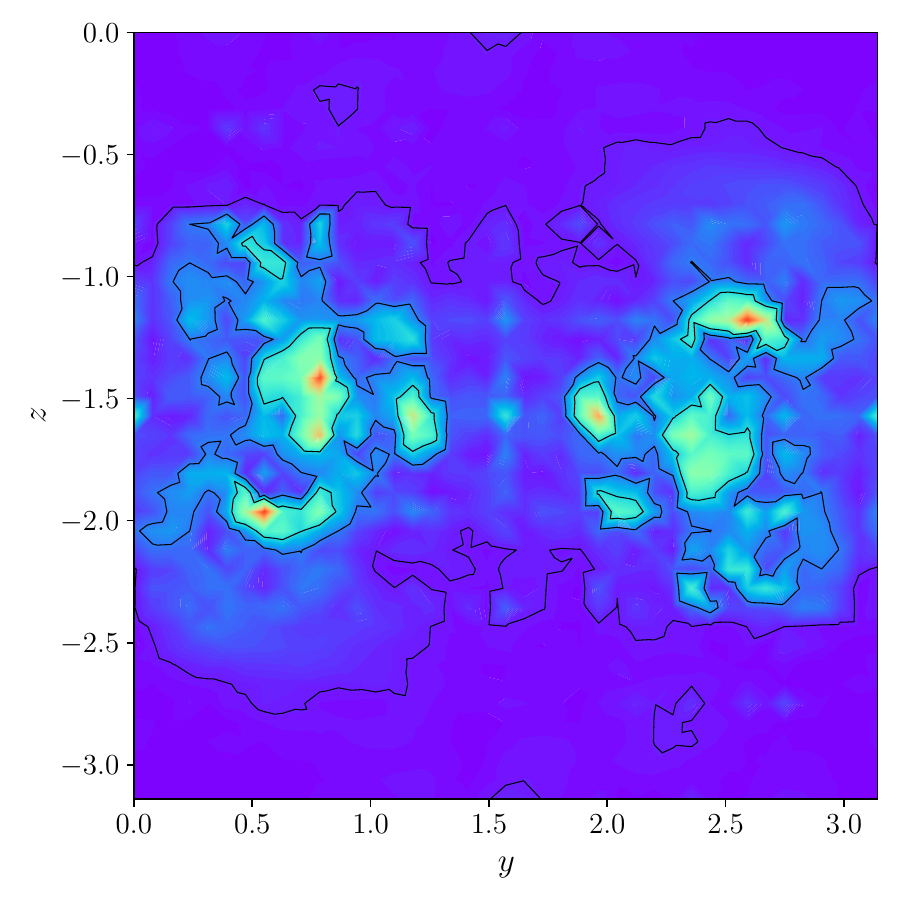}
    \caption{Without oversampling, $n_{\text{equid.}}^{1\text{D}}=\text{p}+1$}
\end{subfigure}
\caption{Filled contours of vorticity magnitude $\left|\bm{\omega}\right|$ at time $t^{*}=9$ in the plane $x^{*}=0$ (one
quadrant shown due to symmetry) for the viscous TGV at $\text{Re}_{\infty}=1600$ using p$5$ $c_{DG}$ NSFR.IR-GL with $96^{3}$ DOF}
\label{fig:iles_vorticity_contour_oversampling_effects}
\end{figure}

\indent In this section, we investigate the performance of NSFR schemes for under-resolved simulations of the TGV; i.e. at DOF less than that of the DNS. From this section onwards, the $256^3$ DOF p$7$ $c_{DG}$ NSFR.IR-GL result will be used as the reference DNS in the plots presented. The resolutions considered in this section are presented below in Table \ref{tab:iles_dofs_table}. 
\begin{table}[H]
\centering
\begin{tabular}{c|c|c|c|c|c}
    \hline
    p & \# Elements & \# Numerical DOF, $N_{\text{el.}}(\text{p}+1)^{3}$ & \# Unique DOF, $N_{\text{el.}}\text{p}^{3}$ & $\kappa_{c}$ & $\kappa_{\text{grid}}$\\
    \hline
    $5$ & $16^{3}$ & $96^{3}$ & $80^{3}$ & $40$ & $8$\\
    $7$ & $8^{3}$ & $64^{3}$ & $56^{3}$ & $28$ & $4$\\
    $5$ & $8^{3}$ & $48^{3}$ & $40^{3}$ & $20$ & $4$\\
    $5$ & $4^{3}$ & $24^{3}$ & $20^{3}$ & $10$ & $4$\\
    \hline
\end{tabular}
\caption{Under-resolved resolutions/DOF considered and their corresponding cutoff wavenumbers}
\label{tab:iles_dofs_table}
\end{table}
The $c_{DG}$ NSFR.IR-GL scheme (detailed in Table \ref{tab:dns_schemes_table}) is run for all resolutions in Table \ref{tab:iles_dofs_table}. The temporal evolution of dissipation rate and enstrophy are presented in Fig.~\ref{fig:cDG-nsfr-convergence-quantities}(a) and (b) respectively, demonstrating that the entropy conserving scheme is stable for any grid without any added dissipation. 
At the resolution $96^3$ DOF (p$5$), the dissipation rate is very close to that of the DNS until reaching the peak around $t^{*}\approx8$, where the under-resolved nature of the simulation becomes apparent throughout the entire decaying phase of the flow. As the DNS is well-tracked throughout the entire transitional phase of the flow from initially laminar to a fully turbulent flow, the $96^3$ DOF (p$5$) resolution has been chosen for the remaining iLES investigations in the subsequent sections of this work. 
For this reason, we have included the reference DNS result (p$7$, $32^{3}$ elements) projected to $96^3$ DOF (p$2$, $32^{3}$ elements) for comparison on all unsteady quantity plots. This result is denoted as the \textit{projected DNS} ($96^3$ DOF, p$2$) in the legends. 
For the lower resolutions ($64^3$, $48^3$, and $24^3$ DOF), the under-resolved nature is apparent much earlier, around $t^{*}<5$ for this quantity. 
Similar trends are observed for the time evolution of enstrophy in Fig.~\ref{fig:cDG-nsfr-convergence-quantities}(b), we see that a higher resolution allows for longer sustained capture of enstrophy with respect to the DNS result. By comparing the $96^3$ DOF (p$5$) and $64^3$ DOF (p$7$) results, we see that the higher polynomial degree is more effective at capturing this quantity on a per-DOF basis. 
\newline\indent The TKE spectra at $t^{*}=9$ with and without oversampling are shown in Fig.~\ref{fig:cDG-nsfr-convergence-tke_spectra}(a) and (b), respectively. From the TKE spectra based on the oversampled flow field shown in Fig.~\ref{fig:cDG-nsfr-convergence-tke_spectra}(a), each resolution considered generally captures all length scales up to $\kappa_{c}$ with a slight deficit of TKE approaching the smallest resolved scales. In addition, the largest resolved scales were not as well captured for the $48^3$ and $24^3$ DOF resolutions compared to the $96^{3}$ and $64^{3}$ DOF. 
However, it is important to note that without oversampling the velocity flow field, the generated TKE spectra are quite different. 
Figure~\ref{fig:cDG-nsfr-convergence-tke_spectra}(b) demonstrates that for all resolutions, there is now a surplus of TKE in the smallest resolved scales, as also observed by \citet{carton2014assessment}. 
To verify the physical correctness of oversampling, we can compare the instantaneous kinetic energy $E_{K}$ obtained at $t^{*}=9$ from Eq.(\ref{eq:volumetric_averaged_integrated_kinetic_energy}) to the kinetic energy obtained by integrating the TKE spectra up until $\kappa_{c}$, defined as~$\tilde{E}_{K}$:
\begin{equation}\label{eq:KE_from_TKE_integral}
    \tilde{E}_{K} = \int_{\displaystyle\text{min}(\kappa)}^{\displaystyle\kappa_{c}} E(\kappa,t)d\kappa,
\end{equation}
where the integral is computed using the trapezoidal rule. The absolute relative percent error,
\begin{equation}
    \epsilon_{E_{K}}=\frac{\left|\tilde{E}_{K}-E_{K}\right|}{E_{K}},
\end{equation}
is tabulated below for the $256^{3}$ DOF p$7$ (DNS), $96^{3}$ DOF p$5$, and $64^{3}$ DOF p$7$ resolutions.
\begin{table}[H]
\centering
\begin{tabular}{c|c|c|c}
    \hline
    Legend name & $E_{K}$ & $\epsilon_{E_{K}}$ with oversampling & $\epsilon_{E_{K}}$ without oversampling\\
    \hline
    $256^{3}$ DOF, p$7$ & $8.642452e-02$ & $0.082685$\% & $0.148560$\%\\
    $96^{3}$ DOF, p$5$ & $8.573783e-02$ & $0.236968$\% & $1.100619$\%\\
    $64^{3}$ DOF, p$7$ & $8.231268e-02$ & $0.522544$\% & $5.026020$\%\\
    \hline
\end{tabular}
\caption{Percent error in kinetic energy computed from TKE spectra with and without oversampling at $t^{*}=9$}
\label{tab:ke_from_tke_spectra_oversampling_effects}
\end{table}
We see that the error $\epsilon_{E_{K}}$ is significantly diminished when oversampling is performed on the velocity field used to compute the TKE spectra, thereby further strengthening the argument that the oversampling procedure is necessary. To further illustrate the effects of oversampling, the instantaneous vorticity magnitude contours based on the flow field with and without oversampling for the $96^3$ DOF case are shown in Fig.~\ref{fig:iles_vorticity_contour_oversampling_effects}(a) and (b), respectively. We clearly see that without oversampling, the flow field representation is quite coarse and does not capture regions of high vorticity as well. Note that from this point onwards, the unresolved region will be omitted from all TKE spectra plots. 

\subsubsection{De-aliasing strategies}\label{subsubsec:dealiasing_comparison}
\begin{figure}[H]
\begin{subfigure}{.495\textwidth}
    \centering
    \includegraphics[width=0.99\linewidth]{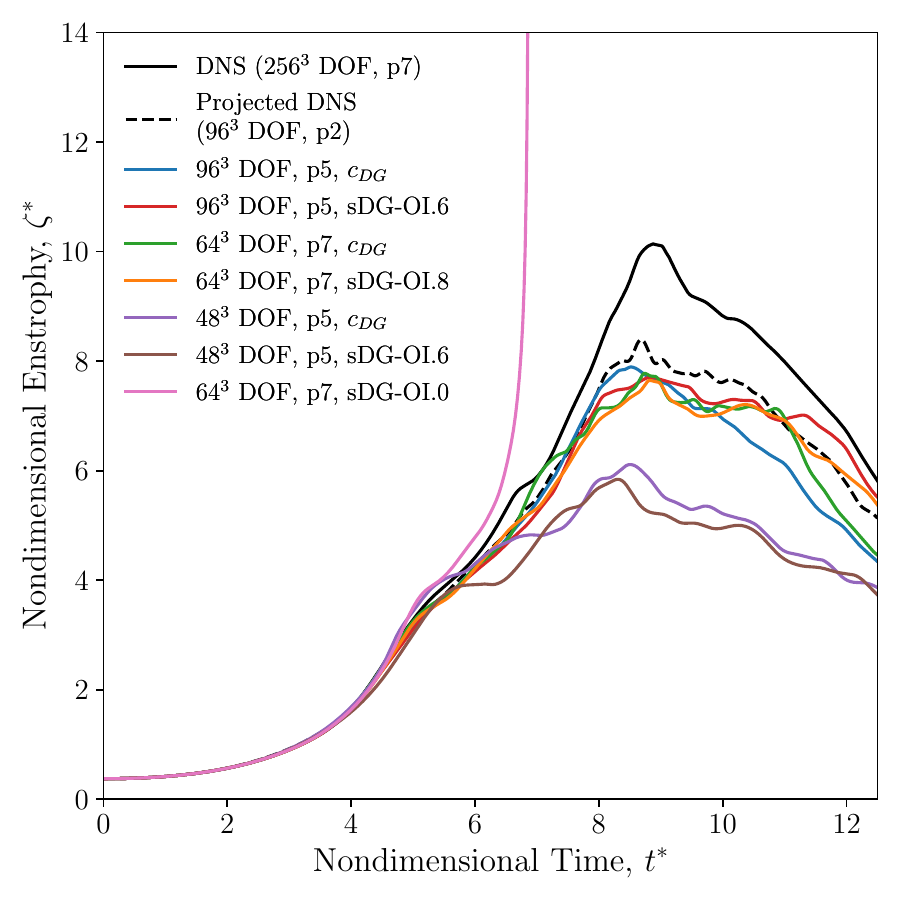}
    \caption{Enstrophy vs time}
\end{subfigure}%
\begin{subfigure}{.495\textwidth}
    \centering
    \includegraphics[width=0.99\linewidth]{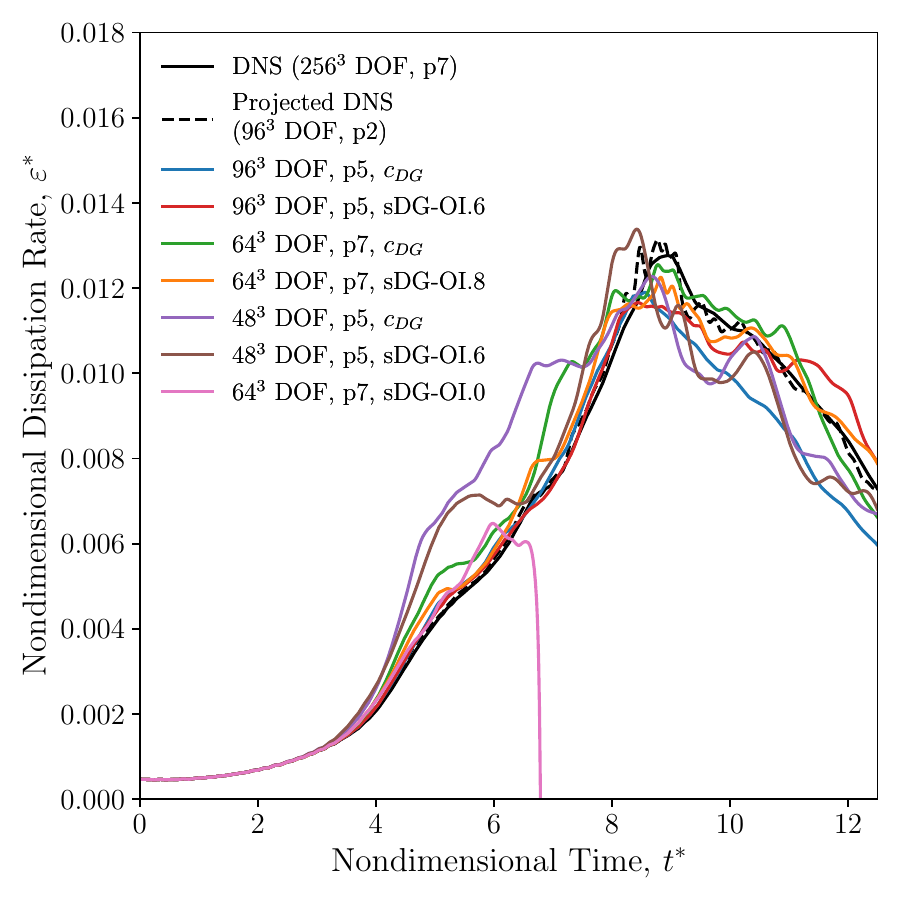}
    \caption{Dissipation rate vs time}
\end{subfigure}
\begin{subfigure}{.495\textwidth}
    \centering
    \includegraphics[width=0.99\linewidth]{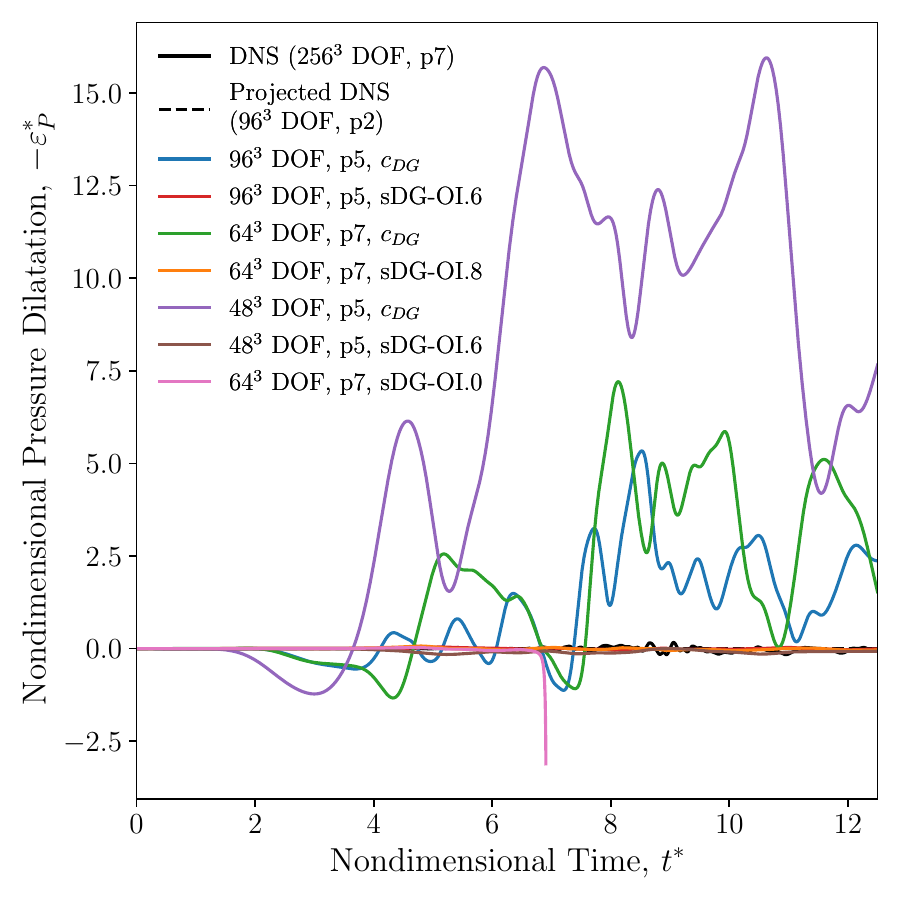}
    \caption{Pressure dilatation vs time}
\end{subfigure}%
\begin{subfigure}{.495\textwidth}
    \centering
    \includegraphics[width=0.99\linewidth]{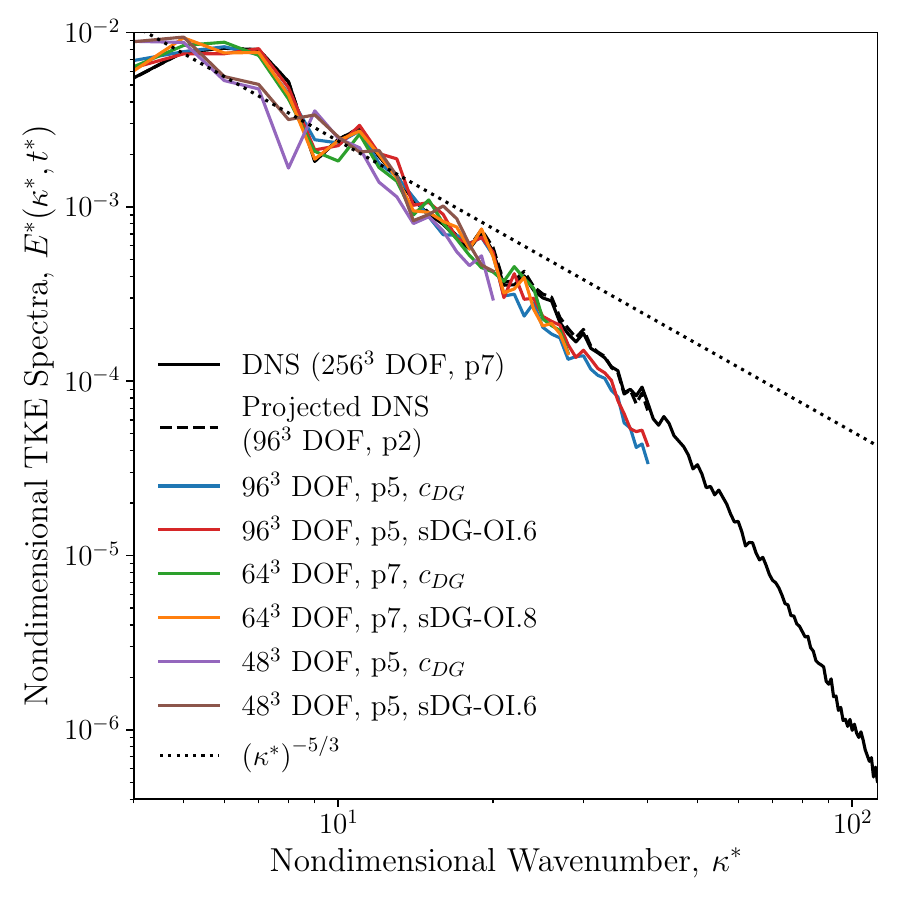}
    \caption{Instantaneous TKE spectra at $t^{*}=9$ using $96^{3},~64^{3},~48^{3}$ DOF plotted until the respective cut-off wavenumbers: $\kappa_{c}=40,~28,~20$}
\end{subfigure}%
\caption{Temporal evolution of enstrophy, dissipation rate, and pressure dilatation along with the TKE spectra at $t^{*}=9$ for the viscous TGV at $\text{Re}_{\infty}=1600$ at various DOF using strong~DG-Roe-GL-OI (denoted as sDG-OI.6 for p$5$, and sDG-OI.8 for p$7$) and $c_{DG}$~NSFR.IR-GL (denoted as $c_{DG}$)} 
\label{fig:split_form_vs_over_integration_convergence-quantities}
\end{figure}

High-order methods for ILES are prone to aliasing issues due to the nonlinearity of the flux. In this section, we will compare two de-aliasing strategies: (1) over-integration, and (2) the use of a split-form. The first strategy uses additional quadrature nodes for numerical integration to be more accurate given the nonlinearity of the flux \cite{kirby2003aliasing}. The approach is the most common and easiest to implement, however, it incurs an increased computational cost by the requirement of evaluations at additional quadrature nodes, e.g. for the compressible Navier-Stokes equations this requires twice the number of quadrature points $N_{\text{quad}}=2\left(p+1\right)$ \cite{beck2014high}.
The second strategy is to reformulate the convective flux as a split formulation that discretely satisfies the integration-by-parts (IBP) property \cite{carpenter2016entropy}. This has the advantage of avoiding the need for additional quadrature nodes, however, the implementation is more complex and requires the evaluation of a two-point flux. 
To compare these two strategies, we have chosen to compare the performance of the classical strong DG with over-integration (strong DG-Roe-GL-OI) scheme, and the $c_{DG}$ NSFR.IR-GL scheme which has a split formulation and does not include Riemann solver dissipation; see Table \ref{tab:dns_schemes_table} for more details on the schemes. 
In addition, classical strong DG without over-integration (sDG-OI.0) has been included for the $64^3$ DOF p$7$ resolution to demonstrate that a linearly stable scheme can be unstable in the absence of over-integration.  
The time evolution of enstrophy, dissipation rate, pressure dilatation, and the instantaneous TKE spectra at $t^{*}=9$ is compared for the two schemes in Fig.~\ref{fig:split_form_vs_over_integration_convergence-quantities}(a-d). The results at all resolutions presented in Table \ref{tab:iles_dofs_table} are included and plotted up to $t^{*}=12.5$ to focus on the key flow phases (transition and initial turbulent decay). From the temporal evolution plots Fig.~\ref{fig:split_form_vs_over_integration_convergence-quantities}(a-c), we see that the sDG-OI.0 scheme is unstable as expected, with the instability occurring during the transitional phase at around $t^{*}=6$.
The time evolution of enstrophy in Fig.~\ref{fig:split_form_vs_over_integration_convergence-quantities}(a) is similar for both schemes throughout the transitional phase of the flow at $96^3$ DOF, with larger differences seen after the peak values have been reached. However, larger differences arise throughout the transition phase as the resolution gets coarser as seen by comparing the $64^3$ and $48^3$ DOF results. At all resolutions, the peak enstrophy captured by the NSFR scheme is higher than that by the over-integrated scheme. This is expected as NSFR is less dissipative by construction with the absence of upwind dissipation (e.g. Roe) and over-integration. 
Similarly for the dissipation rate, both schemes track the DNS quite well throughout the transition phase at $96^3$ DOF. While at $64^3$ DOF, the dissipation rate for the over-integrated scheme is slightly more accurate throughout transition. At $48^3$ DOF, both schemes exhibit phases where the dissipation rate is severely over-predicted throughout transition simply due to the coarse nature of this DOF. 
From the pressure dilatation results, we see that for all resolutions considered, large oscillations are present with the NSFR scheme while absent from the classical strong DG scheme. This result is expected as it was shown in Section \ref{subsec:DNS} that only the schemes with Roe dissipation did not exhibit such large pressure dilatation oscillations. It was shown that adding Roe dissipation to the p$3$ NSFR scheme completely dampened the oscillations. The NSFR schemes in this section are without upwinding, while the classical strong DG scheme has upwind-type dissipation. 
Lastly, the instantaneous TKE spectra in Fig.~\ref{fig:split_form_vs_over_integration_convergence-quantities}(d) reveals that both schemes yield similar spectra at all resolutions. At $96^3$ and $64^3$ DOF, both schemes capture all flow scales except those near the respective cut-off wavenumbers. At $48^3$ DOF, the flow is severely under-resolved and the distribution of TKE differs from the DNS at all wavenumbers. 

\begin{figure}[H]
\begin{subfigure}{.495\textwidth}
    \centering
    \includegraphics[width=0.99\linewidth]{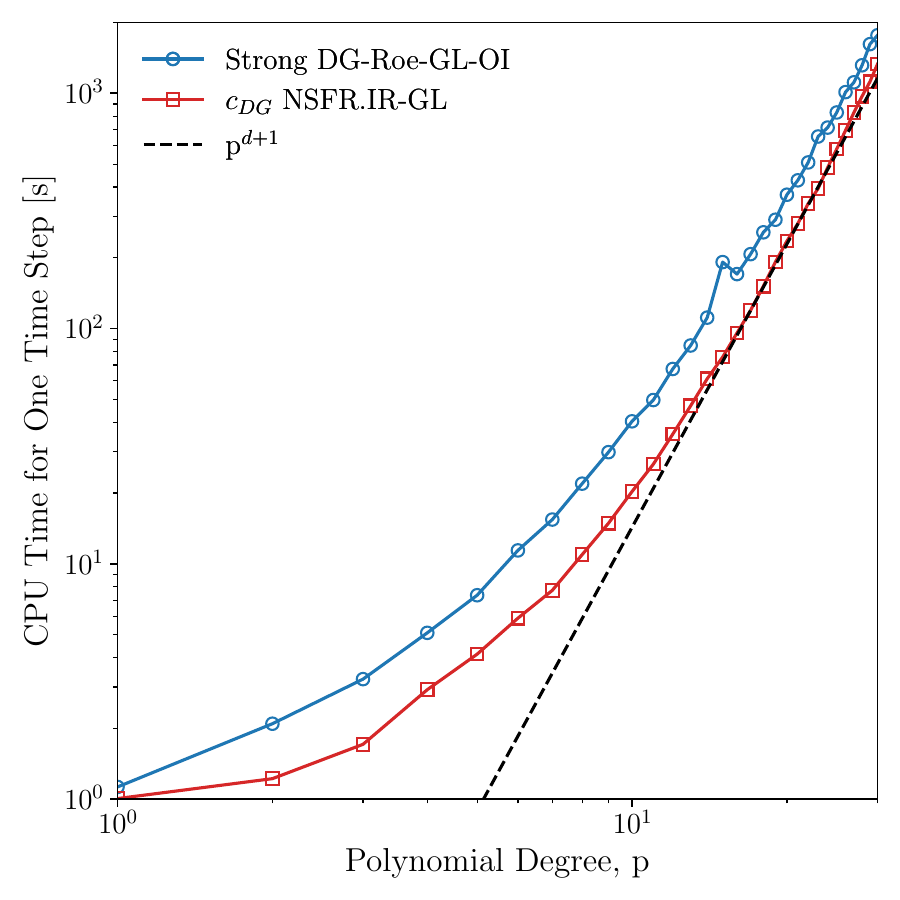}
    \caption{Scaling}
\end{subfigure}%
\begin{subfigure}{.495\textwidth}
    \centering
    \includegraphics[width=0.99\linewidth]{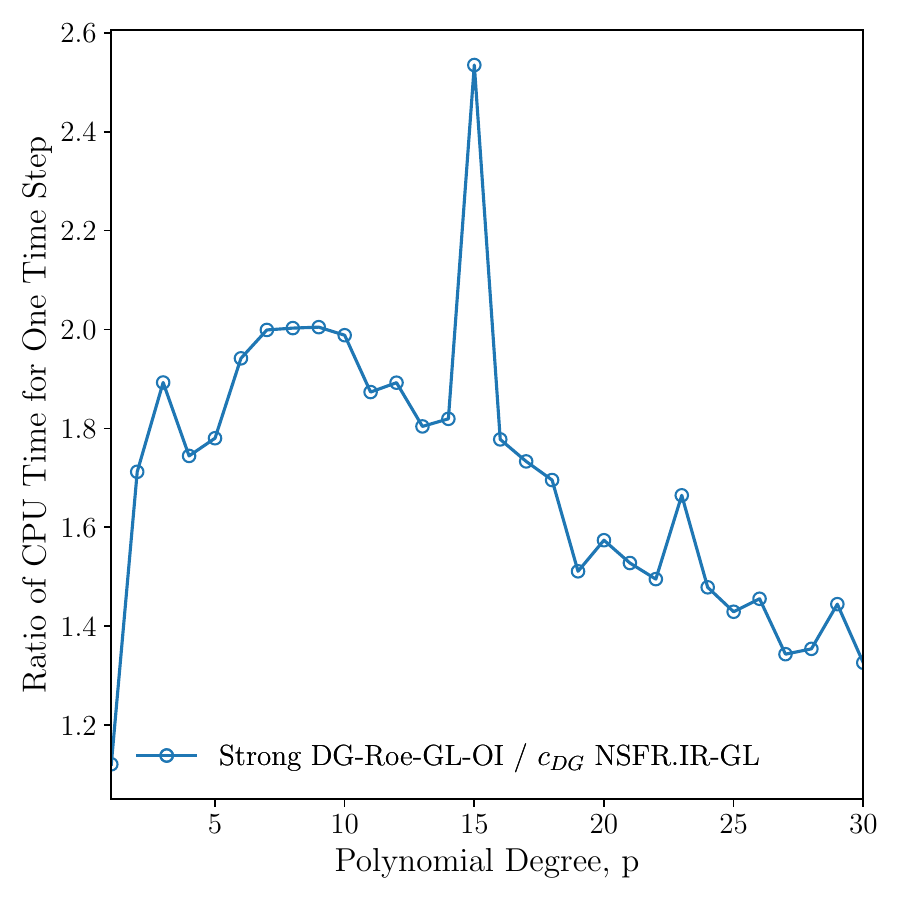}
    \caption{Ratio}
\end{subfigure}
\caption{Scaling of CPU time for one time step of viscous TGV at $\text{Re}_{\infty}=1600$ with $4^3$ elements on 16 CPUs using strong~DG-Roe-GL-OI and $c_{DG}$~NSFR.IR-GL vs polynomial degree}
\label{fig:cpu_cost_oi_vs_sf}
\end{figure}

Next, we compare the computational cost of the two schemes at different polynomial orders $p\in[1,12]$. To do this, we recorded the CPU time elapsed over a single step in time for the TGV problem at each $p$ using $4^3$ uniform elements. This was done 10 times for all $p$, generating 10 sets of timing values that were then averaged. The scaling result of CPU time for one-time step using SSP-RK3 vs polynomial degree is presented in Fig.~\ref{fig:cpu_cost_oi_vs_sf}(a). We see that the correct scaling orders are obtained as also shown by \citet{cicchino2024scalable}. For the useful/common range of $p\in[5,12]$ in Fig.~\ref{fig:cpu_cost_oi_vs_sf}(b), over-integration is approximately twice the CPU time of the split-form scheme. 
Regarding computational resources, each set of timing values was obtained using 16 MPI processes run on 16 cores, each with 3048Mb of memory. The computing resource used is the \href{https://docs.alliancecan.ca/wiki/Narval/en#Node_characteristics}{Narval cluster of the Digital Research Alliance of Canada} which has AMD EPYC Rome processors. Furthermore, all unnecessary calculations (e.g. on-the-fly post-processing) were suppressed in the timing results.

\subsubsection{Collocated vs Uncollocated NSFR}\label{sec:collocation_effects}
\begin{figure}[H]
\begin{subfigure}{.495\textwidth}
    \centering
    \includegraphics[width=0.99\linewidth]{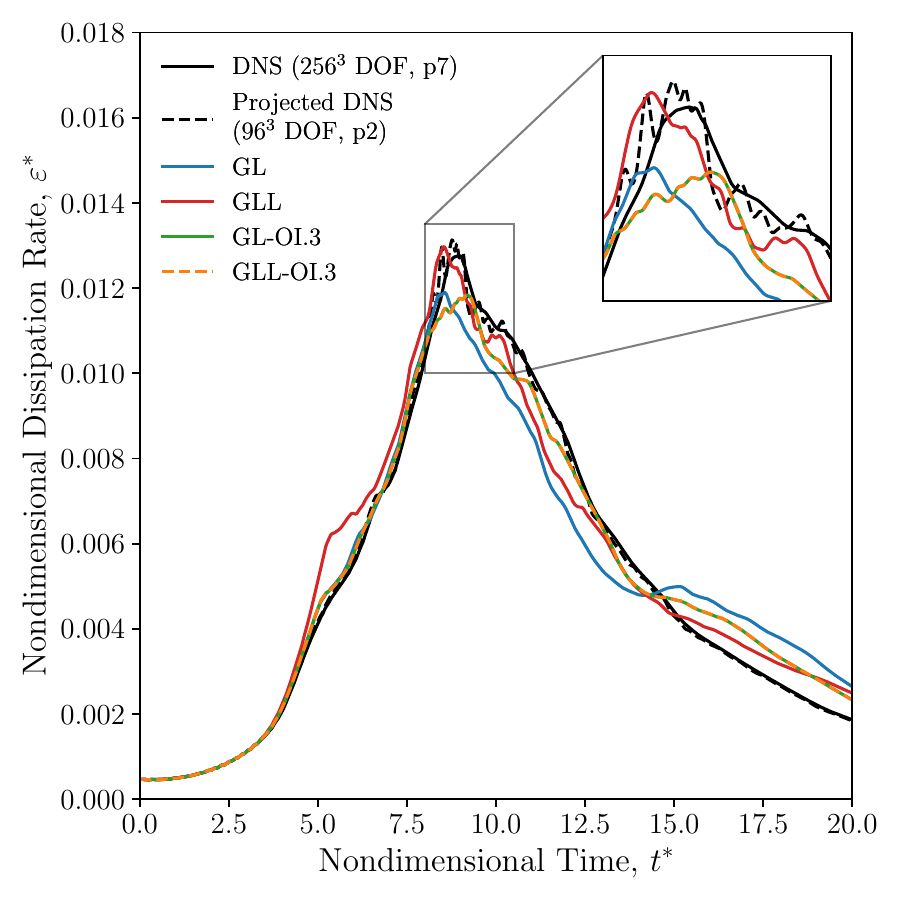}
    \caption{Dissipation rate}
\end{subfigure}%
\begin{subfigure}{.495\textwidth}
    \centering
    \includegraphics[width=0.99\linewidth]{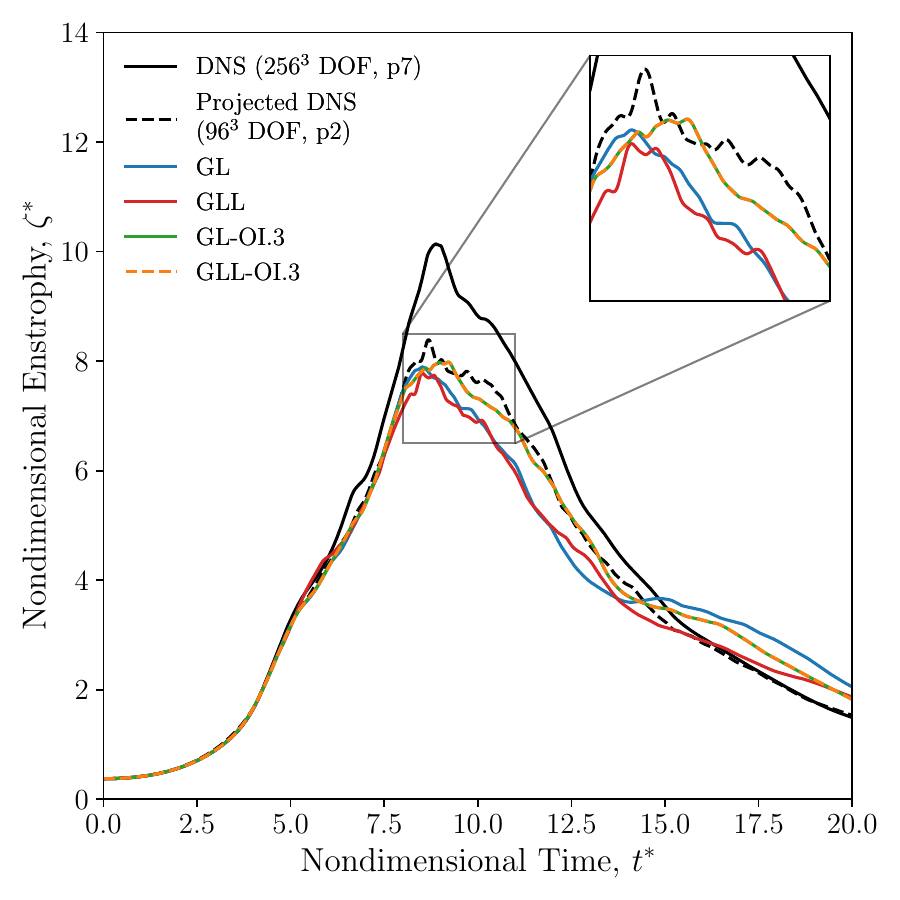}
    \caption{Enstrophy vs time}
\end{subfigure}
\begin{subfigure}{.495\textwidth}
    \centering
    \includegraphics[width=0.99\linewidth]{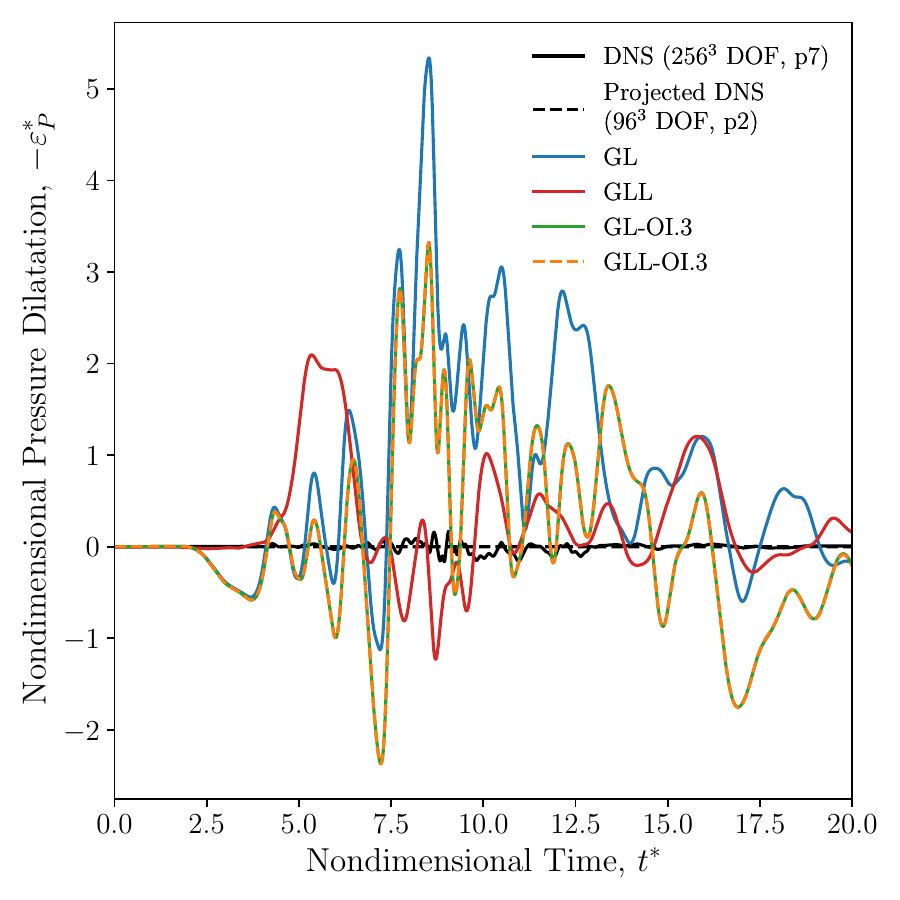}
    \caption{Pressure dilatation}
\end{subfigure}%
\begin{subfigure}{.495\textwidth}
    \centering
    \includegraphics[width=0.99\linewidth]{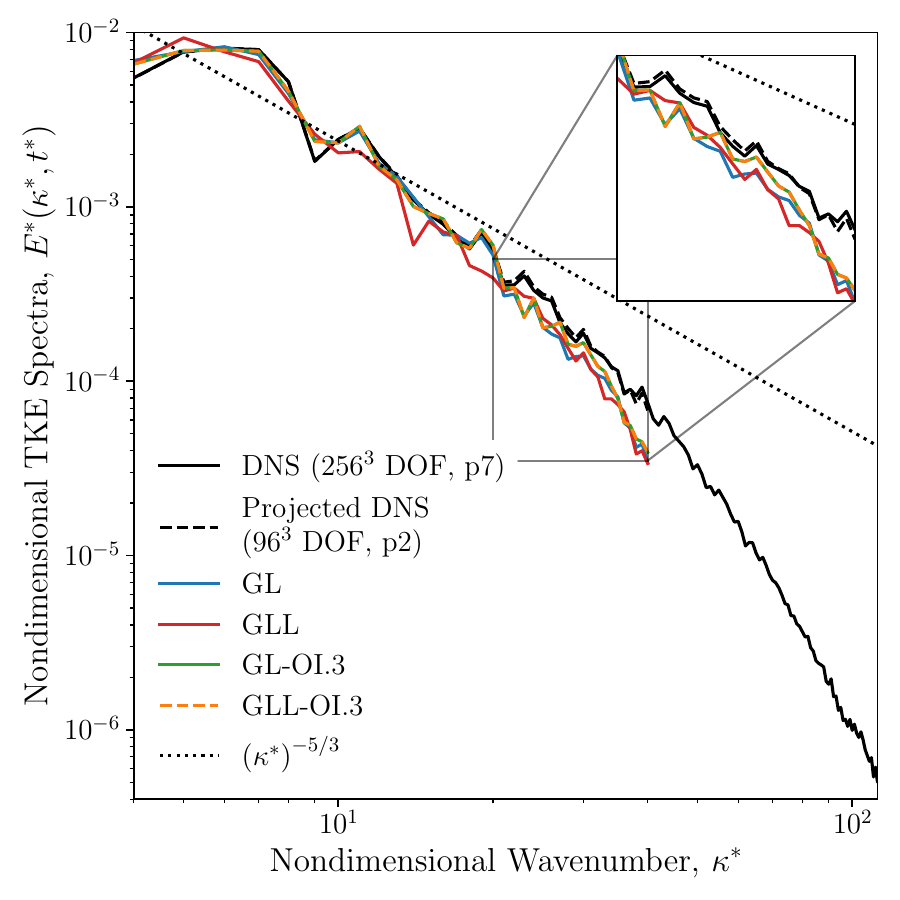}
    \caption{Instantaneous TKE spectra at $t^{*}=9$ plotted until the cut-off wavenumber $\kappa_{c}=40$}
\end{subfigure}
\caption{Temporal evolution of dissipation rate, enstrophy, and pressure dilatation along with the TKE spectra at $t^{*}=9$ for the viscous TGV at $\text{Re}_{\infty}=1600$ with $96^3$ DOF at p$5$ with collocated NSFR ($c_{DG}$ NSFR.IR-GLL, denoted as simply GLL) and uncollocated NSFR ($c_{DG}$ NSFR.IR-GL, denoted as simply GL), with over-integration set to 3 for both schemes (denoted as GLL-OI.3 and GL-OI.3, respectively)}
\label{fig:collocated_vs_uncollocated_96_p5}
\end{figure}

We now consider the use of different flux nodes: Gauss-Lobatto-Legendre (GLL), and Gauss-Legendre (GL). Since our solution nodes and support points for our basis functions (Lagrange polynomials) are fixed as GLL in this work, the choice of different flux nodes allows us to compare a collocated ($c_{DG}$ NSFR.IR-GLL, also known as DG-SEM) NSFR scheme to an uncollocated ($c_{DG}$ NSFR.IR-GL) NSFR scheme. 
For the compressible Navier-Stokes equations, $2(\text{p}+1)$ quadrature points are required per direction in each element for exact integration of cubic nonlinearities present in the viscous flux \cite{kirby2003aliasing}. For a scheme without over-integration, the number of points is half, i.e. $(\text{p}+1)$. With this number of quadrature points, GL and GLL quadrature is exact for integration of polynomial of order $2\text{p}+1$ and $2\text{p}-1$, respectively \cite{zwanenburg2016equivalence}. As a scheme with GL flux nodes has a higher integration accuracy than that with GLL flux nodes, we have chosen to also compare these schemes with partial over-integration, e.g. $n_{\text{quad}}<2(p+1)$ or $n_{\text{quad}}=(p+1)+n_{\text{quad-extra}}$ where $n_{\text{quad-extra}}<p+1$. 
Since the schemes are using p$5$, we have chosen $n_{\text{quad-extra}}=3$; these will be denoted as $c_{DG}$ NSFR.IR-GLL-OI.3 and $c_{DG}$ NSFR.IR-GL-OI.3.
The results for the temporal evolution of the dissipation rate, enstrophy, pressure dilatation, and instantaneous TKE spectra at $t^{*}=9$ are presented in Fig.~\ref{fig:collocated_vs_uncollocated_96_p5}(a-d), respectively. 
We see that $c_{DG}$ NSFR.IR-GLL (denoted as GLL) strongly deviates from the expected dissipation rate around $t^{*}\approx5$ due to the lower integration accuracy. This error is propagated throughout the remaining flow phases, as reflected by the instantaneous TKE spectra at $t^{*}=9$ seen in Fig.~\ref{fig:collocated_vs_uncollocated_96_p5}(d). The TKE spectra of $c_{DG}$ NSFR.IR-GLL at $t^{*}=9$ exhibits larger discrepancies at lower wavenumbers, e.g. $\kappa^{*}\in(10,20)$, with respect to the DNS spectra. 
These effects are indeed due to a lack of sufficient integration strength as the schemes with $n_{\text{quad-extra}}=3$ (denoted as GLL-OI.3 and GL-OI.3 in the legend) yield the same result for all quantities presented; demonstrating that with sufficient integration accuracy, the solution is independent of the choice of flux nodes. We see that $c_{DG}$ NSFR.IR-GL (denoted as GL) has the computational advantage of exhibiting sufficient integration strength without any over-integration as the results are similar to GL-OI.3 and GLL-OI.3 throughout transition. 
From the time evolution of enstrophy in Fig.~\ref{fig:collocated_vs_uncollocated_96_p5}(b), all results are similar throughout the transitional phase, however, the schemes with partial over-integration seem to track the projected DNS more closely throughout the first half of the decaying phase $t^{*}<15$. 
In terms of the pressure dilatation, the collocated scheme exhibits much lower fluctuations compared to the uncollocated scheme and those with partial over-integration in Fig.~\ref{fig:collocated_vs_uncollocated_96_p5}(c) for all time other than around $t^{*}\approx5$. This suggests that these fluctuations may be due to errors associated with interpolating the flux to nodes on element faces as done for uncollocated or over-integrated schemes. 

\subsubsection{FR Correction Parameter Accuracy}\label{sec:correction_parameter_tgv}
\begin{figure}[H]
\begin{subfigure}{.495\textwidth}
    \centering
    \includegraphics[width=0.99\linewidth]{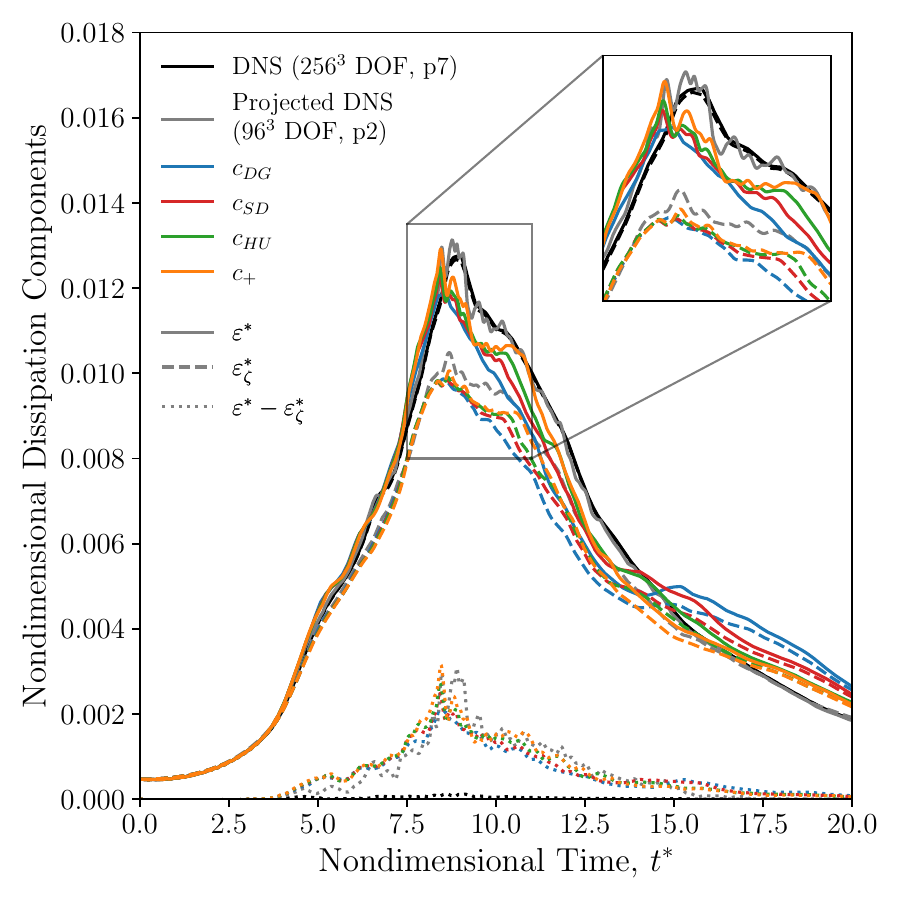}
    \caption{Dissipation components}
\end{subfigure}%
\begin{subfigure}{.495\textwidth}
    \centering
    \includegraphics[width=0.99\linewidth]{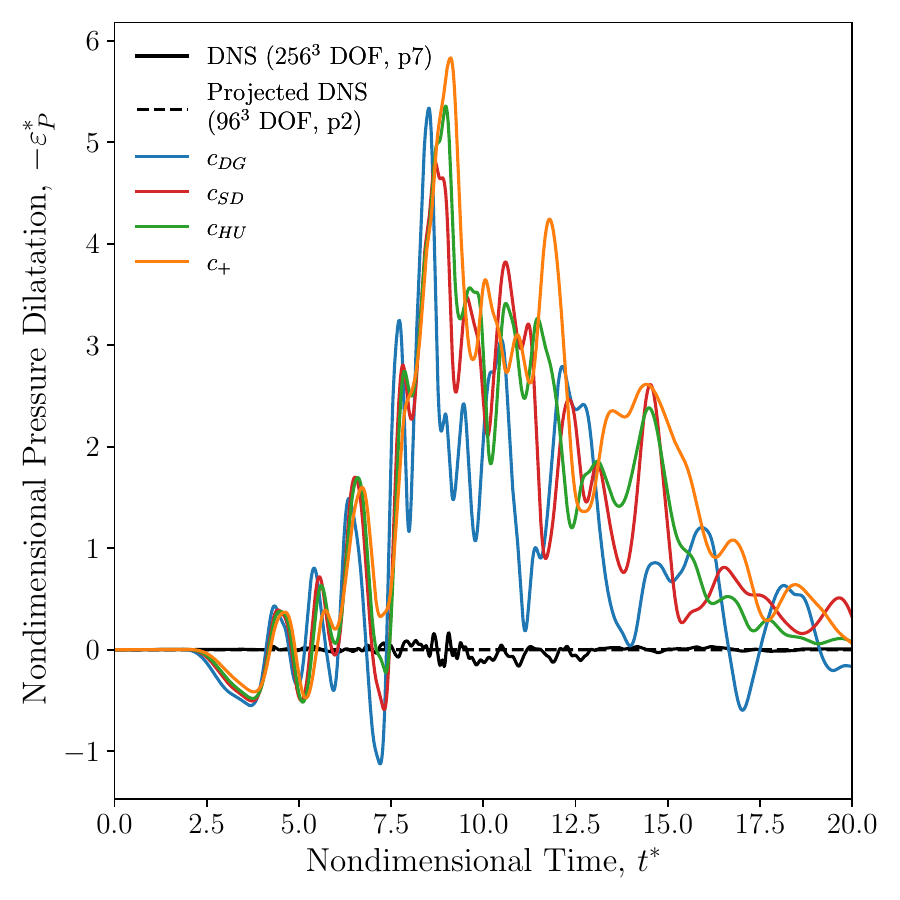}
    \caption{Pressure dilatation}
\end{subfigure}
\begin{subfigure}{.99\textwidth}
    \centering
    \includegraphics[width=0.495\linewidth]{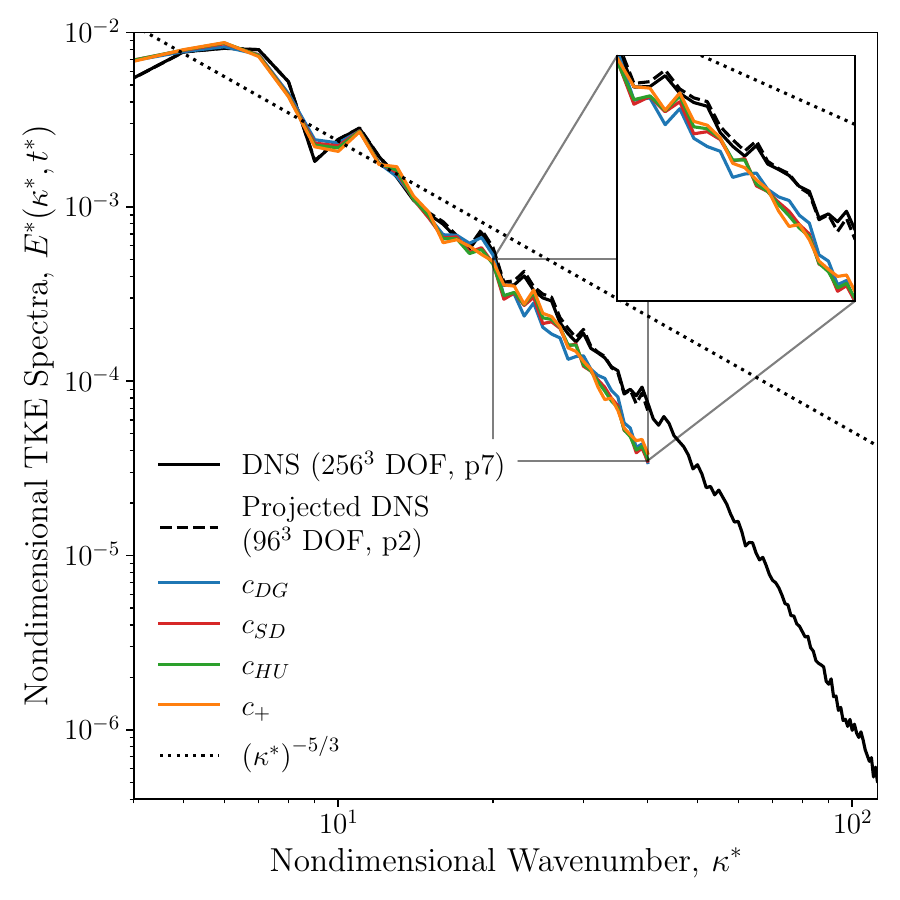}
    \caption{Instantaneous TKE spectra at $t^{*}=9$ plotted until the cut-off wavenumber $\kappa_{c}=40$}
\end{subfigure}%
\caption{Temporal evolution of dissipation components and pressure dilatation, along with the TKE spectra at $t^{*}=9$ for the viscous TGV at $\text{Re}_{\infty}=1600$ with $96^3$ DOF at p$5$ for $c_{DG}$, $c_{SD}$, $c_{HU}$, and $c_{+}$ NSFR.IR-GL}
\label{fig:correction_parameter_accuracy_96}
\end{figure}

As shown by \citet{allaneau2011connections,zwanenburg2016equivalence}, the modified mass matrix $\tilde{\bm{M}}=\bm{M}+\bm{K}$ characteristic to FR methods effectively acts as a linear filtering operator applied on the DG residual. Since $\bm{K}$ is scaled by the correction parameter $c$, this parameter can be directly used to tune the linear filtering operation. For values of $c\in\left[c_{-},0\right)$, the filter will amplify the highest mode of the residual, where $c_{-}$ is the minimum value of $c$ for which the scheme is stable \cite{allaneau2011connections}. For $c=0$, we recover the unfiltered DG method, this value is denoted as $c_{DG}$. Whereas for positive values $0<c\leq c_{+}$, the filter will damp the highest mode of the DG residual; where $c_{+}$ is the largest stable value of $c$ determined by \citet{castonguay2012high} which was determined for $\text{p}\in\left[2,5\right]$. 
To investigate the effects of $c$, we consider $c_{DG}$, $c_{+}$, $c_{SD}$ \cite{allaneau2011connections} recovering the stable spectral difference (SD) scheme, and $c_{HU}$ \cite{allaneau2011connections} recovering Hyunh's $g_{2}$ scheme \cite{huynh2007flux}; where the latter two are both positive values. 
The time evolution of dissipation components, pressure dilatation, and the instantaneous TKE spectra for the $c_{i}$ NSFR.IR-GL schemes where $c_{i}=\left(c_{DG},c_{+},c_{SD},c_{HU}\right)$ are presented in Fig.~\ref{fig:correction_parameter_accuracy_96}(a), (b), and (c) respectively. For all schemes, the dissipation components are similar throughout the transient phase until the peak values are reached. At this point, the highest peak is that of $c_{+}$, followed by $c_{HU}$, then $c_{SD}$, and lastly $c_{DG}$ highlighting the added dissipation associated with larger $c$ values. In terms of the TKE spectra, all schemes produce similar results: capturing the large and mid scales while exhibiting a deficit of TKE at the smallest resolved scales. 
As for the pressure dilatation, Fig.~\ref{fig:correction_parameter_accuracy_96}(b) demonstrates that no choice of correction parameter is effective at damping the pressure dilatation oscillations.

\begin{figure}[H]
\begin{subfigure}{.495\textwidth}
    \centering
    \includegraphics[width=0.99\linewidth]{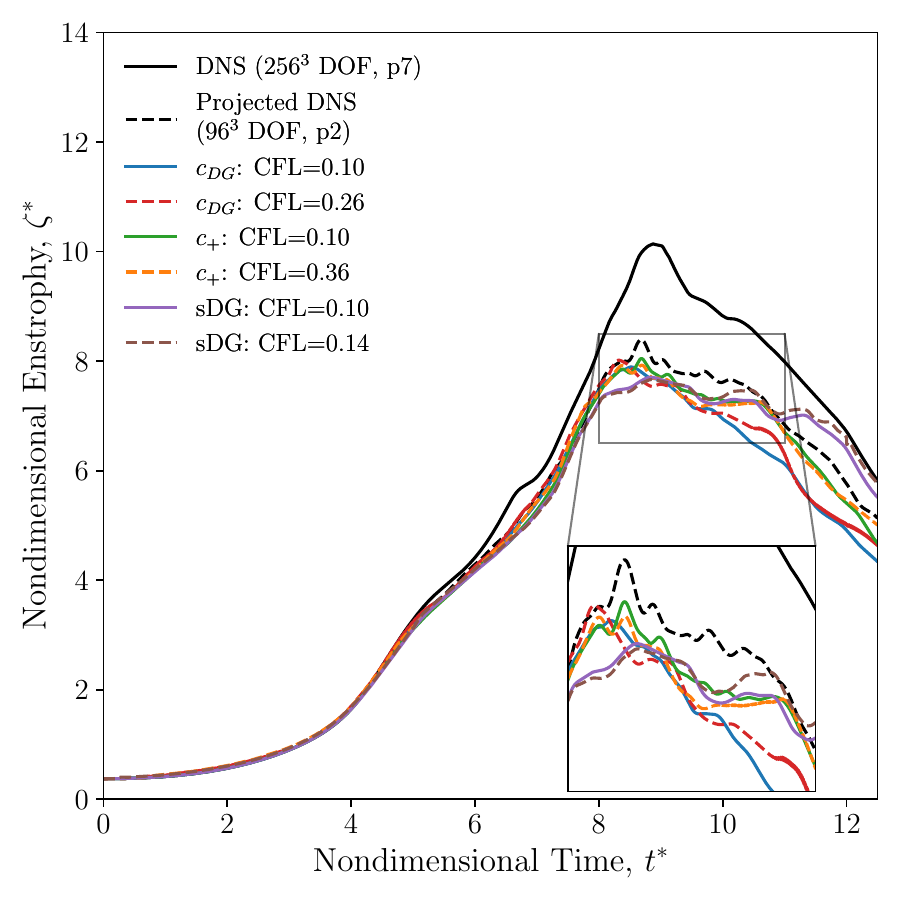}
    \caption{Enstrophy vs time}
\end{subfigure}%
\begin{subfigure}{.495\textwidth}
    \centering
    \includegraphics[width=0.99\linewidth]{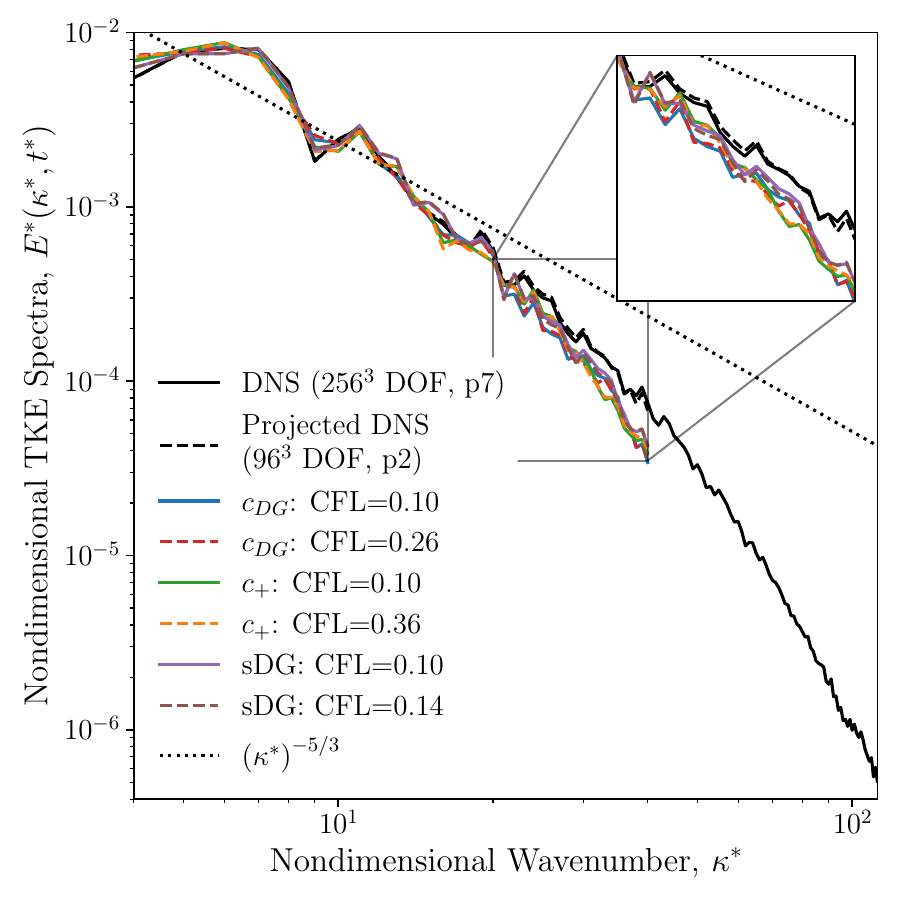}
    \caption{Instantaneous TKE spectra at $t^{*}=9$ plotted until the cut-off wavenumber $\kappa_{c}=40$}
\end{subfigure}%
\caption{Temporal evolution of enstrophy and the TKE spectra at $t^{*}=9$ for the viscous TGV at $\text{Re}_{\infty}=1600$ with $96^3$ DOF at p$5$ for $c_{DG}$ NSFR.IR-GL~(denoted as $c_{DG}$), $c_{+}$ NSFR.IR-GL~(denoted as $c_{+}$), and strong~DG-Roe-GL-OI.$6$~(denoted as sDG) at their respective CFL limits and at the baseline CFL$=0.1$}
\label{fig:correction_parameter_cfl_advantage}
\end{figure}

Furthermore, a value of $c>0$ (e.g. $c_{+}$) has the advantage of allowing for larger explicit time-steps while maintaining the expected orders of accuracy compared to $c_{DG}$, with $c_{+}$ being at the upper limit of this advantage. Since the overall result is similar for all correction parameters, let us compare the largest allowable explicit time-step quantified by the CFL limit for p$5$ NSFR.IR-GL with $c_{DG}$, $c_{+}$, and p$5$ standard DG denoted as strong DG-Roe-GL-OI.$6$. Typically the CFL limit is determined as the maximum allowable CFL that will guarantee a numerically stable simulation. However, in this work, we imposed additional criteria to ensure the solution is indeed physically consistent. Since there are no external sources of energy applied to the closed system and the flow is nearly incompressible, the kinetic energy must always be decreasing; consistent with the DNS result. The physically consistent check ensures that the integrated kinetic energy over the domain $E_{K}$ is always decreasing as time advances, i.e. $-\varepsilon\leq0~\forall~t$. Table~\ref{tab:cfl_limits} lists the stable CFL numbers with 
the physically consistent check for each scheme considered. We see that 
$c_{DG}$ and $c_{+}$ allow for a CFL value that is about $1.9$ and $2.6$ times greater than that of standard DG, respectively. The value allowed by $c_{+}$ is about $38\%$ higher than that of $c_{DG}$. The schemes ran at their respective CFL limits produced very similar results compared to the baseline CFL of $0.10$ throughout the transitional phase as seen in Fig.~\ref{fig:correction_parameter_cfl_advantage}, with small discrepancies seen in the decaying phase of the flow. Therefore, by using $c_{+}$ we can obtain a solution similar to that of $c_{DG}$ and strong DG at about half the computational cost.

\begin{table}[H]
    \centering
    \begin{tabular}{c||c}
    \hfill & CFL Limit\\\hline
    $c_{DG}$ NSFR.IR-GL & 0.26\\\hline
    $c_{+}$ NSFR.IR-GL & 0.36\\\hline
    Strong DG-Roe-GL-OI.$6$ & 0.14\\\hline
    \end{tabular}
\caption{CFL limits of p$5$ $c_{DG}$, $c_{+}$ NSFR.IR-GL, and strong~DG-Roe-GL-OI.$6$ for the simulation of the viscous TGV at $\text{Re}_{\infty}=1600$ using $96^3$ DOF}
\label{tab:cfl_limits}
\end{table}

\subsubsection{Two-Point Numerical Flux Choice}\label{sec:tpflux}
To investigate the effects of the two-point numerical flux choice, the $c_{DG}$ NSFR scheme with GL flux nodes was run for four different two-point fluxes: 
\begin{enumerate}
    \item Ismail-Roe (IR) \cite{ismail2009affordable} that is entropy-conserving (EC), denoted as $c_{DG}$ NSFR.IR-GL scheme
    \item Kennedy-Gruber (KG) \cite{kennedy2008reduced} that is kinetic-energy-preserving (KEP), denoted as $c_{DG}$ NSFR.KG-GL
    \item Chandrashekar (CH) \cite{chandrashekar2013kinetic} that is both EC and KEP, denoted as $c_{DG}$ NSFR.CH-GL
    \item Chandrashekar modified by Ranocha $\left(\text{CH}_{\text{RA}}\right)$ \cite{ranocha2021preventing} that is EC, KEP, and pressure-equilibrium-preserving (PEP), denoted as $c_{DG}$ NSFR.$\text{CH}_{\text{RA}}$-GL
\end{enumerate}
The computational cost of all fluxes is similar, however, the KG two-point flux incurs the lowest cost as it does not require any logarithmic averaging given by Eq.(\ref{eq:log_mean}) unlike the IR, CH, and $\text{CH}_{\text{RA}}$ fluxes. The nearly indistinguishable TKE spectra results for the different schemes are shown below in Fig.~\ref{fig:two-point-flux-spectra}(a). Demonstrating that for this viscous flow case, KEP, EC, and PEP fluxes render equivalent results. Although not shown, the same observation was also found for time evolution of kinetic energy $E_{K}$, dissipation rate $\varepsilon$, enstrophy $\zeta$, and effective viscosity. The only quantity for which the slightest of differences can be seen is the pressure dilatation around $t^{*}=12.5,~15,~20$ in Fig.~\ref{fig:two-point-flux-spectra}(b); however these difference are considered negligible. 

\begin{figure}[H]
\begin{subfigure}{.495\textwidth}
    \centering
    \includegraphics[width=0.99\linewidth]{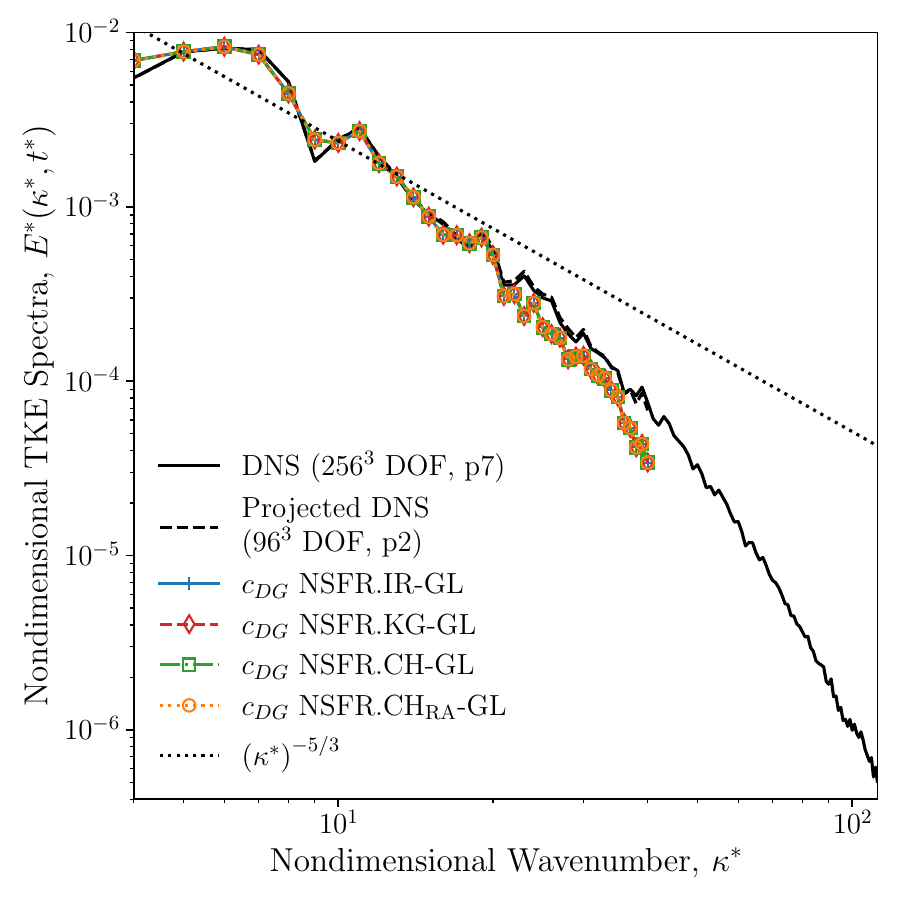}
    \caption{Instantaneous TKE spectra at $t^{*}=9$ plotted until the cut-off wavenumber $\kappa_{c}=40$}
\end{subfigure}%
\begin{subfigure}{.495\textwidth}
    \centering
    \includegraphics[width=0.99\linewidth]{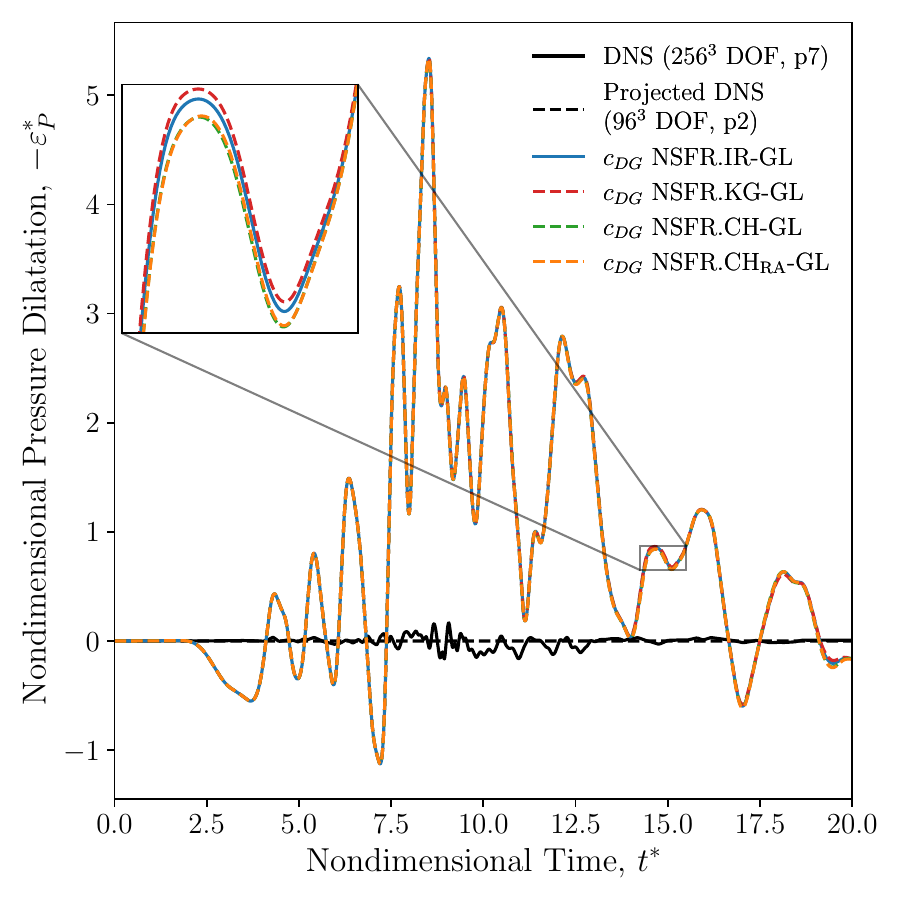}
    \caption{Pressure dilatation}
\end{subfigure}
\caption{Viscous TGV at $\text{Re}_{\infty}=1600$ with $96^3$ DOF at p$5$ for uncollocated $c_{DG}$ NSFR with different two-point numerical fluxes}
\label{fig:two-point-flux-spectra}
\end{figure}

\subsubsection{Added Riemann Solver Dissipation}
\begin{figure}[H]
\begin{subfigure}{.495\textwidth}
    \centering
    \includegraphics[width=0.99\linewidth]{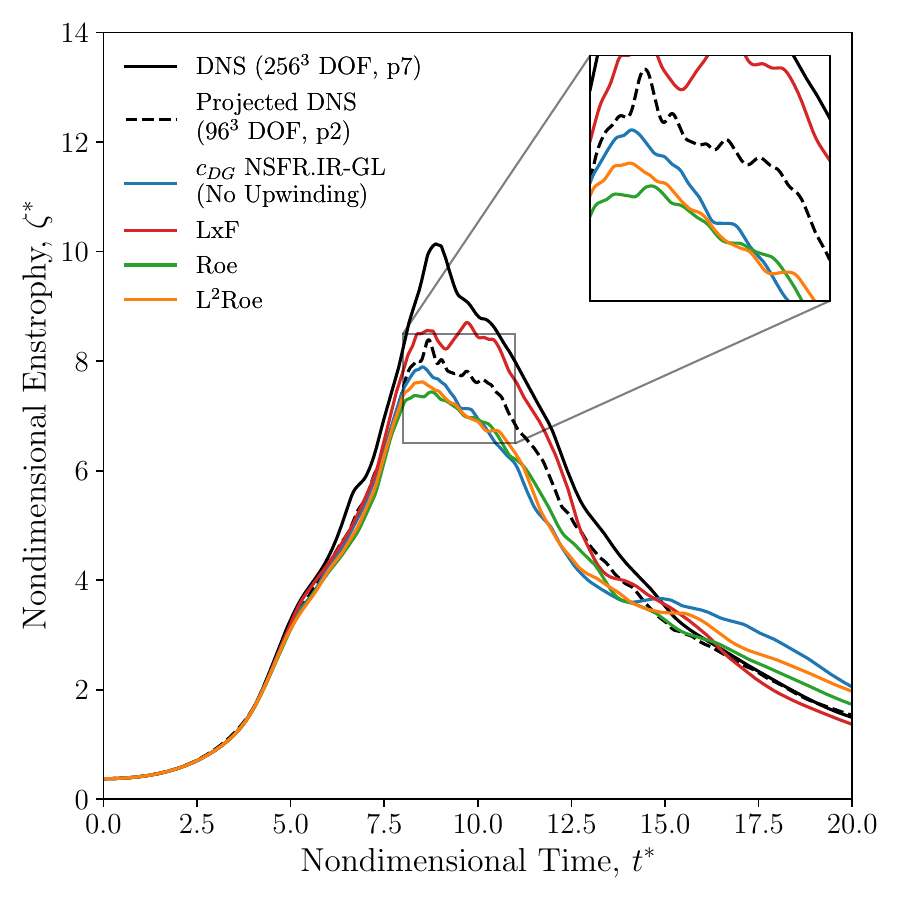}
    \caption{Enstrophy vs time}
\end{subfigure}%
\begin{subfigure}{.495\textwidth}
    \centering
    \includegraphics[width=0.99\linewidth]{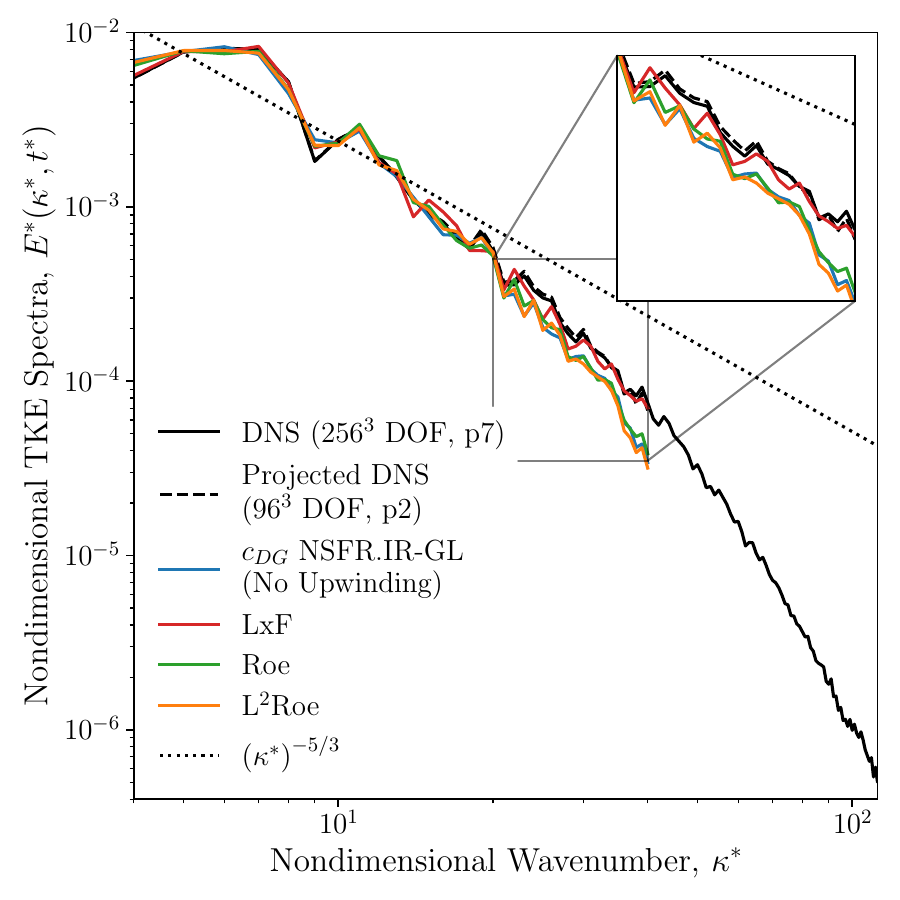}
    \caption{Instantaneous TKE spectra at $t^{*}=9$ plotted until the cut-off wavenumber $\kappa_{c}=40$}
\end{subfigure}
\begin{subfigure}{.99\textwidth}
    \centering
    \includegraphics[width=0.495\linewidth]{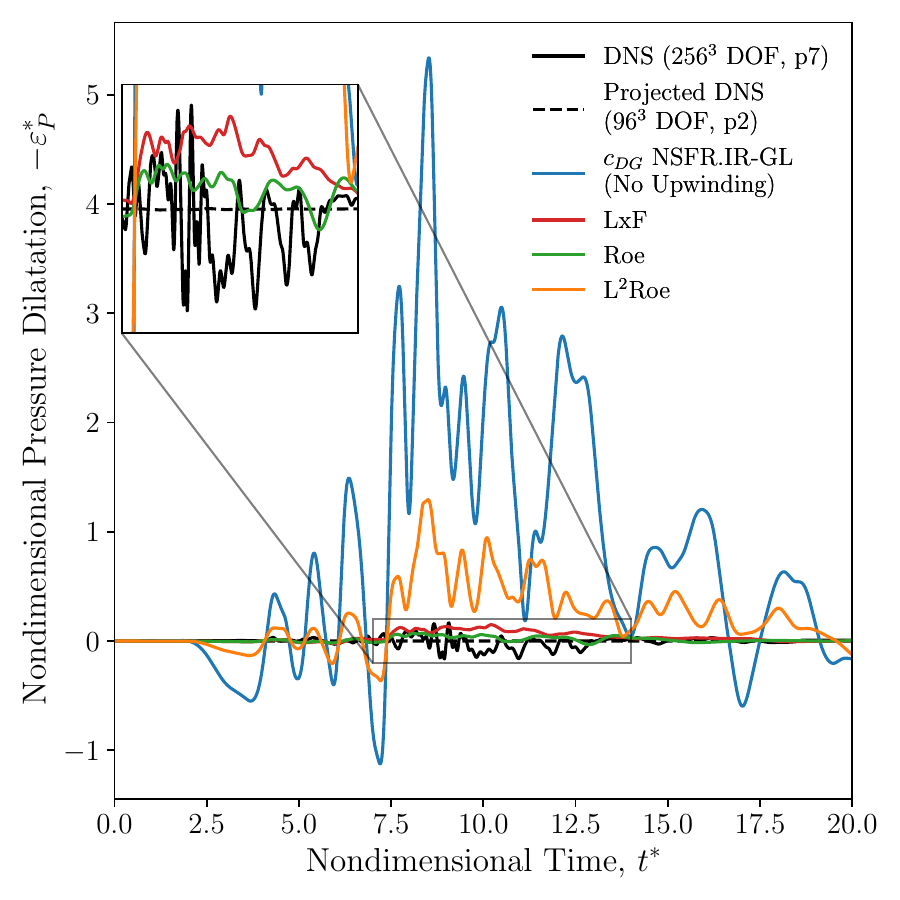}
    \caption{Pressure dilatation}
\end{subfigure}%
\caption{Temporal evolution of enstrophy and pressure dilatation, along with the TKE spectra at $t^{*}=9$ for the viscous TGV at $\text{Re}_{\infty}=1600$ with $96^3$ DOF at p$5$ with added upwind dissipation mechanisms for an uncollocated NSFR scheme}
\label{fig:upwind-quantities}
\end{figure}

\begin{figure}[H]
\begin{subfigure}{.495\textwidth}
    \centering
    \includegraphics[width=0.99\linewidth]{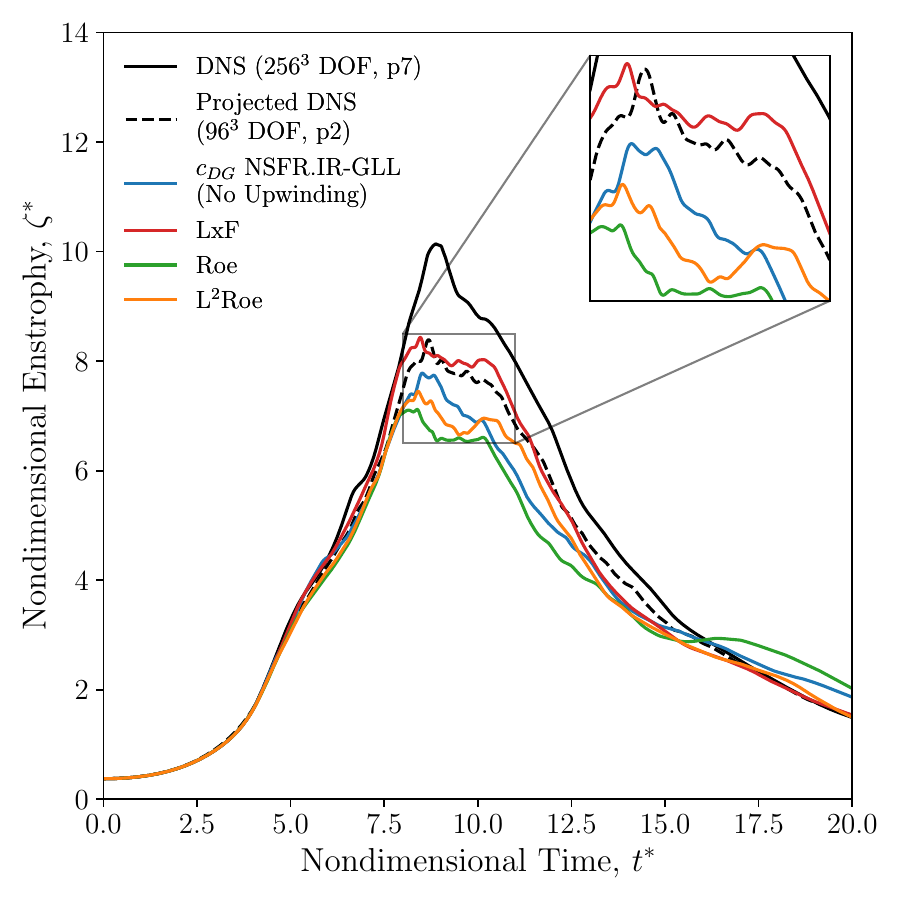}
    \caption{Enstrophy vs time}
\end{subfigure}%
\begin{subfigure}{.495\textwidth}
    \centering
    \includegraphics[width=0.99\linewidth]{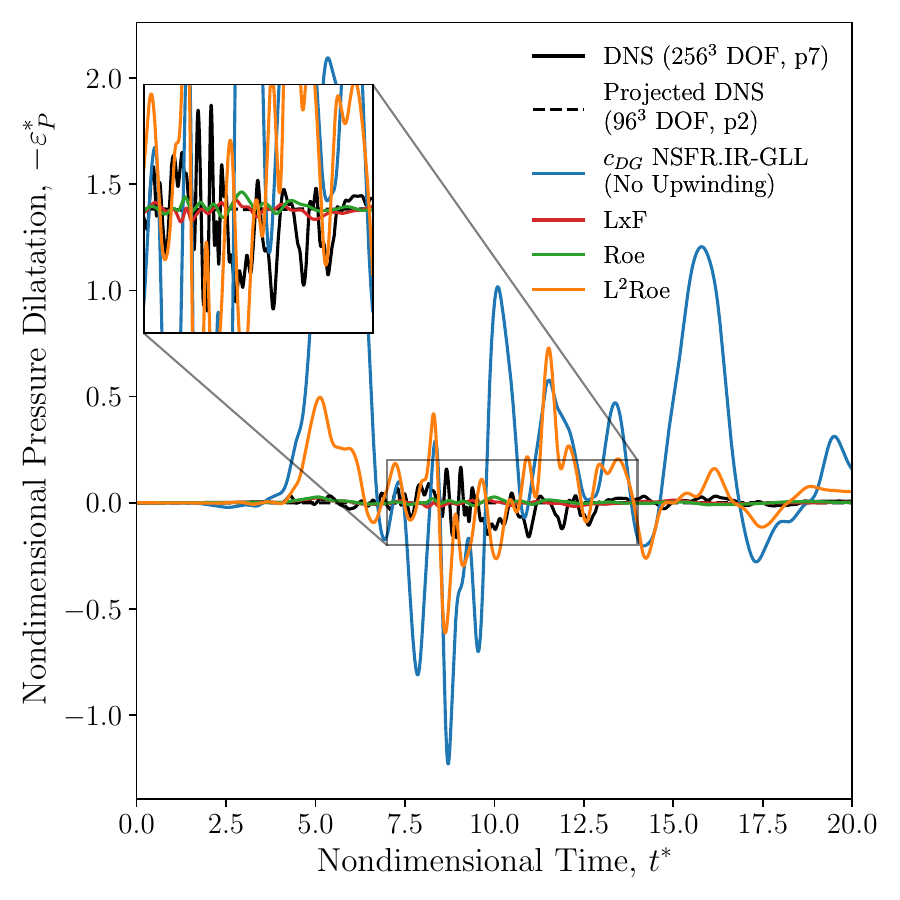}
    \caption{Pressure dilatation}
\end{subfigure}
\caption{Temporal evolution of enstrophy and pressure dilatation for the viscous TGV at $\text{Re}_{\infty}=1600$ with $96^3$ DOF at p$5$ with added upwind dissipation mechanisms for a collocated NSFR scheme}
\label{fig:upwind-quantities-gll}
\end{figure}

To investigate the effects of adding different Riemann solver dissipation, i.e. the upwinding term, to the two-point flux, the $c_{DG}$ NSFR scheme with GL flux nodes was run for three different added upwind-type dissipation separately: Lax-Friedrichs (LxF), Roe, and low-dissipation Roe (L$^{2}$Roe). From the time evolution of enstrophy in Fig.~\ref{fig:upwind-quantities}(a), we see that adding Roe-type dissipation decreases the peak resolved enstrophy with the L$^{2}$Roe being less dissipative by the slightly higher peak enstrophy. Overall, the results for the two Roe-type upwinding are quite similar, with nearly identical TKE spectra at $t^{*}=9$ to the baseline scheme without upwinding in Fig.~\ref{fig:upwind-quantities}(b). However, with added LxF upwinding, the scheme exhibits anti-dissipative behaviour characterized by a higher peak enstrophy compared to the projected DNS and higher TKE at the smallest resolved scales, approaching that of the projected DNS. Taking a look at the time evolution of pressure dilatation $\varepsilon_{P}$ in Fig.~\ref{fig:upwind-quantities}(c), we see that adding upwind dissipation significantly dampens the oscillations seen by the baseline scheme, with the LxF and Roe almost completely attenuating these oscillations. Earlier it was shown that a collocated scheme yielded significantly smaller $\varepsilon_{P}$ oscillations. Therefore, results for collocated NSFR with added upwind dissipation are presented in Fig.~\ref{fig:upwind-quantities-gll}. Remarkably, we see that the resulting pressure dilatation oscillations are now drastically smaller compared to not only the collocated baseline dissipation-free scheme but also to the DNS result of uncollocated NSFR.  

\subsubsection{Added SGS Model Dissipation}
\begin{figure}[H]
\begin{subfigure}{.495\textwidth}
    \centering
    \includegraphics[width=0.99\linewidth]{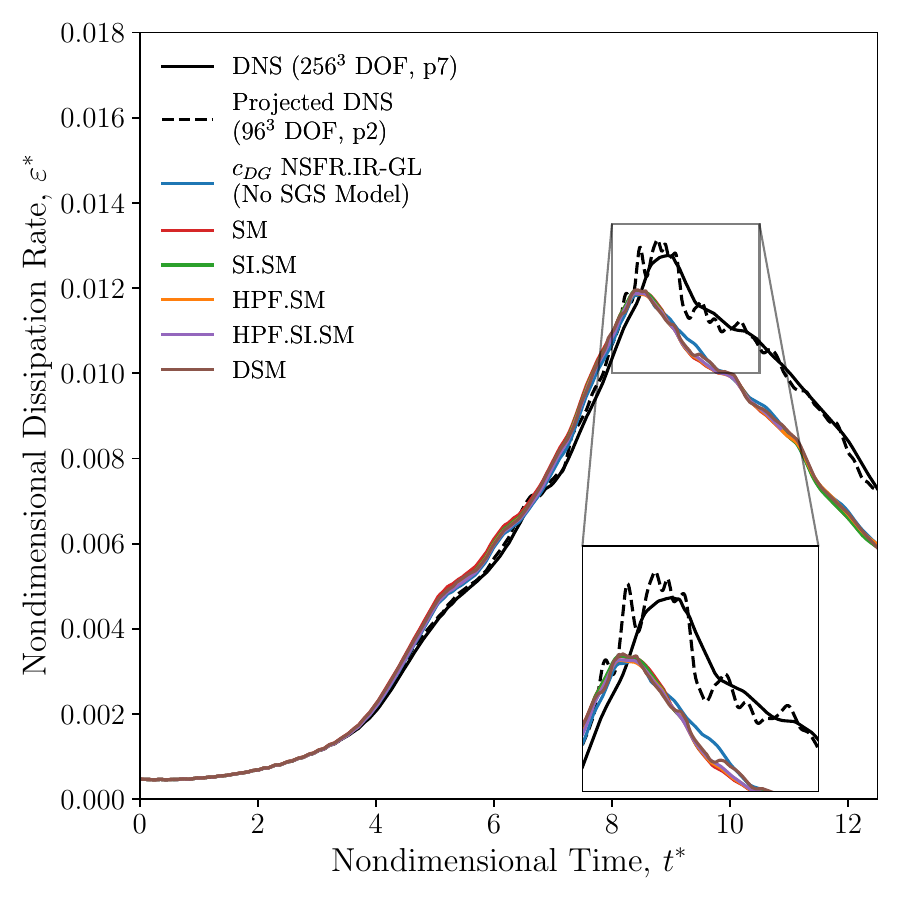}
    \caption{Dissipation rate vs time}
\end{subfigure}%
\begin{subfigure}{.495\textwidth}
    \centering
    \includegraphics[width=0.99\linewidth]{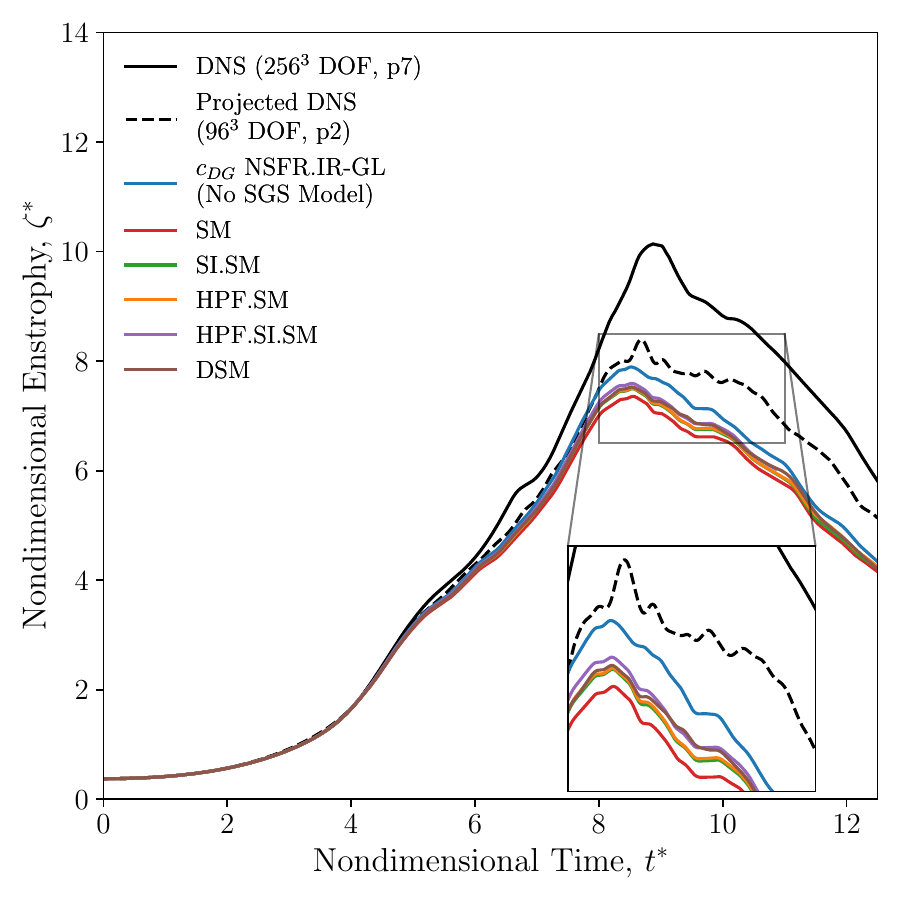}
    \caption{Enstrophy vs time}
\end{subfigure}
\begin{subfigure}{.495\textwidth}
    \centering
    \includegraphics[width=0.99\linewidth]{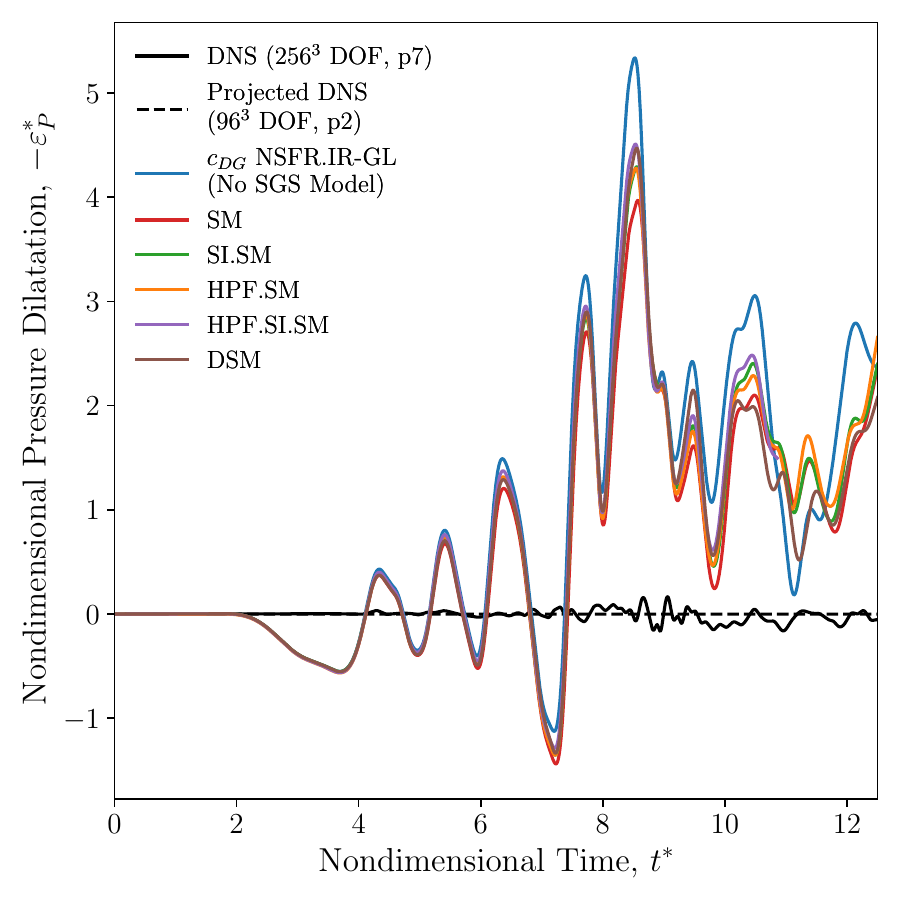}
    \caption{Pressure dilatation}
\end{subfigure}%
\begin{subfigure}{.495\textwidth}
    \centering
    \includegraphics[width=0.99\linewidth]{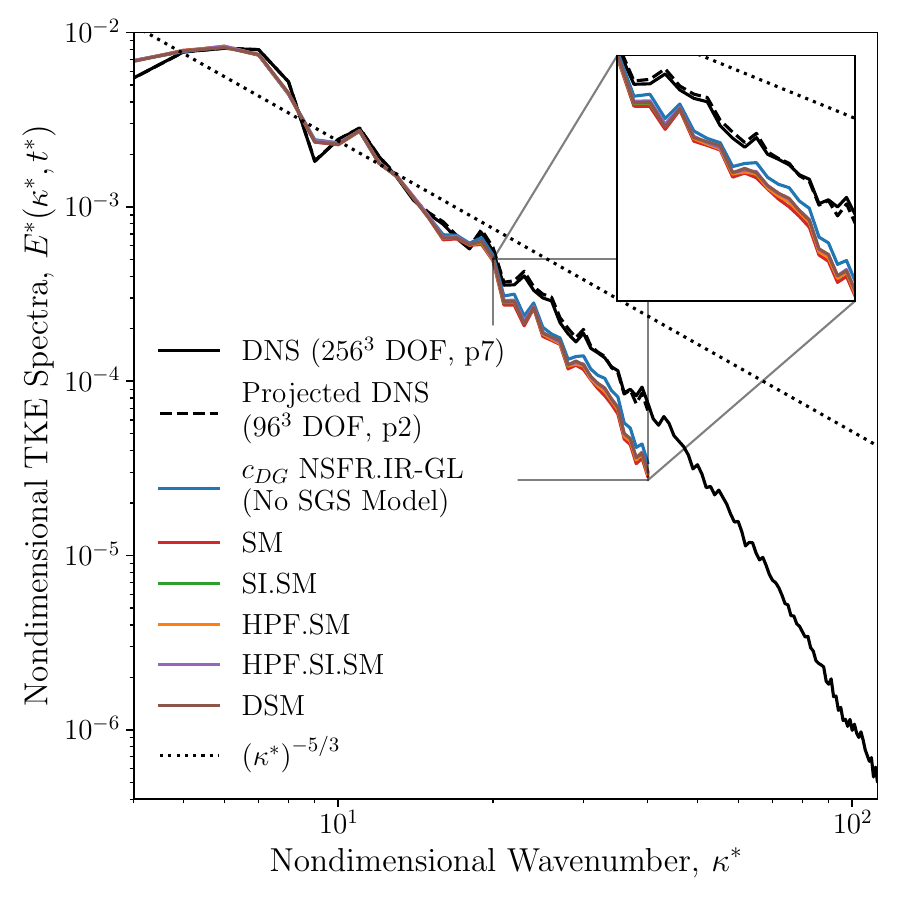}
    \caption{Instantaneous TKE spectra at $t^{*}=9$ plotted until the cut-off wavenumber $\kappa_{c}=40$}
\end{subfigure}
\caption{Temporal evolution of dissipation rate, enstrophy and pressure dilatation, along with the TKE spectra at $t^{*}=9$ for the viscous TGV at $\text{Re}_{\infty}=1600$ with $96^3$ DOFs using p$5$ $c_{DG}$ NSFR.IR-GL with different SGS models}
\label{fig:96_p5_selected_sgs_models_gl}
\end{figure}

To investigate the effects of added SGS models, we consider the $c_{DG}$ NSFR.IR-GL scheme with the addition of the following SGS models: 
\begin{enumerate}
    \item Classical Smagorinsky model (SM) 
    \item Shear-improved Smagorinsky model (SI.SM)
    \item High-pass filtered classical Smagorinsky model (HPF.SM)
    \item High-pass filtered shear-improved Smagorinsky model (HPF.SI.SM)
    \item Dynamic Smagorinsky model (DSM)
\end{enumerate}
In this work, all classical and shear-improved Smagorinsky results use a standard model constant of $C_{S}=0.1$, $\text{p}_{L}=3$ for high-pass filtered (HPF) SGS models, and the model constant for the dynamic Smagorinsky is clipped at $C_{max}=0.1$ so that it is never more dissipative than the classical Smagorinsky model. 
For the dynamic Smagorinsky model, a test-filter polynomial degree $\text{p}_{TF}=3$ is used for the Galerkin projection operations in computing the dynamic model constant. 
Furthermore, the low Reynolds number correction (LRNC) was applied to all SGS models. 
The time evolution of dissipation rate, enstrophy, pressure dilatation, and instantaneous TKE spectra at $t^{*}=9$ are shown in Fig.~\ref{fig:96_p5_selected_sgs_models_gl}(a-d), respectively. 
The dissipation rate of the baseline scheme appears to be only marginally increased with the addition of the SGS model dissipation. 
However, the peak enstrophy is noticeably reduced by the added dissipation. 
The addition of the classical Smagorinsky model yields the largest reduction in peak enstrophy, as expected since this is the most dissipative model. 
For the remaining models, enstrophy is found to be quite similar, with the HPF.SI.SM model being the least dissipative of the turbulence models considered. 
Similar to the enstrophy, the SGS models reduce the amplitude of the pressure dilatation fluctuations but do not successfully attenuate the oscillations to an order of magnitude comparable to that of the DNS. 
In terms of the TKE spectra, the SGS models dissipate TKE at the smallest resolved scales $\kappa^{*}\in(25,40)$ as expected, without affecting the larger resolved scales at smaller wavenumbers. 
Given that there is already a deficit of TKE at the highest resolved wavenumbers with respect to the DNS for the baseline scheme without an SGS model, the SGS models considered in this work are not recommended for the LES of transitional turbulent free-shear flows using NSFR. 

\begin{figure}[H]
\begin{subfigure}{.495\textwidth}
    \centering
    \includegraphics[width=0.99\linewidth]{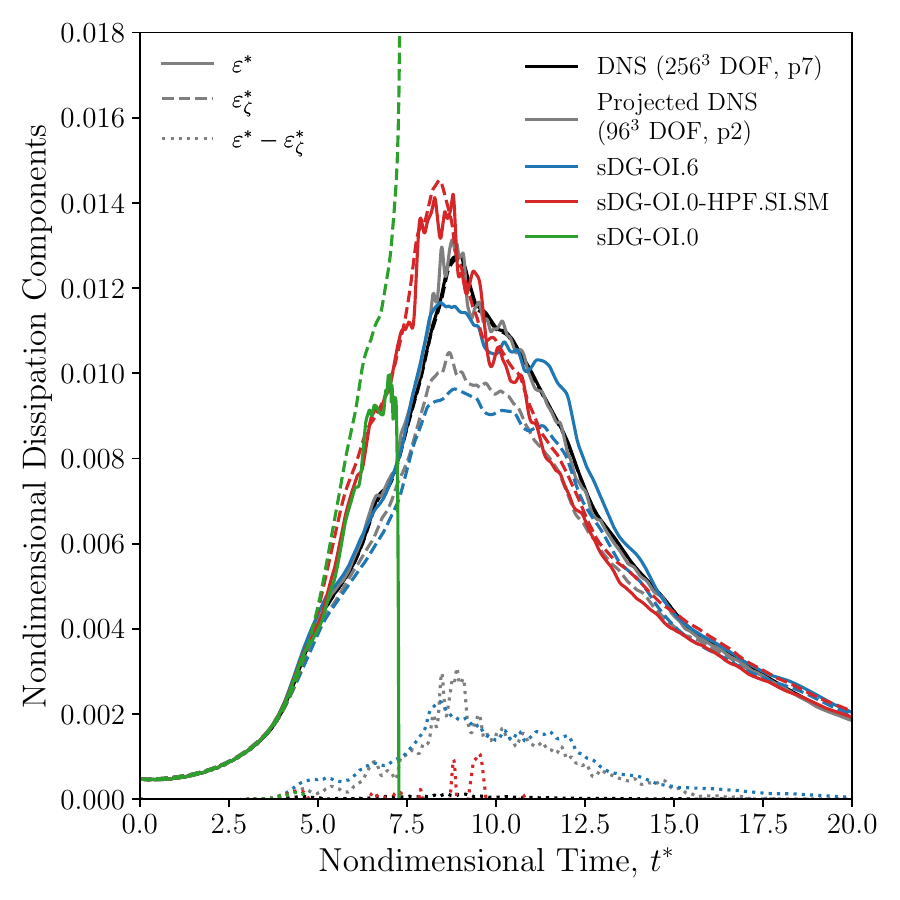}
    \caption{Dissipation components vs time}
\end{subfigure}%
\begin{subfigure}{.495\textwidth}
    \centering
    \includegraphics[width=0.99\linewidth]{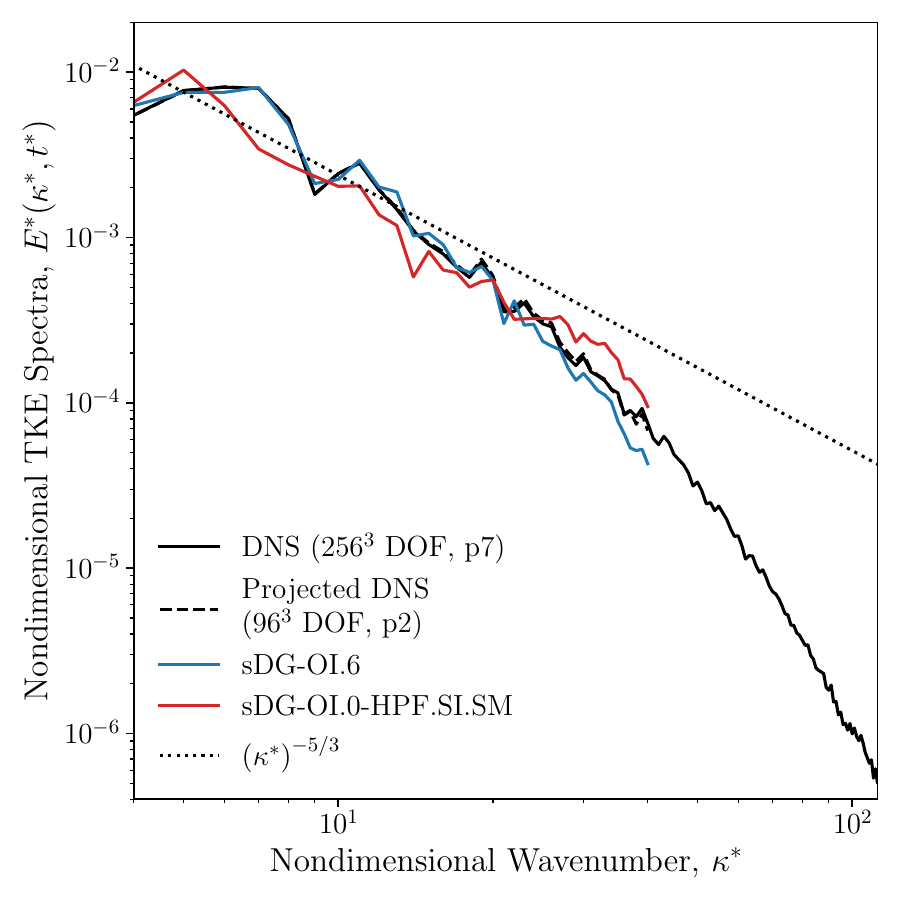}
    \caption{Instantaneous TKE spectra at $t^{*}=9$ plotted until the cut-off wavenumber $\kappa_{c}=40$}
\end{subfigure}%
\caption{Temporal evolution of dissipation components and the TKE spectra at $t^{*}=9$ for the viscous TGV at $\text{Re}_{\infty}=1600$ with $96^3$ DOFs using p$5$ strong DG-Roe-GLL-OI.0 (denoted as sDG-OI.0) with the addition of an SGS model to stabilize the scheme (denoted as sDG-OI.0-HPF.SI.SM) compared to p$5$ strong DG-Roe-GLL-OI.6 (denoted as sDG-OI.6)}
\label{fig:sDG_gll_sgs_model_stabilization}
\end{figure}

\indent To further demonstrate the benefits of having a scheme that is stable without over-integration, we ran p$5$ collocated strong DG without over-integration with the best performing SGS model (p$5$ strong DG-Roe-GLL-OI.0-HPF.SI.SM).
From the dissipation component results in Fig.~\ref{fig:sDG_gll_sgs_model_stabilization}(a) we see that the collocated strong DG scheme without over-integration is unstable, and that the addition of the SGS model provides sufficient dissipation to stabilize the unstable scheme. However, this does not lead to an accurate solution, as demonstrated by the high values of dissipation rates throughout the transition phase and by the instantaneous TKE spectra at $t^{*}=9$ in Fig.~\ref{fig:sDG_gll_sgs_model_stabilization}(b). 
In contrast, the over-integrated scheme (p$5$ strong DG-Roe-GLL-OI.6) follows the reference DNS much closer. This highlights that a stable scheme without any de-aliasing can lead to inaccuracies. 

\subsection{The Comte-Bellot and Corsin experiment}
\subsubsection{Description \& Computational Details}
\begin{figure}[H]
\begin{subfigure}{.495\textwidth}
    \centering
    \includegraphics[width=0.99\linewidth]{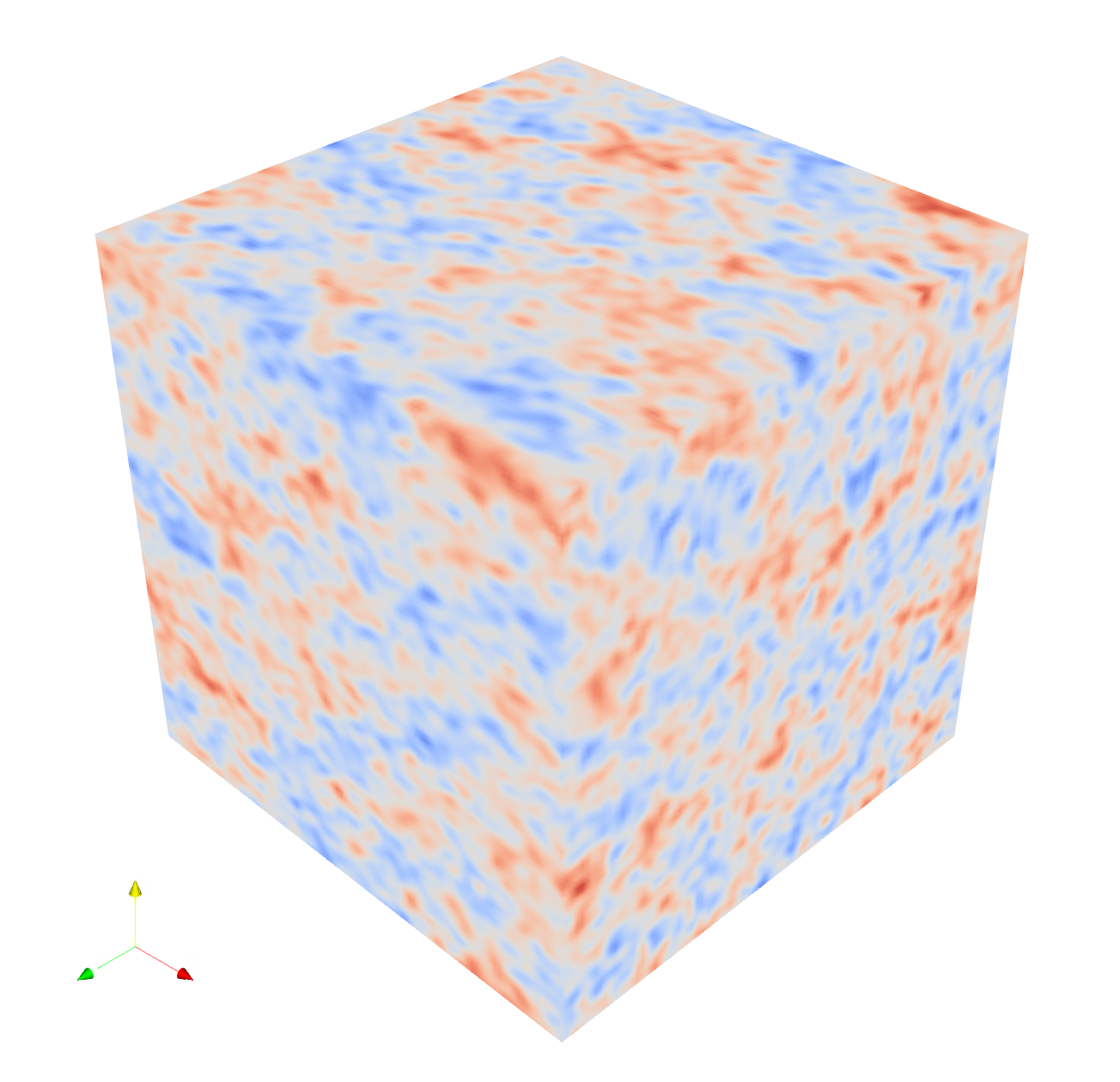}
    \caption{Mach number contours}
\end{subfigure}%
\begin{subfigure}{.495\textwidth}
    \centering
    \includegraphics[width=0.99\linewidth]{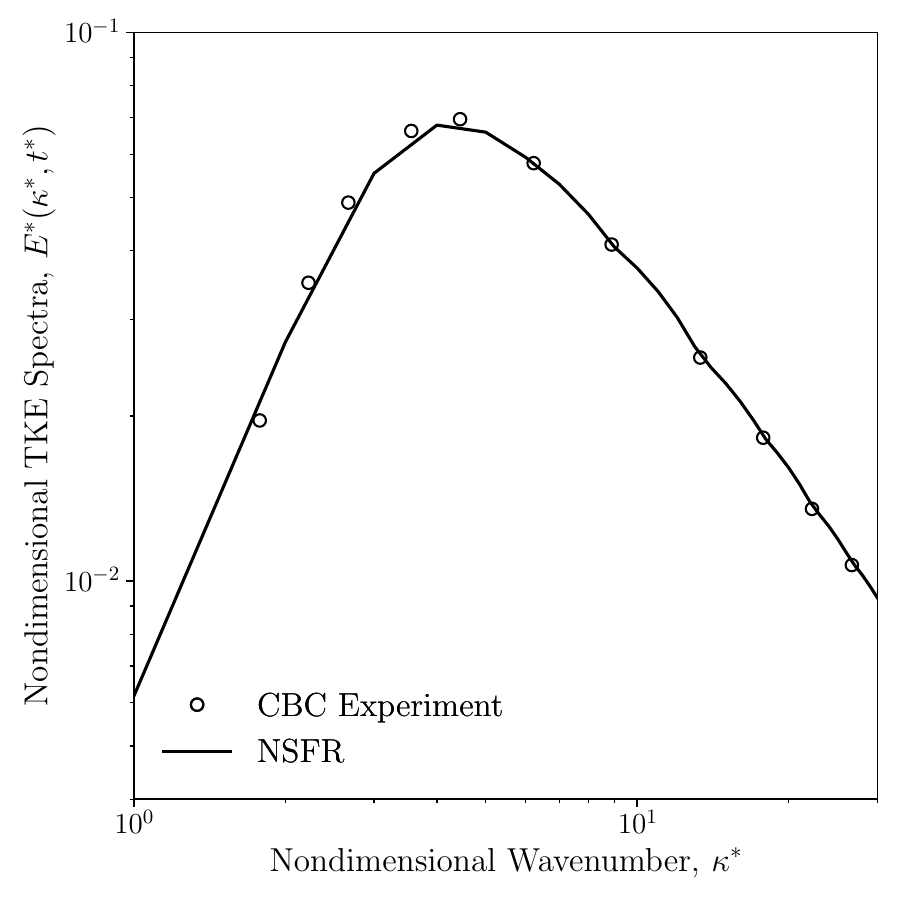}
    \caption{Turbulent Kinetic Energy Spectra}
\end{subfigure}
\caption{Flow field initialization for the initial Comte-Bellot and Corsin (CBC) experiment using NSFR with $128^{3}$ DOF}
\label{fig:dhit-initialization}
\end{figure}

The Comte-Bellot and Corsin (CBC) experiment is the study of decaying homogeneous isotropic turbulence (DHIT) generated by a grid \cite{comte1971simple}. In the CBC experiment, the mean flow of $\overline{V}=10~\text{m}/\text{s}$ is passed through a wire grid with a mesh size of $M=5.08\times10^{-2}~\text{m}$ to generate turbulence at a mesh-based Reynolds number of $\text{Re}_{M}=34000$. The flow can be considered as DHIT in the frame of reference of the mean flow, i.e. following a fluid parcel traveling at the mean velocity. The experimental results for the TKE spectra demonstrate the classical integral scales, inertial range, and dissipative scales making it a good validation case. These spectra were extracted at non-dimensional times: $t\overline{V}/M=42,98,171$; where the Taylor microscale Reynolds number was observed to decay from $\text{Re}_{\lambda}=71.6$ to $\text{Re}_{\lambda}=60.6$ between the first and last spectra extraction times. \newline
\indent The results from this experiment have been used widely for the validation of numerical schemes, e.g. sixth-order flux difference splitting (FDS)~\cite{jefferson2010impingement}, high-order correction procedure via reconstruction (CPR)~\cite{vermeire2016implicit}, and an implicit turbulence modelling finite-volume approach termed the adaptive local deconvolution method (ALDM)~\cite{hickel2009implicit} to name a few. As the experimental TKE spectra are available for three different times, the spectra measured at the earliest time can be used to initialize the simulation's flow field, and then the computational spectra can be compared to that of the experiment at the two corresponding later times. The details of how this is done will now be presented.\newline
% computational details
\indent Taking the free-stream density as the sea-level value $\rho_{\infty}=1.225\text{kg/m}^{3}$, the free-stream dynamic viscosity is determined:
\begin{equation}\label{eq:DHIT_reynolds_number}
    \Re_{M}=34000=\frac{\rho_{\infty}\overline{V}M}{\mu_{\infty}} \rightarrow \mu_{\infty} = \frac{\rho_{\infty}\overline{V}M}{\Re_{M}} = 1.8302941\times10^{-5}~\text{kg}/\text{m}/\text{s}.
\end{equation}
This viscosity value corresponds to a temperature of $T_{\infty}=296.70727~\text{K}$ using Sutherland's law for viscosity. Using ideal gas law, this yields a pressure of $P=\rho_{\infty}R_{\text{air}}T_{\infty}=104335.58~\text{Pa}\approx1.04~\text{bar}$ where the gas constant for air is $R_{\text{air}}=287.057~\text{J}/\text{kg}/\text{K}$. Interpolating the Prandtl number for $T_{\infty}$ from EngineeringToolbox for values measured at a pressure of $1~\text{bar}$ yields a Prandtl number of $\text{Pr}=0.70760$. 
\newline\indent We follow the standard non-dimensionalization procedure of \citet{misra1996evaluation}: $V_{\infty}=\sqrt{3U_{0}^{'2}/2}$ and $L=l/2\pi$, where the velocity at the first measuring station from the CBC experiment is $\sqrt{U_{0}^{'2}}=22.2~\text{cm}/\text{s}$. In addition, the experimental spectra are scaled such that $l=11M$, allowing the computational domain to contain roughly four integral scales, as described in \citet{misra1996evaluation}. From Eq.(\ref{eq:nondim_parameters}), this yields nondimensional parameters $\Re_{\infty}=1618.417$
%$\Re_{\infty}=1618.416650320742$
and $M_{\infty}=7.874\times10^{-4}$. %$M_{\infty}=0.0007873837059678$. 
\newline\indent To obtain the velocity fluctuation field, the initial CBC spectra (i.e. measured at $t\overline{V}/M=42$) is first nondimensionalized by $L_{\infty}$ and $V_{\infty}$, e.g. $k_{CBC}^{*}=k_{CBC} L_{\infty}$ and $E_{CBC}^{*}=E_{CBC}/(V_{\infty}^{2}L_{\infty})$, as in \citet{misra1996evaluation}. Next, the non-dimensional velocity fluctuation field $\text{v}_{i}^{'*}$ is obtained from this spectra using the \texttt{TurboGenPY} code \cite{saad2017scalable} to perform the Fourier transform into physical space. 
\newline\indent Finally, the non-dimensional flow field is initialized as (uniform pressure, uniform density):
\begin{align}
    \rho^{*}\left(x_{i}^{*},0\right)&=1,\\[1mm]
    \text{v}^{*}_{i}\left(x_{i}^{*},0\right) &= \overline{\text{v}}^{*}_{i} + \text{v}_{i}^{'*}\left(x_{i}^{*},0\right),\\[1mm]
    P^{*}\left(x_{i}^{*},0\right) &= \frac{P}{\rho_{\infty}V_{\infty}^{2}},
\end{align}
within a triply-periodic non-dimensional domain $\bm{x}^{*}\in\left[0,~2\pi\right]^{3}$ where the non-dimensional mean velocity vector is $\overline{\textbf{v}}^{*} = \left[\overline{V}/V_{\infty},~0,~0\right]$, where $\overline{V}$ is obtained from Eq.(\ref{eq:DHIT_reynolds_number}). In this work, we take the first measured spectra time of $t\overline{V}/M=42$ to correspond to $t=0$, and by following the non-dimensionalization outlined in Section \ref{sec:dimensionless_gov_eqxns}, the CBC measured spectra times $t\overline{V}/M=42,98,171$ thus correspond to $t^{*}=0,1,2$. In this work, the domain is discretized with $32^3$ elements for a uniform mesh. The NSFR scheme run in this case is p$3$ $c_{DG}$ NSFR.IR-GLL yielding $128^{3}$ DOF (or $96^3$ unique DOF), and a nondimensional cut-off wavenumber $\kappa_{c}^{*}=48$. Lastly, a CFL number of $0.2$ using the standard advection-based CFL condition of Eq.(\ref{eq:approximate_grid_spacing}) was chosen. Mach number contours and the TKE spectra of the initialized flow field are shown in Fig.~\ref{fig:dhit-initialization}(a) and (b), respectively.

\subsubsection{Results}
\begin{figure}[H]
\begin{subfigure}{.4925\textwidth}
    \centering
    \includegraphics[width=0.99\linewidth]{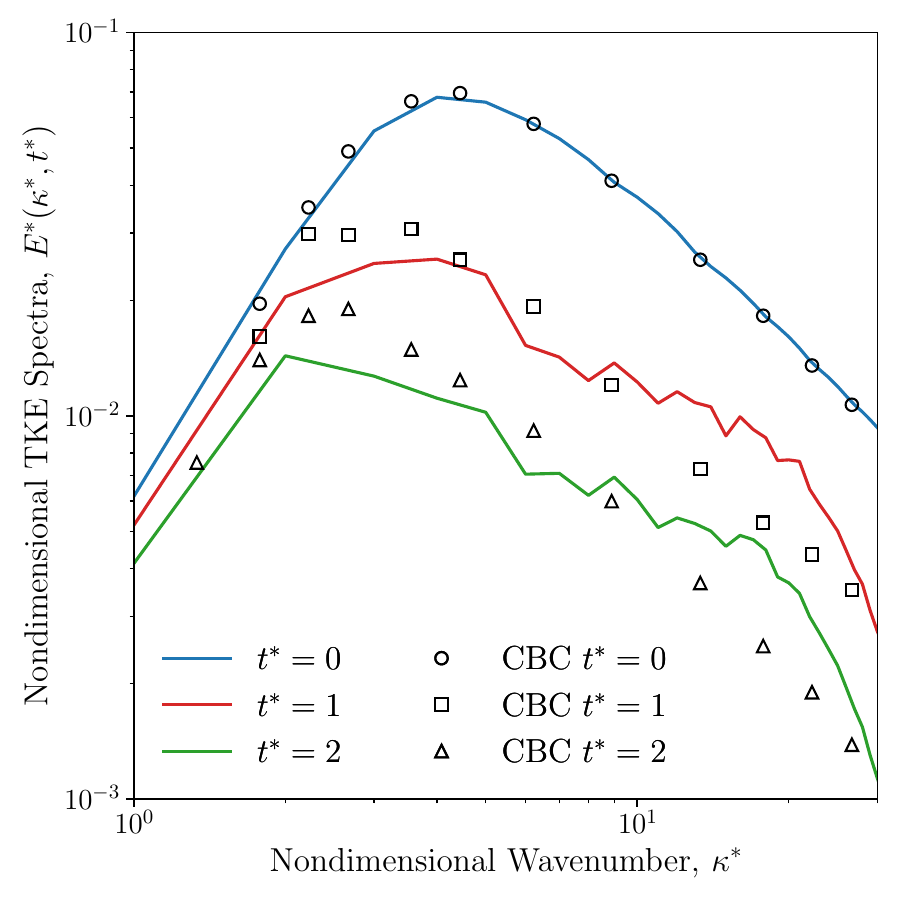}
    \caption{Transient with comparison at all experiment measurement times}
\end{subfigure}%
\begin{subfigure}{.005\textwidth}
    \hfill
\end{subfigure}%
\begin{subfigure}{.4925\textwidth}
    \centering
    \includegraphics[width=0.99\linewidth]{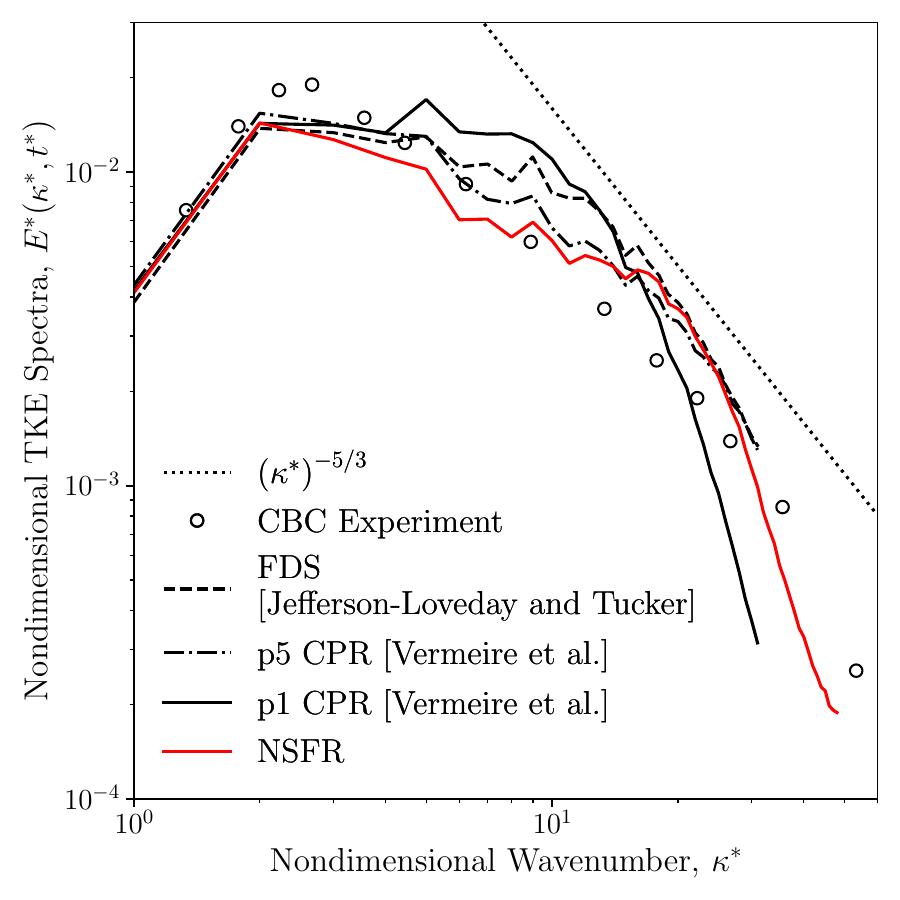}
    \caption{Instantaneous TKE spectra at $t^{*}=2$ where the FDS, p$5$ CPR, p$1$ CPR, and NSFR results are plotted until the cut-off wavenumber $\kappa_{c}=31,~31,~31,~48$ respectively}
\end{subfigure}
\caption{Nondimensional TKE spectra of DHIT using p$3$ $c_{DG}$ NSFR.IR-GLL (denoted as NSFR) with $128^{3}$ DOF with transient on the left, and comparison at final time on the right}
\label{fig:dhit-results}
\end{figure}

The transient nature of the flow is presented in Fig.~\ref{fig:dhit-results}(a) with the instantaneous TKE spectra plotted at times $t^{*}=0,~1,~2$ along with the corresponding CBC experiment measurement data. We see that as time progresses from $t^{*}=0$ to $t^{*}=2$, the flow is naturally decaying with a cascade of energy from large scales to small scales by the downwards shift and progressive reduction in slope with time. In addition, the spectra results are in good agreement with the experiment. 
\newline\indent The spectra at the final time $t^{*}=2$ are compared to the computational results of \citet{jefferson2010impingement} using a sixth-order flux difference splitting (FDS) scheme with $64^3$ DOF, as well as the computational results of \citet{vermeire2016implicit} using a p$5$ correction procedure via reconstruction (CPR) scheme with $78^3$ DOF (or $65^3$ unique DOF) and p$1$ CPR scheme with $126^3$ DOF (or $63^3$ unique DOF) in Fig.~\ref{fig:dhit-results}(b). To quantify the performance of the schemes to capture the experiment-measured TKE spectra at $t^{*}=2$, similar to Section~\ref{subsec:DNS}, we can compare the kinetic energy obtained by integrating the TKE spectra up until $\kappa_{c}$ as in Eq.(\ref{eq:KE_from_TKE_integral}). For this case, the absolute relative percent error is defined as,
\begin{equation}\label{eq:integrate_tke_cbc}
    \epsilon_{E_{K}}=\frac{\left|\tilde{E}_{K}-\tilde{E}^{CBC}_{K}\right|}{\tilde{E}^{CBC}_{K}},
\end{equation}
where $\tilde{E}^{CBC}_{K}$ is the kinetic energy obtained by integrating the CBC experiment's TKE spectra to the cutoff wavenumber of the scheme we are comparing to. Equation (\ref{eq:integrate_tke_cbc}) is computed using the trapezoidal rule, which is sufficient for integrating the $C^{0}$ piecewise continuous TKE spectra, using the same integration bounds and points for each term in the equation. The results are tabulated below.  Note that although Eq.(\ref{eq:cutoff_wavenumber}) would yield a $\kappa_{c}$ of $32$ and $32.5$ for the FDS and CPR schemes respectively, the available data was limited to $\kappa_{c}=31$. From the results in Table~\ref{tab:ke_from_tke_spectra_cbc_errors}, the kinetic energy obtained by the NSFR scheme is much closer to the experimental value. For the same cut-off wavenumber $\kappa_{c}=31$, error for the NSFR scheme, denoted as NSFR$^{\star}$ in Table~\ref{tab:ke_from_tke_spectra_cbc_errors}, is less than half of that for the p$5$ CPR scheme, nearly a quarter of that for the FDS scheme, and fives times less than the p$1$ CPR scheme at a comparable DOF. At the true cut-off wavenumber, the NSFR scheme exhibits the lowest relative percent error. These lower errors seen by the NSFR scheme results are expected to be partially due to the higher unique DOF (i.e. effective resolution) $N_{\text{el.}}\text{p}^{3}$ used for the NSFR scheme, and the use of over-sampling in the generation of the TKE spectra. 
\begin{table}[H]
\centering
\begin{tabular}{c|c|c|c|c|c|c}
    % \hline
    Scheme & \# Numerical DOF, $N_{\text{el.}}(\text{p}+1)^{3}$ & \# Unique DOF, $N_{\text{el.}}\text{p}^{3}$ & $\kappa_{c}$ & $\tilde{E}_{K}^{CBC}$ & $\tilde{E}_{K}$ & $\epsilon_{E_{K}}$\\[1mm]
    \hline
    FDS & $64^{3}$ & $64^{3}$ & $31$ & $0.145$ & $0.188$ & $29.81$\%\\
    p$5$ CPR & $78^{3}$ & $65^{3}$ & $31$ & $0.145$ & $0.168$ & $15.55$\%\\
    p$1$ CPR & $126^{3}$ & $63^{3}$ & $31$ & $0.145$ & $0.188$ & $29.75$\%\\
    NSFR & $128^{3}$ & $96^{3}$ & $48$ & $0.158$ & $0.160$ & $1.36$\%\\
    NSFR$^{\star}$ & $128^{3}$ & $96^{3}$ & $31$ & $0.145$ & $0.153$ & $5.30$\%\\
    \hline
\end{tabular}
\caption{Percent error in kinetic energy compared to the CBC experiment for the FDS, CPR, and NSFR schemes}
\label{tab:ke_from_tke_spectra_cbc_errors}
\end{table}

\section{Conclusion}
\indent The performance of the NSFR schemes \cite{cicchino2022nonlinearly,cicchino2022provably} for resolving viscous turbulent flows has been investigated through simulations of subsonic free-shear turbulent flows. 
The schemes were extensively verified for a DNS of the viscous TGV problem. 
All quantities investigated were in agreement with the available results in the literature. 
It was found that the oversampling procedure was necessary to match the reference TKE spectra, eliminating an apparent pile-up of TKE at the smallest resolved scales.
It was shown that oversampling does indeed yield more physically consistent spectra by comparing the total TKE to the instantaneous kinetic energy of the flow. 
The pressure dilatation based dissipation rate for the nonlinearly stable schemes is consistent with literature when computed from the kinetic energy budget terms, while spurious oscillations are seen when the term is directly computed. 
The magnitude of these oscillations is significantly lower with a collocated NSFR scheme, and are effectively attenuated with addition of upwind-dissipation to the two-point convective numerical flux. 
Therefore, these spurious oscillations are believed to be associated with the treatment of the face terms in nonlinearly stable schemes. 
The performance of the schemes for under-resolved simulations was assessed. First, it was demonstrated that if the split-form is implemented with sum factorization for both tensor and Hadamard products, then entropy-stable schemes can be more cost-effective than over-integration. 
Second, for the TGV flow case, increasing the FR correction parameter $c$ ensures that NSFR is stable and accurate for ILES while allowing for larger time-steps.
Finally, the impact of various dissipation mechanisms on the accuracy of the solution for NSFR schemes was demonstrated. 
It was found that the choice of the two-point numerical flux does not impact the solution and the use of modern eddy-viscosity-based sub-grid scale (SGS) models does not yield improvements for the iLES of the TGV problem. 
The use of a collocated NSFR scheme with added Roe-type upwind dissipation to the convective two-point numerical flux is an effective strategy for eliminating oscillations in the pressure dilatation based dissipation rate. 
The schemes were further demonstrated for the simulation of a DHIT flow case based on the CBC experiment. The obtained TKE spectra results were in good agreement with the spectra measured from the experiment, demonstrating the lowest error in total TKE compared to other computational results in the literature. 
To develop an NSFR approach that is suitable to a wide range of turbulent flows, evaluating the performance of the schemes for the simulation of wall-bounded flows and shock-turbulence interaction problems is ongoing research.

\section*{Acknowledgments}
\indent Julien Brillon thanks the Vadasz Family Foundation, McGill Engineering Doctoral Award (MEDA), and Natural Sciences and Engineering Research Council of Canada (NSERC) Discovery Grant. The authors would also like to thank Jean-Baptiste Chapelier and Alexander Cicchino for helpful discussions throughout the course of this work. 

\newpage
\bibliographystyle{model1-num-names}
\bibliography{refs}

\begin{thebibliography}{92}
\expandafter\ifx\csname natexlab\endcsname\relax\def\natexlab#1{#1}\fi
\providecommand{\url}[1]{\texttt{#1}}
\providecommand{\href}[2]{#2}
\providecommand{\path}[1]{#1}
\providecommand{\DOIprefix}{doi:}
\providecommand{\ArXivprefix}{arXiv:}
\providecommand{\URLprefix}{URL: }
\providecommand{\Pubmedprefix}{pmid:}
\providecommand{\doi}[1]{\href{http://dx.doi.org/#1}{\path{#1}}}
\providecommand{\Pubmed}[1]{\href{pmid:#1}{\path{#1}}}
\providecommand{\bibinfo}[2]{#2}
\ifx\xfnm\relax \def\xfnm[#1]{\unskip,\space#1}\fi
%Type = Article
\bibitem[{Wang et~al.(2013)Wang, Fidkowski, Abgrall, Bassi, Caraeni, Cary,
  Deconinck, Hartmann, Hillewaert, Huynh et~al.}]{wang2013high}
\bibinfo{author}{Z.~J. Wang}, \bibinfo{author}{K.~Fidkowski},
  \bibinfo{author}{R.~Abgrall}, \bibinfo{author}{F.~Bassi},
  \bibinfo{author}{D.~Caraeni}, \bibinfo{author}{A.~Cary},
  \bibinfo{author}{H.~Deconinck}, \bibinfo{author}{R.~Hartmann},
  \bibinfo{author}{K.~Hillewaert}, \bibinfo{author}{H.~T. Huynh}, et~al.,
\newblock \bibinfo{title}{High-order {CFD} methods: current status and
  perspective},
\newblock \bibinfo{journal}{Int. J. Numer. Meth. Fl.} \bibinfo{volume}{72}
  (\bibinfo{year}{2013}) \bibinfo{pages}{811--845}.
%Type = Techreport
\bibitem[{Reed and Hill(1973)}]{reed1973triangular}
\bibinfo{author}{W.~H. Reed}, \bibinfo{author}{T.~Hill},
  \bibinfo{title}{Triangular mesh methods for the neutron transport equation},
  \bibinfo{type}{Technical Report}, Los Alamos Scientific Lab., N. Mex.(USA),
  \bibinfo{year}{1973}.
%Type = Book
\bibitem[{Hesthaven and Warburton(2007)}]{hesthaven2007nodal}
\bibinfo{author}{J.~S. Hesthaven}, \bibinfo{author}{T.~Warburton},
  \bibinfo{title}{Nodal discontinuous {G}alerkin methods: algorithms, analysis,
  and applications}, \bibinfo{publisher}{Springer Science \& Business Media},
  \bibinfo{year}{2007}.
%Type = Book
\bibitem[{Kopriva(2009)}]{kopriva2009implementing}
\bibinfo{author}{D.~A. Kopriva}, \bibinfo{title}{Implementing spectral methods
  for partial differential equations: {A}lgorithms for scientists and
  engineers}, \bibinfo{publisher}{Springer Science \& Business Media},
  \bibinfo{year}{2009}.
%Type = Inproceedings
\bibitem[{Huynh(2007)}]{huynh2007flux}
\bibinfo{author}{H.~T. Huynh},
\newblock \bibinfo{title}{A flux reconstruction approach to high-order schemes
  including discontinuous {G}alerkin methods},
\newblock in: \bibinfo{booktitle}{18th AIAA Computational Fluid Dynamics
  Conference}, \bibinfo{year}{2007}, p. \bibinfo{pages}{4079}.
%Type = Article
\bibitem[{Wang and Gao(2009)}]{wang2009unifying}
\bibinfo{author}{Z.~J. Wang}, \bibinfo{author}{H.~Gao},
\newblock \bibinfo{title}{A unifying lifting collocation penalty formulation
  including the discontinuous {G}alerkin, spectral volume/difference methods
  for conservation laws on mixed grids},
\newblock \bibinfo{journal}{J. Comput. Phys.} \bibinfo{volume}{228}
  (\bibinfo{year}{2009}) \bibinfo{pages}{8161--8186}.
%Type = Article
\bibitem[{Jameson et~al.(2012)Jameson, Vincent, and
  Castonguay}]{jameson2012non}
\bibinfo{author}{A.~Jameson}, \bibinfo{author}{P.~E. Vincent},
  \bibinfo{author}{P.~Castonguay},
\newblock \bibinfo{title}{On the non-linear stability of flux reconstruction
  schemes},
\newblock \bibinfo{journal}{J. Sci. Comput.} \bibinfo{volume}{50}
  (\bibinfo{year}{2012}) \bibinfo{pages}{434--445}.
%Type = Article
\bibitem[{Vincent et~al.(2011)Vincent, Castonguay, and
  Jameson}]{vincent2011new}
\bibinfo{author}{P.~E. Vincent}, \bibinfo{author}{P.~Castonguay},
  \bibinfo{author}{A.~Jameson},
\newblock \bibinfo{title}{A new class of high-order energy stable flux
  reconstruction schemes},
\newblock \bibinfo{journal}{J. Sci. Comput.} \bibinfo{volume}{47}
  (\bibinfo{year}{2011}) \bibinfo{pages}{50--72}.
%Type = Book
\bibitem[{Castonguay(2012)}]{castonguay2012high}
\bibinfo{author}{P.~Castonguay}, \bibinfo{title}{High-order energy stable flux
  reconstruction schemes for fluid flow simulations on unstructured grids},
  \bibinfo{publisher}{Stanford University}, \bibinfo{year}{2012}.
%Type = Article
\bibitem[{Haga et~al.(2011)Haga, Gao, and Wang}]{haga2011high}
\bibinfo{author}{T.~Haga}, \bibinfo{author}{H.~Gao}, \bibinfo{author}{Z.~J.
  Wang},
\newblock \bibinfo{title}{A high-order unifying discontinuous formulation for
  the {N}avier-{S}tokes equations on 3{D} mixed grids},
\newblock \bibinfo{journal}{Math. Model. Nat. Pheno.} \bibinfo{volume}{6}
  (\bibinfo{year}{2011}) \bibinfo{pages}{28--56}.
%Type = Article
\bibitem[{Castonguay et~al.(2013)Castonguay, Williams, Vincent, and
  Jameson}]{castonguay2013energy}
\bibinfo{author}{P.~Castonguay}, \bibinfo{author}{D.~M. Williams},
  \bibinfo{author}{P.~E. Vincent}, \bibinfo{author}{A.~Jameson},
\newblock \bibinfo{title}{Energy stable flux reconstruction schemes for
  advection-diffusion problems},
\newblock \bibinfo{journal}{Comput. Method Appl. M.} \bibinfo{volume}{267}
  (\bibinfo{year}{2013}) \bibinfo{pages}{400--417}.
%Type = Article
\bibitem[{Kopriva and Kolias(1996)}]{kopriva1996conservative}
\bibinfo{author}{D.~A. Kopriva}, \bibinfo{author}{J.~H. Kolias},
\newblock \bibinfo{title}{A conservative staggered-grid {C}hebyshev multidomain
  method for compressible flows},
\newblock \bibinfo{journal}{J. Comput. Phys.} \bibinfo{volume}{125}
  (\bibinfo{year}{1996}) \bibinfo{pages}{244--261}.
%Type = Article
\bibitem[{Liu et~al.(2006)Liu, Vinokur, and Wang}]{liu2006spectral}
\bibinfo{author}{Y.~Liu}, \bibinfo{author}{M.~Vinokur}, \bibinfo{author}{Z.~J.
  Wang},
\newblock \bibinfo{title}{Spectral difference method for unstructured grids
  {I}: basic formulation},
\newblock \bibinfo{journal}{J. Comput. Phys.} \bibinfo{volume}{216}
  (\bibinfo{year}{2006}) \bibinfo{pages}{780--801}.
%Type = Article
\bibitem[{Allaneau and Jameson(2011)}]{allaneau2011connections}
\bibinfo{author}{Y.~Allaneau}, \bibinfo{author}{A.~Jameson},
\newblock \bibinfo{title}{Connections between the filtered discontinuous
  {G}alerkin method and the flux reconstruction approach to high order
  discretizations},
\newblock \bibinfo{journal}{Comput. Method Appl. M.} \bibinfo{volume}{200}
  (\bibinfo{year}{2011}) \bibinfo{pages}{3628--3636}.
%Type = Article
\bibitem[{Zwanenburg and Nadarajah(2016)}]{zwanenburg2016equivalence}
\bibinfo{author}{P.~Zwanenburg}, \bibinfo{author}{S.~Nadarajah},
\newblock \bibinfo{title}{Equivalence between the energy stable flux
  reconstruction and filtered discontinuous {G}alerkin schemes},
\newblock \bibinfo{journal}{J. Comput. Phys.} \bibinfo{volume}{306}
  (\bibinfo{year}{2016}) \bibinfo{pages}{343--369}.
%Type = Article
\bibitem[{Taylor and Green(1937)}]{taylor1937mechanism}
\bibinfo{author}{G.~I. Taylor}, \bibinfo{author}{A.~E. Green},
\newblock \bibinfo{title}{Mechanism of the production of small eddies from
  large ones},
\newblock \bibinfo{journal}{Proceedings of the Royal Society of London. Series
  A-Mathematical and Physical Sciences} \bibinfo{volume}{158}
  (\bibinfo{year}{1937}) \bibinfo{pages}{499--521}.
%Type = Article
\bibitem[{Drikakis et~al.(2007)Drikakis, Fureby, Grinstein, and
  Youngs}]{drikakis2007simulation}
\bibinfo{author}{D.~Drikakis}, \bibinfo{author}{C.~Fureby},
  \bibinfo{author}{F.~F. Grinstein}, \bibinfo{author}{D.~Youngs},
\newblock \bibinfo{title}{Simulation of transition and turbulence decay in the
  {T}aylor-{G}reen vortex},
\newblock \bibinfo{journal}{J. Turbul.}  (\bibinfo{year}{2007})
  \bibinfo{pages}{N20}.
%Type = Article
\bibitem[{Gassner and Beck(2013)}]{gassner2013accuracy}
\bibinfo{author}{G.~J. Gassner}, \bibinfo{author}{A.~D. Beck},
\newblock \bibinfo{title}{On the accuracy of high-order discretizations for
  underresolved turbulence simulations},
\newblock \bibinfo{journal}{Theor. Comp. Fluid Dyn.} \bibinfo{volume}{27}
  (\bibinfo{year}{2013}) \bibinfo{pages}{221--237}.
%Type = Inproceedings
\bibitem[{Diosady and Murman(2013)}]{diosady2013design}
\bibinfo{author}{L.~T. Diosady}, \bibinfo{author}{S.~M. Murman},
\newblock \bibinfo{title}{Design of a variational multiscale method for high
  {R}eynolds number compressible flows},
\newblock in: \bibinfo{booktitle}{21st AIAA Computational Fluid Dynamics
  Conference}, \bibinfo{year}{2013}, p. \bibinfo{pages}{2870}.
%Type = Inproceedings
\bibitem[{Diosady and Murman(2014)}]{diosady2014dns}
\bibinfo{author}{L.~T. Diosady}, \bibinfo{author}{S.~M. Murman},
\newblock \bibinfo{title}{{DNS} of flows over periodic hills using a
  discontinuous {G}alerkin spectral-element method},
\newblock in: \bibinfo{booktitle}{44th AIAA Fluid Dynamics Conference},
  \bibinfo{year}{2014}, p. \bibinfo{pages}{2784}.
%Type = Inproceedings
\bibitem[{Chapelier et~al.(2012)Chapelier, Plata, Renac, and
  Martin}]{chapelier2012final}
\bibinfo{author}{J.~Chapelier}, \bibinfo{author}{M.~D. L.~L. Plata},
  \bibinfo{author}{F.~Renac}, \bibinfo{author}{E.~Martin},
\newblock \bibinfo{title}{Final abstract for {ONERA} {T}aylor-{G}reen {DG}
  participation},
\newblock in: \bibinfo{booktitle}{1st International Workshop On High-Order CFD
  Methods}, \bibinfo{year}{2012}, pp. \bibinfo{pages}{7--8}.
%Type = Phdthesis
\bibitem[{Chapelier(2013)}]{chapelier2013developpement}
\bibinfo{author}{J.-B. Chapelier}, \bibinfo{title}{D{\'e}veloppement et
  {\'e}valuation de la m{\'e}thode de {G}alerkin discontinue pour la simulation
  des grandes {\'e}chelles des {\'e}coulements turbulents}, Ph.D. thesis,
  Universit{\'e} Sciences et Technologies-Bordeaux I, \bibinfo{year}{2013}.
%Type = Article
\bibitem[{Chapelier et~al.(2016)Chapelier, De~La Llave~Plata, and
  Lamballais}]{chapelier2016development}
\bibinfo{author}{J.-B. Chapelier}, \bibinfo{author}{M.~De~La Llave~Plata},
  \bibinfo{author}{E.~Lamballais},
\newblock \bibinfo{title}{Development of a multiscale {LES} model in the
  context of a modal discontinuous {G}alerkin method},
\newblock \bibinfo{journal}{Comput. Method Appl. M.} \bibinfo{volume}{307}
  (\bibinfo{year}{2016}) \bibinfo{pages}{275--299}.
%Type = Article
\bibitem[{Carton~de Wiart et~al.(2014)Carton~de Wiart, Hillewaert, Duponcheel,
  and Winckelmans}]{carton2014assessment}
\bibinfo{author}{C.~Carton~de Wiart}, \bibinfo{author}{K.~Hillewaert},
  \bibinfo{author}{M.~Duponcheel}, \bibinfo{author}{G.~Winckelmans},
\newblock \bibinfo{title}{Assessment of a discontinuous {G}alerkin method for
  the simulation of vortical flows at high {R}eynolds number},
\newblock \bibinfo{journal}{International Journal for Numerical Methods in
  Fluids} \bibinfo{volume}{74} (\bibinfo{year}{2014})
  \bibinfo{pages}{469--493}.
%Type = Inproceedings
\bibitem[{Johnsen et~al.(2013)Johnsen, Varadan, and
  Van~Leer}]{johnsen2013three}
\bibinfo{author}{E.~Johnsen}, \bibinfo{author}{S.~Varadan},
  \bibinfo{author}{B.~Van~Leer},
\newblock \bibinfo{title}{A three-dimensional recovery-based discontinuous
  {G}alerkin method for turbulence simulations},
\newblock in: \bibinfo{booktitle}{51st AIAA Aerospace Sciences Meeting
  including the New Horizons Forum and Aerospace Exposition},
  \bibinfo{year}{2013}, p. \bibinfo{pages}{515}.
%Type = Article
\bibitem[{Vermeire et~al.(2016)Vermeire, Nadarajah, and
  Tucker}]{vermeire2016implicit}
\bibinfo{author}{B.~C. Vermeire}, \bibinfo{author}{S.~Nadarajah},
  \bibinfo{author}{P.~G. Tucker},
\newblock \bibinfo{title}{Implicit large eddy simulation using the high-order
  correction procedure via reconstruction scheme},
\newblock \bibinfo{journal}{Int. J. Numer. Meth. Fl.} \bibinfo{volume}{82}
  (\bibinfo{year}{2016}) \bibinfo{pages}{231--260}.
%Type = Article
\bibitem[{Navah et~al.(2020)Navah, de~la Llave~Plata, and
  Couaillier}]{navah2020high}
\bibinfo{author}{F.~Navah}, \bibinfo{author}{M.~de~la Llave~Plata},
  \bibinfo{author}{V.~Couaillier},
\newblock \bibinfo{title}{A high-order multiscale approach to turbulence for
  compact nodal schemes},
\newblock \bibinfo{journal}{Computer methods in applied mechanics and
  engineering} \bibinfo{volume}{363} (\bibinfo{year}{2020})
  \bibinfo{pages}{112885}.
%Type = Article
\bibitem[{Bull and Jameson(2015)}]{bull2015simulation}
\bibinfo{author}{J.~R. Bull}, \bibinfo{author}{A.~Jameson},
\newblock \bibinfo{title}{Simulation of the {T}aylor-{G}reen vortex using
  high-order flux reconstruction schemes},
\newblock \bibinfo{journal}{AIAA Journal} \bibinfo{volume}{53}
  (\bibinfo{year}{2015}) \bibinfo{pages}{2750--2761}.
%Type = Article
\bibitem[{Kirby and Karniadakis(2003)}]{kirby2003aliasing}
\bibinfo{author}{R.~M. Kirby}, \bibinfo{author}{G.~E. Karniadakis},
\newblock \bibinfo{title}{De-aliasing on non-uniform grids: algorithms and
  applications},
\newblock \bibinfo{journal}{J. Comput. Phys.} \bibinfo{volume}{191}
  (\bibinfo{year}{2003}) \bibinfo{pages}{249--264}.
%Type = Article
\bibitem[{Gassner et~al.(2016)Gassner, Winters, and Kopriva}]{gassner2016split}
\bibinfo{author}{G.~J. Gassner}, \bibinfo{author}{A.~R. Winters},
  \bibinfo{author}{D.~A. Kopriva},
\newblock \bibinfo{title}{Split form nodal discontinuous {G}alerkin schemes
  with summation-by-parts property for the compressible {E}uler equations},
\newblock \bibinfo{journal}{J. Comput. Phys.} \bibinfo{volume}{327}
  (\bibinfo{year}{2016}) \bibinfo{pages}{39--66}.
%Type = Article
\bibitem[{Moura et~al.(2017)Moura, Mengaldo, Peir{\'o}, and
  Sherwin}]{moura2017eddy}
\bibinfo{author}{R.~C. Moura}, \bibinfo{author}{G.~Mengaldo},
  \bibinfo{author}{J.~Peir{\'o}}, \bibinfo{author}{S.~J. Sherwin},
\newblock \bibinfo{title}{On the eddy-resolving capability of high-order
  discontinuous {G}alerkin approaches to implicit {LES}/under-resolved {DNS} of
  {E}uler turbulence},
\newblock \bibinfo{journal}{J. Comput. Phys.} \bibinfo{volume}{330}
  (\bibinfo{year}{2017}) \bibinfo{pages}{615--623}.
%Type = Article
\bibitem[{Tadmor(1984)}]{tadmor1984skew}
\bibinfo{author}{E.~Tadmor},
\newblock \bibinfo{title}{Skew-selfadjoint form for systems of conservation
  laws},
\newblock \bibinfo{journal}{J. Math. Anal. Appl.} \bibinfo{volume}{103}
  (\bibinfo{year}{1984}) \bibinfo{pages}{428--442}.
%Type = Article
\bibitem[{Harten(1983)}]{harten1983symmetric}
\bibinfo{author}{A.~Harten},
\newblock \bibinfo{title}{On the symmetric form of systems of conservation laws
  with entropy},
\newblock \bibinfo{journal}{J. Comput. Phys.} \bibinfo{volume}{49}
  (\bibinfo{year}{1983}).
%Type = Article
\bibitem[{Lefloch et~al.(2002)Lefloch, Mercier, and Rohde}]{lefloch2002fully}
\bibinfo{author}{P.~G. Lefloch}, \bibinfo{author}{J.-M. Mercier},
  \bibinfo{author}{C.~Rohde},
\newblock \bibinfo{title}{Fully discrete, entropy conservative schemes of
  arbitrary order},
\newblock \bibinfo{journal}{SIAM Journal on Numerical Analysis}
  \bibinfo{volume}{40} (\bibinfo{year}{2002}) \bibinfo{pages}{1968--1992}.
%Type = Article
\bibitem[{Fisher et~al.(2013)Fisher, Carpenter, Nordstr{\"o}m, Yamaleev, and
  Swanson}]{fisher2013discretely}
\bibinfo{author}{T.~C. Fisher}, \bibinfo{author}{M.~H. Carpenter},
  \bibinfo{author}{J.~Nordstr{\"o}m}, \bibinfo{author}{N.~K. Yamaleev},
  \bibinfo{author}{C.~Swanson},
\newblock \bibinfo{title}{Discretely conservative finite-difference
  formulations for nonlinear conservation laws in split form: {T}heory and
  boundary conditions},
\newblock \bibinfo{journal}{Journal of Computational Physics}
  \bibinfo{volume}{234} (\bibinfo{year}{2013}) \bibinfo{pages}{353--375}.
%Type = Article
\bibitem[{Fern{\'a}ndez et~al.(2014)Fern{\'a}ndez, Hicken, and
  Zingg}]{fernandez2014review}
\bibinfo{author}{D.~C. D.~R. Fern{\'a}ndez}, \bibinfo{author}{J.~E. Hicken},
  \bibinfo{author}{D.~W. Zingg},
\newblock \bibinfo{title}{Review of summation-by-parts operators with
  simultaneous approximation terms for the numerical solution of partial
  differential equations},
\newblock \bibinfo{journal}{Comput. Fluids} \bibinfo{volume}{95}
  (\bibinfo{year}{2014}) \bibinfo{pages}{171--196}.
%Type = Article
\bibitem[{Sv{\"a}rd and Nordstr{\"o}m(2014)}]{svard2014review}
\bibinfo{author}{M.~Sv{\"a}rd}, \bibinfo{author}{J.~Nordstr{\"o}m},
\newblock \bibinfo{title}{Review of summation-by-parts schemes for
  initial-boundary-value problems},
\newblock \bibinfo{journal}{J. Comput. Phys.} \bibinfo{volume}{268}
  (\bibinfo{year}{2014}) \bibinfo{pages}{17--38}.
%Type = Article
\bibitem[{Gassner(2013)}]{gassner2013skew}
\bibinfo{author}{G.~J. Gassner},
\newblock \bibinfo{title}{A skew-symmetric discontinuous {G}alerkin spectral
  element discretization and its relation to {SBP}-{SAT} finite difference
  methods},
\newblock \bibinfo{journal}{SIAM J. Sci. Comput.} \bibinfo{volume}{35}
  (\bibinfo{year}{2013}) \bibinfo{pages}{A1233--A1253}.
%Type = Article
\bibitem[{Crean et~al.(2018)Crean, Hicken, Fern{\'a}ndez, Zingg, and
  Carpenter}]{crean2018entropy}
\bibinfo{author}{J.~Crean}, \bibinfo{author}{J.~E. Hicken},
  \bibinfo{author}{D.~C. D.~R. Fern{\'a}ndez}, \bibinfo{author}{D.~W. Zingg},
  \bibinfo{author}{M.~H. Carpenter},
\newblock \bibinfo{title}{Entropy-stable summation-by-parts discretization of
  the {E}uler equations on general curved elements},
\newblock \bibinfo{journal}{J. Comput. Phys.} \bibinfo{volume}{356}
  (\bibinfo{year}{2018}) \bibinfo{pages}{410--438}.
%Type = Article
\bibitem[{Chan et~al.(2019)Chan, Del Rey~Fernández, and
  Carpenter}]{chan2019efficient}
\bibinfo{author}{J.~Chan}, \bibinfo{author}{D.~C. Del Rey~Fernández},
  \bibinfo{author}{M.~H. Carpenter},
\newblock \bibinfo{title}{Efficient entropy stable {G}auss collocation
  methods},
\newblock \bibinfo{journal}{SIAM J. Sci. Comput.} \bibinfo{volume}{41}
  (\bibinfo{year}{2019}) \bibinfo{pages}{A2938--A2966}.
%Type = Article
\bibitem[{Fernandez et~al.(2019)Fernandez, Carpenter, Dalcin, Fredrich, Rojas,
  Winters, Gassner, Zampini, and Parsani}]{fernandez2019entropy}
\bibinfo{author}{D.~Fernandez}, \bibinfo{author}{M.~H. Carpenter},
  \bibinfo{author}{L.~Dalcin}, \bibinfo{author}{L.~Fredrich},
  \bibinfo{author}{D.~Rojas}, \bibinfo{author}{A.~R. Winters},
  \bibinfo{author}{G.~J. Gassner}, \bibinfo{author}{S.~Zampini},
  \bibinfo{author}{M.~Parsani},
\newblock \bibinfo{title}{Entropy stable p-nonconforming discretizations with
  the summation-by-parts property for the compressible {E}uler equations},
\newblock \bibinfo{journal}{arXiv preprint arXiv:1909.12536}
  (\bibinfo{year}{2019}).
%Type = Article
\bibitem[{Flad and Gassner(2017)}]{flad2017use}
\bibinfo{author}{D.~Flad}, \bibinfo{author}{G.~Gassner},
\newblock \bibinfo{title}{On the use of kinetic energy preserving {DG}-schemes
  for large eddy simulation},
\newblock \bibinfo{journal}{J. Comput. Phys.} \bibinfo{volume}{350}
  (\bibinfo{year}{2017}) \bibinfo{pages}{782--795}.
%Type = Article
\bibitem[{Fern{\'a}ndez et~al.(2020)Fern{\'a}ndez, Carpenter, Dalcin, Zampini,
  and Parsani}]{fernandez2020entropy}
\bibinfo{author}{D.~C. D.~R. Fern{\'a}ndez}, \bibinfo{author}{M.~H. Carpenter},
  \bibinfo{author}{L.~Dalcin}, \bibinfo{author}{S.~Zampini},
  \bibinfo{author}{M.~Parsani},
\newblock \bibinfo{title}{Entropy stable h/p-nonconforming discretization with
  the summation-by-parts property for the compressible {E}uler and
  {N}avier--{S}tokes equations},
\newblock \bibinfo{journal}{SN Partial Differential Equations and Applications}
  \bibinfo{volume}{1} (\bibinfo{year}{2020}) \bibinfo{pages}{9}.
%Type = Article
\bibitem[{Parsani et~al.(2021)Parsani, Boukharfane, Nolasco, Fern{\'a}ndez,
  Zampini, Hadri, and Dalcin}]{parsani2021high}
\bibinfo{author}{M.~Parsani}, \bibinfo{author}{R.~Boukharfane},
  \bibinfo{author}{I.~R. Nolasco}, \bibinfo{author}{D.~C. D.~R. Fern{\'a}ndez},
  \bibinfo{author}{S.~Zampini}, \bibinfo{author}{B.~Hadri},
  \bibinfo{author}{L.~Dalcin},
\newblock \bibinfo{title}{High-order accurate entropy-stable discontinuous
  collocated {G}alerkin methods with the summation-by-parts property for
  compressible {CFD} frameworks: Scalable {SSDC} algorithms and flow solver},
\newblock \bibinfo{journal}{J. Comput. Phys.} \bibinfo{volume}{424}
  (\bibinfo{year}{2021}) \bibinfo{pages}{109844}.
%Type = Article
\bibitem[{Rojas et~al.(2021)Rojas, Boukharfane, Dalcin, Fern{\'a}ndez, Ranocha,
  Keyes, and Parsani}]{rojas2021robustness}
\bibinfo{author}{D.~Rojas}, \bibinfo{author}{R.~Boukharfane},
  \bibinfo{author}{L.~Dalcin}, \bibinfo{author}{D.~C. D.~R. Fern{\'a}ndez},
  \bibinfo{author}{H.~Ranocha}, \bibinfo{author}{D.~E. Keyes},
  \bibinfo{author}{M.~Parsani},
\newblock \bibinfo{title}{On the robustness and performance of entropy stable
  collocated discontinuous {G}alerkin methods},
\newblock \bibinfo{journal}{J. Comput. Phys.} \bibinfo{volume}{426}
  (\bibinfo{year}{2021}) \bibinfo{pages}{109891}.
%Type = Inproceedings
\bibitem[{Al~Jahdali et~al.(2021)Al~Jahdali, Boukharfane, Dalcin, and
  Parsani}]{al2021optimized}
\bibinfo{author}{R.~Al~Jahdali}, \bibinfo{author}{R.~Boukharfane},
  \bibinfo{author}{L.~Dalcin}, \bibinfo{author}{M.~Parsani},
\newblock \bibinfo{title}{Optimized explicit {R}unge-{K}utta schemes for
  entropy stable discontinuous collocated methods applied to the {E}uler and
  {N}avier--{S}tokes equations},
\newblock in: \bibinfo{booktitle}{AIAA Scitech 2021 Forum},
  \bibinfo{year}{2021}, p. \bibinfo{pages}{0633}.
%Type = Article
\bibitem[{Cicchino et~al.(2022)Cicchino, Nadarajah, and
  Fern{\'a}ndez}]{cicchino2022nonlinearly}
\bibinfo{author}{A.~Cicchino}, \bibinfo{author}{S.~Nadarajah},
  \bibinfo{author}{D.~C. D.~R. Fern{\'a}ndez},
\newblock \bibinfo{title}{Nonlinearly stable flux reconstruction high-order
  methods in split form},
\newblock \bibinfo{journal}{J. Comput. Phys.} \bibinfo{volume}{458}
  (\bibinfo{year}{2022}) \bibinfo{pages}{111094}.
%Type = Article
\bibitem[{Shi-Dong and Nadarajah(2021)}]{shi2021full}
\bibinfo{author}{D.~Shi-Dong}, \bibinfo{author}{S.~Nadarajah},
\newblock \bibinfo{title}{Full-space approach to aerodynamic shape
  optimization},
\newblock \bibinfo{journal}{Comput. Fluids} \bibinfo{volume}{218}
  (\bibinfo{year}{2021}) \bibinfo{pages}{104843}.
%Type = Article
\bibitem[{Shi-Dong et~al.(2022)Shi-Dong, Nadarajah, Cicchino, Brillon, Blais,
  MacLean, Thakur, and Pethrick}]{philip2022code}
\bibinfo{author}{D.~Shi-Dong}, \bibinfo{author}{S.~Nadarajah},
  \bibinfo{author}{A.~Cicchino}, \bibinfo{author}{J.~Brillon},
  \bibinfo{author}{D.~Blais}, \bibinfo{author}{K.~MacLean},
  \bibinfo{author}{P.~Thakur}, \bibinfo{author}{C.~Pethrick},
\newblock \bibinfo{title}{Philip}  (\bibinfo{year}{2022}).
%Type = Article
\bibitem[{Bangerth et~al.(2007)Bangerth, Hartmann, and
  Kanschat}]{bangerth2007deal}
\bibinfo{author}{W.~Bangerth}, \bibinfo{author}{R.~Hartmann},
  \bibinfo{author}{G.~Kanschat},
\newblock \bibinfo{title}{deal.{II}—a general-purpose object-oriented finite
  element library},
\newblock \bibinfo{journal}{ACM Transactions on Mathematical Software (TOMS)}
  \bibinfo{volume}{33} (\bibinfo{year}{2007}) \bibinfo{pages}{24--es}.
%Type = Article
\bibitem[{Stokes(2007)}]{stokes2007theories}
\bibinfo{author}{G.~G. Stokes},
\newblock \bibinfo{title}{On the theories of the internal friction of fluids in
  motion, and of the equilibrium and motion of elastic solids}
  (\bibinfo{year}{2007}).
%Type = Article
\bibitem[{Sutherland(1893)}]{sutherland1893lii}
\bibinfo{author}{W.~Sutherland},
\newblock \bibinfo{title}{Lii. the viscosity of gases and molecular force},
\newblock \bibinfo{journal}{The London, Edinburgh, and Dublin Philosophical
  Magazine and Journal of Science} \bibinfo{volume}{36} (\bibinfo{year}{1893})
  \bibinfo{pages}{507--531}.
%Type = Misc
\bibitem[{Pope(2001)}]{pope2001turbulent}
\bibinfo{author}{S.~B. Pope}, \bibinfo{title}{Turbulent flows},
  \bibinfo{year}{2001}.
%Type = Article
\bibitem[{de~la Llave~Plata et~al.(2019)de~la Llave~Plata, Lamballais, and
  Naddei}]{plata2019performance}
\bibinfo{author}{M.~de~la Llave~Plata}, \bibinfo{author}{E.~Lamballais},
  \bibinfo{author}{F.~Naddei},
\newblock \bibinfo{title}{On the performance of a high-order multiscale {DG}
  approach to {LES} at increasing {R}eynolds number},
\newblock \bibinfo{journal}{Comput. Fluids} \bibinfo{volume}{194}
  (\bibinfo{year}{2019}) \bibinfo{pages}{104306}.
%Type = Article
\bibitem[{Schranner et~al.(2015)Schranner, Domaradzki, Hickel, and
  Adams}]{schranner2015assessing}
\bibinfo{author}{F.~S. Schranner}, \bibinfo{author}{J.~A. Domaradzki},
  \bibinfo{author}{S.~Hickel}, \bibinfo{author}{N.~A. Adams},
\newblock \bibinfo{title}{Assessing the numerical dissipation rate and
  viscosity in numerical simulations of fluid flows},
\newblock \bibinfo{journal}{Comput. Fluids} \bibinfo{volume}{114}
  (\bibinfo{year}{2015}) \bibinfo{pages}{84--97}.
%Type = Article
\bibitem[{Castiglioni et~al.(2015)Castiglioni, Domaradzki, Krais, Munz,
  Schranner et~al.}]{castiglioni2015characterizaion}
\bibinfo{author}{G.~Castiglioni}, \bibinfo{author}{J.~Domaradzki},
  \bibinfo{author}{N.~B. Krais}, \bibinfo{author}{K.~Munz},
  \bibinfo{author}{F.~Schranner}, et~al.,
\newblock \bibinfo{title}{Characterizaion of numerical dissipation rates in
  numerical simulations performed using discontinuous {G}alerkin methods}
  (\bibinfo{year}{2015}).
%Type = Article
\bibitem[{Duan and Wang(2024)}]{duan2024calibrating}
\bibinfo{author}{Z.~Duan}, \bibinfo{author}{Z.~Wang},
\newblock \bibinfo{title}{Calibrating sub-grid scale models for high-order
  wall-modeled large eddy simulation},
\newblock \bibinfo{journal}{Advances in Aerodynamics} \bibinfo{volume}{6}
  (\bibinfo{year}{2024}) \bibinfo{pages}{5}.
%Type = Article
\bibitem[{Smagorinsky(1963)}]{smagorinsky1963general}
\bibinfo{author}{J.~Smagorinsky},
\newblock \bibinfo{title}{General circulation experiments with the primitive
  equations: {I}. {T}he basic experiment},
\newblock \bibinfo{journal}{Mon. Weather Rev.} \bibinfo{volume}{91}
  (\bibinfo{year}{1963}) \bibinfo{pages}{99--164}.
%Type = Article
\bibitem[{Deardorff(1970)}]{deardorff1970numerical}
\bibinfo{author}{J.~W. Deardorff},
\newblock \bibinfo{title}{A numerical study of three-dimensional turbulent
  channel flow at large reynolds numbers},
\newblock \bibinfo{journal}{J. Fluid Mech.} \bibinfo{volume}{41}
  (\bibinfo{year}{1970}) \bibinfo{pages}{453--480}.
%Type = Article
\bibitem[{L{\'e}v{\^e}que et~al.(2007)L{\'e}v{\^e}que, Toschi, Shao, and
  Bertoglio}]{leveque2007shear}
\bibinfo{author}{E.~L{\'e}v{\^e}que}, \bibinfo{author}{F.~Toschi},
  \bibinfo{author}{L.~Shao}, \bibinfo{author}{J.-P. Bertoglio},
\newblock \bibinfo{title}{Shear-improved smagorinsky model for large-eddy
  simulation of wall-bounded turbulent flows},
\newblock \bibinfo{journal}{Journal of Fluid Mechanics} \bibinfo{volume}{570}
  (\bibinfo{year}{2007}) \bibinfo{pages}{491--502}.
%Type = Article
\bibitem[{Lilly(1992)}]{lilly1992proposed}
\bibinfo{author}{D.~K. Lilly},
\newblock \bibinfo{title}{A proposed modification of the {G}ermano
  subgrid-scale closure method},
\newblock \bibinfo{journal}{Physics of Fluids A: Fluid Dynamics}
  \bibinfo{volume}{4} (\bibinfo{year}{1992}) \bibinfo{pages}{633--635}.
%Type = Book
\bibitem[{Blazek(2015)}]{blazek2001cfd}
\bibinfo{author}{J.~Blazek}, \bibinfo{title}{Computational fluid dynamics:
  principles and applications}, \bibinfo{publisher}{Elsevier},
  \bibinfo{year}{2015}.
%Type = Article
\bibitem[{Meyers and Sagaut(2006)}]{meyers2006model}
\bibinfo{author}{J.~Meyers}, \bibinfo{author}{P.~Sagaut},
\newblock \bibinfo{title}{On the model coefficients for the standard and the
  variational multi-scale smagorinsky model},
\newblock \bibinfo{journal}{Journal of Fluid Mechanics} \bibinfo{volume}{569}
  (\bibinfo{year}{2006}) \bibinfo{pages}{287--319}.
%Type = Article
\bibitem[{Cicchino et~al.(2022)Cicchino, Fern{\'a}ndez, Nadarajah, Chan, and
  Carpenter}]{cicchino2022provably}
\bibinfo{author}{A.~Cicchino}, \bibinfo{author}{D.~C. D.~R. Fern{\'a}ndez},
  \bibinfo{author}{S.~Nadarajah}, \bibinfo{author}{J.~Chan},
  \bibinfo{author}{M.~H. Carpenter},
\newblock \bibinfo{title}{Provably stable flux reconstruction high-order
  methods on curvilinear elements},
\newblock \bibinfo{journal}{J. Comput. Phys.} \bibinfo{volume}{463}
  (\bibinfo{year}{2022}) \bibinfo{pages}{111259}.
%Type = Article
\bibitem[{Cicchino and Nadarajah(2023)}]{cicchino2023discretely}
\bibinfo{author}{A.~Cicchino}, \bibinfo{author}{S.~Nadarajah},
\newblock \bibinfo{title}{Discretely nonlinearly stable weight-adjusted flux
  reconstruction high-order method for compressible flows on curvilinear
  grids},
\newblock \bibinfo{journal}{arXiv preprint arXiv:2312.07725}
  (\bibinfo{year}{2023}).
%Type = Article
\bibitem[{Cicchino and Nadarajah(2024)}]{cicchino2024scalable}
\bibinfo{author}{A.~Cicchino}, \bibinfo{author}{S.~Nadarajah},
\newblock \bibinfo{title}{Scalable evaluation of {H}adamard products with
  tensor product basis for entropy-stable high-order methods},
\newblock \bibinfo{journal}{Journal of Computational Physics}
  (\bibinfo{year}{2024}) \bibinfo{pages}{113134}.
%Type = Phdthesis
\bibitem[{Cicchino(2024)}]{cicchino2024thesis}
\bibinfo{author}{A.~Cicchino}, \bibinfo{title}{Weight-Adjusted Nonlinearly
  Stable Flux Reconstruction High-Order Methods for Compressible Flows in
  Curvilinear Coordinates}, Ph.D. thesis, McGill University,
  \bibinfo{year}{2024}.
%Type = Article
\bibitem[{Quaegebeur et~al.(2019)Quaegebeur, Nadarajah, Navah, and
  Zwanenburg}]{quaegebeur2019cnf}
\bibinfo{author}{S.~Quaegebeur}, \bibinfo{author}{S.~Nadarajah},
  \bibinfo{author}{F.~Navah}, \bibinfo{author}{P.~Zwanenburg},
\newblock \bibinfo{title}{Stability of energy stable flux reconstruction for
  the diffusion problem using compact numerical fluxes},
\newblock \bibinfo{journal}{SIAM J. Sci. Comput.} \bibinfo{volume}{41}
  (\bibinfo{year}{2019}) \bibinfo{pages}{A643--A667}.
%Type = Article
\bibitem[{Quaegebeur and Nadarajah(2019)}]{quaegebeur2019ipbr2f}
\bibinfo{author}{S.~Quaegebeur}, \bibinfo{author}{S.~Nadarajah},
\newblock \bibinfo{title}{Stability of energy stable flux reconstruction for
  the diffusion problem using the interior penalty and {B}assi and {R}ebay {II}
  numerical fluxes for linear triangular elements},
\newblock \bibinfo{journal}{J. Comput. Phys.} \bibinfo{volume}{380}
  (\bibinfo{year}{2019}) \bibinfo{pages}{88--118}.
%Type = Article
\bibitem[{Chan(2018)}]{chan2018discretely}
\bibinfo{author}{J.~Chan},
\newblock \bibinfo{title}{On discretely entropy conservative and entropy stable
  discontinuous {G}alerkin methods},
\newblock \bibinfo{journal}{J. Comput. Phys.} \bibinfo{volume}{362}
  (\bibinfo{year}{2018}) \bibinfo{pages}{346--374}.
%Type = Article
\bibitem[{Ismail and Roe(2009)}]{ismail2009affordable}
\bibinfo{author}{F.~Ismail}, \bibinfo{author}{P.~L. Roe},
\newblock \bibinfo{title}{Affordable, entropy-consistent {E}uler flux functions
  {II}: {E}ntropy production at shocks},
\newblock \bibinfo{journal}{J. Comput. Phys.} \bibinfo{volume}{228}
  (\bibinfo{year}{2009}) \bibinfo{pages}{5410--5436}.
%Type = Article
\bibitem[{Hartmann(2008)}]{hartmann2008numerical}
\bibinfo{author}{R.~Hartmann},
\newblock \bibinfo{title}{Numerical analysis of higher order discontinuous
  galerkin finite element methods}  (\bibinfo{year}{2008}).
%Type = Article
\bibitem[{Shu and Osher(1988)}]{shu1988efficient}
\bibinfo{author}{C.-W. Shu}, \bibinfo{author}{S.~Osher},
\newblock \bibinfo{title}{Efficient implementation of essentially
  non-oscillatory shock-capturing schemes},
\newblock \bibinfo{journal}{J. Comput. Phys.} \bibinfo{volume}{77}
  (\bibinfo{year}{1988}) \bibinfo{pages}{439--471}.
%Type = Article
\bibitem[{Cockburn and Shu(1998)}]{cockburn1998runge}
\bibinfo{author}{B.~Cockburn}, \bibinfo{author}{C.-W. Shu},
\newblock \bibinfo{title}{The {R}unge--{K}utta discontinuous {G}alerkin method
  for conservation laws {V}: multidimensional systems},
\newblock \bibinfo{journal}{J. Comput. Phys.} \bibinfo{volume}{141}
  (\bibinfo{year}{1998}) \bibinfo{pages}{199--224}.
%Type = Inproceedings
\bibitem[{Roe and Pike(1985)}]{roe1985efficient}
\bibinfo{author}{P.~Roe}, \bibinfo{author}{J.~Pike},
\newblock \bibinfo{title}{Efficient construction and utilisation of approximate
  {R}iemann solutions},
\newblock in: \bibinfo{booktitle}{Proc. of the sixth int'l. symposium on
  Computing methods in applied sciences and engineering, VI},
  \bibinfo{year}{1985}, pp. \bibinfo{pages}{499--518}.
%Type = Book
\bibitem[{Toro(2013)}]{toro2013riemann}
\bibinfo{author}{E.~F. Toro}, \bibinfo{title}{Riemann solvers and numerical
  methods for fluid dynamics: a practical introduction},
  \bibinfo{publisher}{Springer Science \& Business Media},
  \bibinfo{year}{2013}.
%Type = Article
\bibitem[{O{\ss}wald et~al.(2016)O{\ss}wald, Siegmund, Birken, Hannemann, and
  Meister}]{osswald2016l2roe}
\bibinfo{author}{K.~O{\ss}wald}, \bibinfo{author}{A.~Siegmund},
  \bibinfo{author}{P.~Birken}, \bibinfo{author}{V.~Hannemann},
  \bibinfo{author}{A.~Meister},
\newblock \bibinfo{title}{{L2R}oe: a low dissipation version of {R}oe's
  approximate {R}iemann solver for low mach numbers},
\newblock \bibinfo{journal}{Int. J. Numer. Meth. Fl.} \bibinfo{volume}{81}
  (\bibinfo{year}{2016}) \bibinfo{pages}{71--86}.
%Type = Article
\bibitem[{Beck et~al.(2014)Beck, Bolemann, Flad, Frank, Gassner, Hindenlang,
  and Munz}]{beck2014high}
\bibinfo{author}{A.~D. Beck}, \bibinfo{author}{T.~Bolemann},
  \bibinfo{author}{D.~Flad}, \bibinfo{author}{H.~Frank}, \bibinfo{author}{G.~J.
  Gassner}, \bibinfo{author}{F.~Hindenlang}, \bibinfo{author}{C.-D. Munz},
\newblock \bibinfo{title}{High-order discontinuous {G}alerkin spectral element
  methods for transitional and turbulent flow simulations},
\newblock \bibinfo{journal}{Int. J. Numer. Meth. Fl.} \bibinfo{volume}{76}
  (\bibinfo{year}{2014}) \bibinfo{pages}{522--548}.
%Type = Article
\bibitem[{Van~Rees et~al.(2011)Van~Rees, Leonard, Pullin, and
  Koumoutsakos}]{van2011comparison}
\bibinfo{author}{W.~M. Van~Rees}, \bibinfo{author}{A.~Leonard},
  \bibinfo{author}{D.~I. Pullin}, \bibinfo{author}{P.~Koumoutsakos},
\newblock \bibinfo{title}{A comparison of vortex and pseudo-spectral methods
  for the simulation of periodic vortical flows at high {R}eynolds numbers},
\newblock \bibinfo{journal}{J. Comput. Phys.} \bibinfo{volume}{230}
  (\bibinfo{year}{2011}) \bibinfo{pages}{2794--2805}.
%Type = Article
\bibitem[{Manzanero et~al.(2018)Manzanero, Ferrer, Rubio, and
  Valero}]{manzanero2018role}
\bibinfo{author}{J.~Manzanero}, \bibinfo{author}{E.~Ferrer},
  \bibinfo{author}{G.~Rubio}, \bibinfo{author}{E.~Valero},
\newblock \bibinfo{title}{On the role of numerical dissipation in stabilising
  under-resolved turbulent simulations using discontinuous {G}alerkin methods},
\newblock \bibinfo{journal}{arXiv preprint arXiv:1805.10519}
  (\bibinfo{year}{2018}).
%Type = Article
\bibitem[{Saad et~al.(2017)Saad, Cline, Stoll, and
  Sutherland}]{saad2017scalable}
\bibinfo{author}{T.~Saad}, \bibinfo{author}{D.~Cline},
  \bibinfo{author}{R.~Stoll}, \bibinfo{author}{J.~C. Sutherland},
\newblock \bibinfo{title}{Scalable tools for generating synthetic isotropic
  turbulence with arbitrary spectra},
\newblock \bibinfo{journal}{AIAA journal} \bibinfo{volume}{55}
  (\bibinfo{year}{2017}) \bibinfo{pages}{327--331}.
%Type = Article
\bibitem[{Dairay et~al.(2017)Dairay, Lamballais, Laizet, and
  Vassilicos}]{dairay2017numerical}
\bibinfo{author}{T.~Dairay}, \bibinfo{author}{E.~Lamballais},
  \bibinfo{author}{S.~Laizet}, \bibinfo{author}{J.~C. Vassilicos},
\newblock \bibinfo{title}{Numerical dissipation vs. subgrid-scale modelling for
  large eddy simulation},
\newblock \bibinfo{journal}{J. Comput. Phys.} \bibinfo{volume}{337}
  (\bibinfo{year}{2017}) \bibinfo{pages}{252--274}.
%Type = Inproceedings
\bibitem[{DeBonis(2013)}]{debonis2013solutions}
\bibinfo{author}{J.~DeBonis},
\newblock \bibinfo{title}{Solutions of the {T}aylor-{G}reen vortex problem
  using high-resolution explicit finite difference methods},
\newblock in: \bibinfo{booktitle}{51st AIAA aerospace sciences meeting
  including the new horizons forum and aerospace exposition},
  \bibinfo{year}{2013}, p. \bibinfo{pages}{382}.
%Type = Inproceedings
\bibitem[{Chapelier et~al.(2012)Chapelier, De~La Llave~Plata, and
  Renac}]{chapelier2012inviscid}
\bibinfo{author}{J.-B. Chapelier}, \bibinfo{author}{M.~De~La Llave~Plata},
  \bibinfo{author}{F.~Renac},
\newblock \bibinfo{title}{Inviscid and viscous simulations of the
  {T}aylor-{G}reen vortex flow using a modal discontinuous {G}alerkin
  approach},
\newblock in: \bibinfo{booktitle}{42nd AIAA Fluid Dynamics Conference and
  Exhibit}, \bibinfo{year}{2012}, p. \bibinfo{pages}{3073}.
%Type = Incollection
\bibitem[{Carpenter et~al.(2016)Carpenter, Fisher, Nielsen, Parsani, Sv{\"a}rd,
  and Yamaleev}]{carpenter2016entropy}
\bibinfo{author}{M.~Carpenter}, \bibinfo{author}{T.~Fisher},
  \bibinfo{author}{E.~Nielsen}, \bibinfo{author}{M.~Parsani},
  \bibinfo{author}{M.~Sv{\"a}rd}, \bibinfo{author}{N.~Yamaleev},
\newblock \bibinfo{title}{Entropy stable summation-by-parts formulations for
  compressible computational fluid dynamics},
\newblock in: \bibinfo{booktitle}{Handbook of Numerical Analysis},
  volume~\bibinfo{volume}{17}, \bibinfo{publisher}{Elsevier},
  \bibinfo{year}{2016}, pp. \bibinfo{pages}{495--524}.
%Type = Article
\bibitem[{Kennedy and Gruber(2008)}]{kennedy2008reduced}
\bibinfo{author}{C.~A. Kennedy}, \bibinfo{author}{A.~Gruber},
\newblock \bibinfo{title}{Reduced aliasing formulations of the convective terms
  within the {N}avier-{S}tokes equations for a compressible fluid},
\newblock \bibinfo{journal}{J. Comput. Phys.} \bibinfo{volume}{227}
  (\bibinfo{year}{2008}) \bibinfo{pages}{1676--1700}.
%Type = Article
\bibitem[{Chandrashekar(2013)}]{chandrashekar2013kinetic}
\bibinfo{author}{P.~Chandrashekar},
\newblock \bibinfo{title}{Kinetic energy preserving and entropy stable finite
  volume schemes for compressible euler and navier-stokes equations},
\newblock \bibinfo{journal}{Commun. Comput. Phys.} \bibinfo{volume}{14}
  (\bibinfo{year}{2013}) \bibinfo{pages}{1252--1286}.
%Type = Article
\bibitem[{Ranocha and Gassner(2021)}]{ranocha2021preventing}
\bibinfo{author}{H.~Ranocha}, \bibinfo{author}{G.~J. Gassner},
\newblock \bibinfo{title}{Preventing pressure oscillations does not fix local
  linear stability issues of entropy-based split-form high-order schemes},
\newblock \bibinfo{journal}{Communications on Applied Mathematics and
  Computation}  (\bibinfo{year}{2021}) \bibinfo{pages}{1--24}.
%Type = Article
\bibitem[{Comte-Bellot and Corrsin(1971)}]{comte1971simple}
\bibinfo{author}{G.~Comte-Bellot}, \bibinfo{author}{S.~Corrsin},
\newblock \bibinfo{title}{Simple eulerian time correlation of full-and
  narrow-band velocity signals in grid-generated,‘isotropic’turbulence},
\newblock \bibinfo{journal}{J. Fluid Mech.} \bibinfo{volume}{48}
  (\bibinfo{year}{1971}) \bibinfo{pages}{273--337}.
%Type = Article
\bibitem[{Jefferson-Loveday and Tucker(2010)}]{jefferson2010impingement}
\bibinfo{author}{R.~Jefferson-Loveday}, \bibinfo{author}{P.~Tucker},
\newblock \bibinfo{title}{{LES} of impingement heat transfer on a concave
  surface},
\newblock \bibinfo{journal}{Numerical Heat Transfer, Part A: Applications}
  \bibinfo{volume}{58} (\bibinfo{year}{2010}) \bibinfo{pages}{247--271}.
%Type = Incollection
\bibitem[{Hickel et~al.(2009)Hickel, Devesa, and Adams}]{hickel2009implicit}
\bibinfo{author}{S.~Hickel}, \bibinfo{author}{A.~Devesa},
  \bibinfo{author}{N.~A. Adams},
\newblock \bibinfo{title}{Implicit turbulence modeling by finite volume
  methods},
\newblock in: \bibinfo{booktitle}{Numerical Simulation of Turbulent Flows and
  Noise Generation: Results of the DFG/CNRS Research Groups FOR 507 and FOR
  508}, \bibinfo{publisher}{Springer}, \bibinfo{year}{2009}, pp.
  \bibinfo{pages}{149--173}.
%Type = Article
\bibitem[{Misra and Lund(1996)}]{misra1996evaluation}
\bibinfo{author}{A.~Misra}, \bibinfo{author}{T.~S. Lund},
\newblock \bibinfo{title}{Evaluation of a vortex-based subgrid stress model
  using {DNS} databases},
\newblock \bibinfo{journal}{Center for Turbulence Research}
  (\bibinfo{year}{1996}) \bibinfo{pages}{359--368}.

\end{thebibliography}

\newpage
\appendix
\section{Derivation of viscous dissipation terms}
\subsection{Compressible Flow}\label{appendix:derivation_viscous_dissipation_rate_terms_compressible}
To prove Eq.(\ref{eq:viscous_dissipation_strain_rate_terms}),
we substitute Eq.(\ref{eq:dimensionless_viscous_stress_tensor}) for the dimensionless viscous stress-tensor $\tau_{ij}$, omit the asterisks for clarity:
\begin{align*}
    \tau_{ij}\pdv{\text{v}_{j}}{x_{i}} & = \frac{2\mu}{\text{Re}_{\infty}}\left(S_{ij}-\frac{1}{3}S_{kk}\delta_{ij}\right)\pdv{\text{v}_{j}}{x_{i}} \\ 
    & = \frac{\mu}{\text{Re}_{\infty}}\left(\pdv{\text{v}_{i}}{x_{j}}+\pdv{\text{v}_{j}}{x_{i}}-\frac{2}{3}\pdv{\text{v}_{k}}{x_{k}}\delta_{ij}\right)\pdv{\text{v}_{j}}{x_{i}} \\
    & = \frac{\mu}{\text{Re}_{\infty}}\left(\left(\pdv{\text{v}_{i}}{x_{j}}+\pdv{\text{v}_{j}}{x_{i}}\right)\pdv{\text{v}_{j}}{x_{i}}-\left(\frac{2}{3}\pdv{\text{v}_{k}}{x_{k}}\delta_{ij}\right)\pdv{\text{v}_{j}}{x_{i}}\right) \\
    & = \frac{\mu}{\text{Re}_{\infty}}\left(\frac{1}{2}\left(\pdv{\text{v}_{i}}{x_{j}}\pdv{\text{v}_{i}}{x_{j}}+2\pdv{\text{v}_{i}}{x_{j}}\pdv{\text{v}_{j}}{x_{i}}+\pdv{\text{v}_{j}}{x_{i}}\pdv{\text{v}_{j}}{x_{i}}\right)-\frac{2}{3}\left(\pdv{\text{v}_{k}}{x_{k}}\right)^{2}\right) \\
    & = \frac{\mu}{\text{Re}_{\infty}}\left(\frac{1}{2}\left(\pdv{\text{v}_{i}}{x_{j}}+\pdv{\text{v}_{j}}{x_{i}}\right)\left(\pdv{\text{v}_{i}}{x_{j}}+\pdv{\text{v}_{j}}{x_{i}}\right)-\frac{2}{3}\left(\pdv{\text{v}_{k}}{x_{k}}\right)^{2}\right) \\
    & = \frac{\mu}{\text{Re}_{\infty}}\left(\frac{1}{2}\left(2S_{ij}\right)\left(2S_{ij}\right)-\frac{2}{3}\left(\pdv{\text{v}_{k}}{x_{k}}\right)^{2}\right) \\ 
    & = \frac{\mu}{\text{Re}_{\infty}}\left(2S_{ij}S_{ij}-\frac{2}{3}\left(\pdv{\text{v}_{k}}{x_{k}}\right)^{2}\right).
\end{align*}
Furthermore, to prove Eq.(\ref{eq:viscous_dissipation_deviatoric_strain_rate_terms}), we can write:
\begin{align*}
    \frac{2\mu}{\text{Re}_{\infty}}S_{ij}^{d}S_{ij}^{d} & = \frac{2\mu}{\text{Re}_{\infty}}\left(S_{ij}-\frac{1}{3}S_{kk}\delta_{ij}\right)\left(S_{ij}-\frac{1}{3}S_{kk}\delta_{ij}\right) \\ 
    & = \frac{2\mu}{\text{Re}_{\infty}}\left(\frac{1}{4}\left(\pdv{\text{v}_{i}}{x_{j}}+\pdv{\text{v}_{j}}{x_{i}}-\frac{2}{3}\pdv{\text{v}_{k}}{x_{k}}\delta_{ij}\right)\left(\pdv{\text{v}_{i}}{x_{j}}+\pdv{\text{v}_{j}}{x_{i}}-\frac{2}{3}\pdv{\text{v}_{k}}{x_{k}}\delta_{ij}\right)\right) \\
    & = \frac{\mu}{2\text{Re}_{\infty}}\left(\pdv{\text{v}_{i}}{x_{j}}\pdv{\text{v}_{i}}{x_{j}}+2\pdv{\text{v}_{i}}{x_{j}}\pdv{\text{v}_{j}}{x_{i}}+\pdv{\text{v}_{j}}{x_{i}}\pdv{\text{v}_{j}}{x_{i}}-\frac{4}{3}\pdv{\text{v}_{k}}{x_{k}}\delta_{ij}\pdv{\text{v}_{i}}{x_{j}}
    -\frac{4}{3}\pdv{\text{v}_{k}}{x_{k}}\delta_{ij}\pdv{\text{v}_{j}}{x_{i}}
    +\frac{4}{9}\left(\pdv{\text{v}_{k}}{x_{k}}\right)^{2}\delta_{ij}\delta_{ij}\right) \\
    & = \frac{\mu}{2\text{Re}_{\infty}}\left(\pdv{\text{v}_{i}}{x_{j}}\pdv{\text{v}_{i}}{x_{j}}+2\pdv{\text{v}_{i}}{x_{j}}\pdv{\text{v}_{j}}{x_{i}}+\pdv{\text{v}_{j}}{x_{i}}\pdv{\text{v}_{j}}{x_{i}}-\frac{8}{3}\left(\pdv{\text{v}_{k}}{x_{k}}\right)^{2}
    +\frac{4}{9}\left(\pdv{\text{v}_{k}}{x_{k}}\right)^{2}\delta_{ij}\delta_{ij}\right)
\end{align*}
For 3D, $\delta_{ij}\delta_{ij}=3$, thus:
\begin{align*}
    \frac{2\mu}{\text{Re}_{\infty}}S_{ij}^{d}S_{ij}^{d} & = \frac{\mu}{2\text{Re}_{\infty}}\left(\pdv{\text{v}_{i}}{x_{j}}\pdv{\text{v}_{i}}{x_{j}}+2\pdv{\text{v}_{i}}{x_{j}}\pdv{\text{v}_{j}}{x_{i}}+\pdv{\text{v}_{j}}{x_{i}}\pdv{\text{v}_{j}}{x_{i}}-\frac{4}{3}\left(\pdv{\text{v}_{k}}{x_{k}}\right)^{2}\right) \\
    & = \frac{\mu}{\text{Re}_{\infty}}\left(\frac{1}{2}\left(\pdv{\text{v}_{i}}{x_{j}}+\pdv{\text{v}_{j}}{x_{i}}\right)\left(\pdv{\text{v}_{i}}{x_{j}}+\pdv{\text{v}_{j}}{x_{i}}\right)-\frac{2}{3}\left(\pdv{\text{v}_{k}}{x_{k}}\right)^{2}\right) \\
    & = \frac{\mu}{\text{Re}_{\infty}}\left(\frac{1}{2}\left(2S_{ij}\right)\left(2S_{ij}\right)-\frac{2}{3}\left(\pdv{\text{v}_{k}}{x_{k}}\right)^{2}\right) \\
    & = \frac{\mu}{\text{Re}_{\infty}}\left(2S_{ij}S_{ij}-\frac{2}{3}\left(\pdv{\text{v}_{k}}{x_{k}}\right)^{2}\right) \\
    & = \tau_{ij}\pdv{\text{v}_{j}}{x_{i}}
\end{align*}
\subsection{Incompressible Flow}\label{appendix:derivation_viscous_dissipation_rate_terms_incompressible}
The dot product of the vorticity can be written as:
\begin{align*}
    \omega_{k}\omega_{k} & = \varepsilon_{ijk}\pdv{\text{v}_{j}}{x_{i}}\varepsilon_{pqk}\pdv{\text{v}_{q}}{x_{p}} \\
    & = \varepsilon_{ijk}\varepsilon_{pqk}\pdv{\text{v}_{j}}{x_{i}}\pdv{\text{v}_{q}}{x_{p}} \\
    & = \varepsilon_{kij}\varepsilon_{kpq}\pdv{\text{v}_{j}}{x_{i}}\pdv{\text{v}_{q}}{x_{p}} \\
    & = \left(\delta_{ip}\delta_{jq}-\delta_{iq}\delta_{jp}\right)\pdv{\text{v}_{j}}{x_{i}}\pdv{\text{v}_{q}}{x_{p}} \\
    & = \pdv{\text{v}_{j}}{x_{i}}\pdv{\text{v}_{j}}{x_{i}} - \pdv{\text{v}_{j}}{x_{i}}\pdv{\text{v}_{i}}{x_{j}}\\
    & = \pdv{\text{v}_{j}}{x_{i}}\left(\pdv{\text{v}_{j}}{x_{i}} - \pdv{\text{v}_{i}}{x_{j}}\right).
\end{align*}
Twice the magnitude of the strain tensor can be written as:
\begin{align*}
    2S_{ij}S_{ij} & = 2\left[\frac{1}{2}\left(\pdv{\text{v}_{i}}{x_{j}}+\pdv{\text{v}_{j}}{x_{i}}\right)\right]\left[\frac{1}{2}\left(\pdv{\text{v}_{i}}{x_{j}}+\pdv{\text{v}_{j}}{x_{i}}\right)\right] \\
    & = \frac{1}{2}\left(\pdv{\text{v}_{i}}{x_{j}}\pdv{\text{v}_{i}}{x_{j}}+2\pdv{\text{v}_{i}}{x_{j}}\pdv{\text{v}_{j}}{x_{i}}+\pdv{\text{v}_{j}}{x_{i}}\pdv{\text{v}_{j}}{x_{i}}\right) \\
    & = \frac{1}{2}\left(2\pdv{\text{v}_{i}}{x_{j}}\pdv{\text{v}_{i}}{x_{j}}+2\pdv{\text{v}_{i}}{x_{j}}\pdv{\text{v}_{j}}{x_{i}}\right) \\
    & = \pdv{\text{v}_{j}}{x_{i}}\left(\pdv{\text{v}_{i}}{x_{j}}+\pdv{\text{v}_{j}}{x_{i}}\right) \\
    & = \pdv{\text{v}_{j}}{x_{i}}\left(\pdv{\text{v}_{j}}{x_{i}}-\pdv{\text{v}_{i}}{x_{j}}\right) + 2\pdv{\text{v}_{j}}{x_{i}}\pdv{\text{v}_{i}}{x_{j}} \\
    & = \omega_{k}\omega_{k} + 2\pdv{\text{v}_{j}}{x_{i}}\pdv{\text{v}_{i}}{x_{j}} \\ 
    & = \omega_{k}\omega_{k} + 2\pdv{}{x_{j}}\left(\text{v}_{i}\pdv{\text{v}_{j}}{x_{i}}\right)
\end{align*}

\end{document}